\documentclass[a4paper,11pt]{article}
\usepackage{jheppub}
\pdfoutput=1
\usepackage[utf8]{inputenc}
\usepackage[T1]{fontenc}
\usepackage{braket}
\usepackage{graphicx}
\usepackage{longtable}
\usepackage{epsfig}
\usepackage{amsmath}
\usepackage{slashed}
\usepackage{amsfonts} 
\usepackage{float} 
\usepackage{amssymb}
\usepackage[table]{xcolor} 
\usepackage{hhline}
\usepackage{pbox}
\usepackage{subfigure}
\usepackage{enumitem}
\usepackage{array, multirow,makecell}
\usepackage{hyperref}
\usepackage{tabularx}
\usepackage{titlesec}
\usepackage{mathrsfs,mathtools}
\usepackage[normalem]{ulem}
\usepackage{natbib}
\usepackage{array,multirow,makecell}
\usepackage{lipsum}
\usepackage{comment}
\usepackage[section]{placeins}
\usepackage{cancel}
\usepackage[normalem]{ulem}
\usepackage{bm}
\usepackage{booktabs}
\usepackage{indentfirst}
\usepackage{orcidlink}
\usepackage[usestackEOL]{stackengine}
\strutlongstacks{T}

\def\l{\lambda}

\def\th{\theta}

\newcommand{\ba}{\begin{array}}
	\newcommand{\ea}{\end{array}}
\newcommand{\ovr}{\overline}

\def\lsim{\raise0.3ex\hbox{$\;<$\kern-0.75em\raise-1.1ex\hbox{$\sim\;$}}}
\def\gsim{\raise0.3ex\hbox{$\;>$\kern-0.75em\raise-1.1ex\hbox{$\sim\;$}}}
\def\be{\begin{equation}}
	\def\ee{\end{equation}}
\def\bea{\begin{eqnarray}}
	\def\eea{\end{eqnarray}}
\def\nn{\nonumber}
\newcommand{\beq}{\begin{equation}}
	\newcommand{\eeq}{\end{equation}}
\newcommand{\beqn}{\begin{eqnarray}}
	\newcommand{\eeqn}{\end{eqnarray}}

\def\lsim{\raise0.3ex\hbox{$\;<$\kern-0.75em\raise-1.1ex\hbox{$\sim\;$}}}
\def\gsim{\raise0.3ex\hbox{$\;>$\kern-0.75em\raise-1.1ex\hbox{$\sim\;$}}}
\def\be{\begin{equation}}
	\def\ee{\end{equation}}
\def\bea{\begin{eqnarray}}
	\def\eea{\end{eqnarray}}
\def\nn{\nonumber}
\title{
	Unveiling the inert Triplet desert region with a pNGB Dark Matter and its Gravitational Wave signatures	}
\author{Pankaj Borah\orcidlink{0000-0003-2715-271X},}
\author{Pradipta Ghosh\orcidlink{0000-0002-6560-4556}}
\affiliation{Department of Physics, Indian Institute of Technology Delhi, Hauz Khas, New Delhi 110 016, India}
\emailAdd{Pankaj.Borah@physics.iitd.ac.in}
\emailAdd{tphyspg@physics.iitd.ac.in}
\abstract{
In this work, we extend the scalar sector of the conventional hyperchargeless inert triplet model (ITM) to include a second dark matter (DM) candidate, which appears to be a pseudo-Nambu-Goldstone boson (pNGB). The usual ITM with an extended scalar sector offers a DM candidate along with novel signatures at different experiments, e.g., colliders, gravitational wave detectors, etc. Nevertheless, hitherto unseen experimental detections have placed stringent constraints on the ITM parameter space. Moreover, triplet masses lighter than $1.9$ TeV, consistent with the existing or upcoming collider sensitivity reach, are already excluded from the DM observable, as they yield an underabundant relic density due to a strong $SU(2)_L$ gauge annihilation. Inclusion of a pNGB DM, via a complex $SU(2)_L$ scalar singlet and through the soft-breaking of a $U(1)$ symmetry, helps to revive the sub-TeV regime of the triplet DM. This resurgence relies on a proficient conversion between the two DM species. Using this inter-conversion, with the triplet DM as the lighter one between the two, we show that it is possible to push the triplet DM contribution to $50\%  - 60\%$ of the total relic density. This offers a significant improvement over the traditional ITM with a single DM candidate, where the same can at most reach $10\% - 20\%$. Besides, the concerned bipartite DM framework also offers the possibility of a first-order phase transition along various constituent field directions. Among these, the one along the real $SU(2)_L$ singlet direction can be a strong one which subsequently yields detectable gravitational wave signals at the upcoming space-based gravitational wave detectors such as LISA, BBO, DECIGO, etc., alongside distinctive and complementary signatures at the various DM and collider quests.
}
\begin{document} 
	\maketitle	
\section{Introduction}
The Standard Model (SM) of particle physics, despite being a successful and well-tested theory, suffers from several shortcomings. Chief among them, perhaps, is the absence of a viable particle Dark Matter (DM) candidate and the inability to explain the Baryon Asymmetry of the Universe (BAU) \cite{Steigman:1976ev,Cohen:1997ac}. The existence of DM is supported by multiple observations, including the flatness of galaxy rotation curves \cite{Zwicky:1933gu,Rubin:1970zza}, observation of the Bullet Cluster \cite{Clowe:2006eq}, and precise measurements of the DM relic density by the Planck \cite{Planck:2018vyg} and WMAP \cite{WMAP:2012fli,WMAP:2012nax} satellites. Likewise, the baryon density of the Universe has been determined with remarkable accuracy \cite{Planck:2018vyg}.

Extending the SM framework is essential to incorporate an experimentally viable DM candidate. A plausible but minimal particle physics motivated DM model involves a Weakly Interacting Massive Particle (WIMP) \cite{Steigman:1984ac,Jungman:1995df,Kolb:1990vq}, which interacts with the SM particles through the weak force, see Refs.~\cite{Arcadi:2017kky,Arcadi:2024ukq} for extensive reviews. However, despite extensive experimental efforts, both direct detection (DD) experiments, such as XENON1T \cite{XENON:2019gfn}, LUX-ZEPLIN (LZ) \cite{LZ:2022lsv,LZCollaboration:2024lux}, DARWIN \cite{DARWIN:2016hyl}, etc., and indirect detection (ID) searches, such as FERMI-LAT \cite{Abdo:2010ex,Fermi-LAT:2011vow,Ackermann:2012nb,Abramowski:2012au,Ackermann:2013uma,Ackermann:2013yva,Ackermann:2015tah,Fermi-LAT:2015att,Fermi-LAT:2015kyq,Fermi-LAT:2016uux},  AMS \cite{AMS:2002yni,AMS:2021nhj},  H.E.S.S \cite{Abramowski:2011hc,Abramowski:2014tra,HESS:2015cda,Abdallah:2016ygi,Abdalla:2016olq}, etc., have yet to yield any conclusive evidence for the WIMP-like DM. These null results place strong constraints on many otherwise appealing beyond the SM (BSM) scenarios with a viable DM candidate. Although the absence of a detection of the DM particles in the DD and the ID previous experiments has not eliminated all possibilities for a single particle DM model, however, it increases the possibility of the existence of a more complex DM sector, similar to the visible sector that consists of multiple types of elementary particles. Additionally, introducing multipartite DM \cite{Cao:2007fy} frameworks can be advantageous in evading the strong DD and ID bounds, as the detection cross-sections are normalised \cite{Cao:2007fy,Aoki:2012ub} by the relative abundances of each component in such cases.

Separately, to account for the observed BAU, Electroweak Baryogenesis (EWBG)\footnote{Interested readers may look at Refs. \cite{Cohen:1993nk,Trodden:1998ym,Riotto:1998bt,Riotto:1999yt,Quiros:1999jp,Dine:2003ax,Cline:2006ts,Morrissey:2012db} for detailed review of the EWBG.} remains a compelling mechanism. This framework requires a first-order electroweak phase transition (EWPT) (see, e.g., Refs.~\cite{Mazumdar:2018dfl,Athron:2023xlk} for a review), which leads to electroweak symmetry breaking (EWSB) and plays a central role in this context. Particularly, the EWPT can provide sufficient out-of-equilibrium conditions to satisfy one of the three Sakharov criteria \cite{Sakharov:1967dj} to generate the baryon asymmetry.

The hyperchargeless ($Y=0$) scalar $SU(2)_L$ inert triplet model (ITM) is a well-studied BSM scenario that naturally accommodates a DM candidate \cite{Ross:1975fq,Cheng:1980qt,Gunion:1989ci,Cirelli:2005uq,Cirelli:2007xd,FileviezPerez:2008bj,Araki:2011hm}. However, due to its sizable $SU(2)_L$ gauge interactions, it faces strong constraints, creating the so-called ``desert region'' \cite{Cirelli:2005uq,Hambye:2009pw} where DM masses up to $\sim 1.9$ TeV \cite{Cirelli:2005uq, Fischer:2011zz,Araki:2011hm, Khan:2016sxm} are excluded due to underabundance. The ITM can also support a successful EWPT, but the relevant parameter space for such transitions typically requires triplet masses below 250 GeV \cite{Niemi:2020hto}, if perturbativity of the scalar quartic couplings is maintained up to a high scale, e.g., the Planck scale. Consequently, achieving simultaneously achieve a viable DM candidate and a successful EWPT within the ITM framework becomes futile. This motivates extending the ITM framework to include multiple DM components in order to accommodate unify both goals concurrently within a single particle physics framework.

Recently, pseudo-Nambu-Goldstone boson (pNGB) \cite{Nambu:1960tm,Nambu:1961tp,Goldstone:1961eq,Goldstone:1962es} DM models \cite{Silveira:1985rk,Barger:2008jx} have gained interest due to their naturally suppressed DD cross-sections, arising from reduced scattering rates at low momentum transfer. This feature was first discussed in Ref.~\cite{Gross:2017dan}, where a cancellation mechanism in Higgs portal was identified, leading to a vanishing tree-level DM-nucleus scattering cross-section in the zero momentum transfer limit. Therefore, it can address the non-observance of DM signals at the DD experiments. The minimal pNGB DM setup (see some recent works, for e.g., Refs. \cite{Gross:2017dan,Huitu:2018gbc,Alanne:2018zjm,Karamitros:2019ewv,Cline:2019okt,Arina:2019tib}) consists of an $SU(2)_L$ complex scalar singlet ($S$) alongside the SM Higgs doublet ($H$) \footnote{Also, see e.g., Ref.~\cite{DiazSaez:2023wli}, in the context of a two-component DM setup.}. In this class of models, a global $U(1)$ symmetry is softly broken to a $\mathbb{Z}_2$ by a DM mass term \cite{Gross:2017dan}, ensuring the DM stability. On the cosmological frontier, such models exhibit EWPT, which are predominantly of second order \cite{Kannike:2019wsn}. However, if the global $U(1)$ symmetry is broken instead to $\mathbb{Z}_3$ \cite{Kannike:2019mzk}, the induced cubic terms can give rise to strong first-order phase transitions (SFOPT)\footnote{See some other plausible non-minimal extensions accommodating an SFOPT, e.g., \cite{Alanne:2020jwx,Ghosh:2024ing}.} in parts of the parameter space, potentially producing gravitational wave (GW)\footnote{See Ref.~\cite{Athron:2023xlk} for an extensive review.} signals detectable at experiments like LISA \cite{eLISA:2013xep,LISA:2017pwj}, BBO \cite{Crowder:2005nr,Corbin:2005ny,Harry:2006fi} or DECIGO \cite{Kudoh:2005as, Yagi:2011wg,Kawamura:2020pcg}. Importantly, in the $\mathbb{Z}_2$-symmetric case, the tree-level DD cross-section vanishes in the zero-momentum-transfer limit \cite{Gross:2017dan}. This does not necessarily hold in general \cite{Alanne:2020jwx}, and loop-level effects can also introduce non-zero contributions \cite{Azevedo:2018exj,Ishiwata:2018sdi,Alanne:2020jwx}. However, such contributions are found to be mild and remain practically unaffected \cite{Azevedo:2018exj,Ishiwata:2018sdi,Kannike:2019mzk,Alanne:2020jwx}.

The preceding discussions make limitations of a $Y=0$ ITM DM framework and the advantage of a pNGB DM model in evading DD bounds very apparent. With these facts, in this work, we propose a minimalistic two-component DM framework by extending the $Y=0$ ITM setup with a complex scalar $SU(2)_L$ singlet $S$, where a global $U(1)$ symmetry is explicitly broken by a $\mathbb{Z}_3$-symmetric term in $S$. The $\mathbb{Z}_3$ symmetry is then spontaneously broken, rendering the imaginary part of $S$, denoted by $\chi$, as the other stable pNGB DM candidate. Both the DM components interact with each other and with the SM particles via Higgs portal, enabled through the mixing between $S$ and the SM Higgs doublet $H$. Notably, the interaction between the two DM species also induces inter-conversion \cite{Liu:2011aa,Belanger:2011ww} between the two DM components, significantly influencing their relic density under certain assumptions. We find that, in the case when the pNGB is the heavier DM candidate, the DM-DM conversion can enhance the scalar triplet DM relic contribution to the Planck measured relic density up to $50-60\%$ within the sub-TeV mass range of the triplet DM. This offers a substantial improvement over the pure $Y=0$ ITM scenario, where the sub-TeV triplet typically contributes only $10-20\%$ of the observed relic density \cite{Cirelli:2005uq,Araki:2011hm, Khan:2016sxm}.

Given the extended scalar sector, we also explore the EWPT and associated GW signals in the early Universe. We observe that the PTs along the SM Higgs direction, in the parameter space compatible with the DM phenomenology, are generally too weak to yield observable GW signatures. On the other hand, the triplet scalar being one of the two DM candidates, it does not acquire a vacuum expectation value (VEV) at zero temperature. Further, we do not consider any transitions along the triplet field direction at non-zero temperatures, although it can still affect the effective potential through loop-level contributions, as do the other particles (including the pNGB DM). As a result, our analysis focuses only on the PT along the direction of the charge-parity ($CP$)-even component of $S$, denoted by $s$. We identify parameter regions where an SFOPT occurs along $s$, yield GW signals detectable by future interferometers, while maintaining consistency with the DM relic abundance, the spin-independent (SI) DD limits from experiments like the XENON1T \cite{XENON:2019gfn}, LZ-2022 \cite{LZ:2022lsv}, LZ-2024 \cite{LZCollaboration:2024lux} and DARWIN (projected) \cite{DARWIN:2016hyl}, ID limits from FERMI-LAT \cite{Fermi-LAT:2016uux,Fermi-LAT:2015kyq,Fermi-LAT:2015att,Ackermann:2015tah,Ackermann:2013yva,Ackermann:2013uma,Abramowski:2012au,Ackermann:2012nb,Fermi-LAT:2011vow,Abdo:2010ex}, AMS \cite{AMS:2021nhj,AMS:2002yni}, H.E.S.S \cite{Abdalla:2016olq,Abdallah:2016ygi,HESS:2015cda,Abramowski:2014tra,Abramowski:2011hc}, etc., and collider searches, besides the other relevant theoretical and experimental constraints. Finally, we assess the detectability of the resulting GW signals using both the conventional power-law integrated (PLI) \cite{Thrane:2013oya} sensitivity curves and the recently proposed peak-integrated sensitivity curves (PISCs) \cite{Alanne:2019bsm,Schmitz:2020syl}. We find that, GWs generated by SFOPTs along the $s$ field direction lie within the reach of upcoming GW detectors such as LISA \cite{eLISA:2013xep,LISA:2017pwj}, BBO \cite{Crowder:2005nr,Corbin:2005ny,Harry:2006fi} and DECIGO \cite{Kudoh:2005as, Yagi:2011wg,Kawamura:2020pcg}. Our analyses highlights the complementarity of the GW observations with the DM searches and collider experiments.

The paper is organised as follows: in Sec. \ref{sec:the-model}, we introduce the proposed two-component DM framework. A comprehensive account of the relevant theoretical and experimental constraints is presented in Sec. \ref{sec:obs-cons}. The results of our numerical analyses of the dark sector, based on a dedicated parameter scan, are discussed in Sec. \ref{sec:DM-pheno}. Section \ref{sec:PT-GW} explores features of the EWPT and the resulting GW signals, along with their potential interplay with the DM sector. This section also provides a detailed examination of the GW detection prospects, incorporating both the conventional PLI method and the newly proposed improved PISCs analyses. Finally, Sec. \ref{sec:conclu} contains the summary and conclusion of our study. Additional technical details are relegated to Appendix \ref{appx:appendixA}.
\section{The Model} \label{sec:the-model}
In this section, we outline our model, which extends the SM scalar sector by including an $SU(2)_L$ scalar triplet $\bm{T}$ with zero hypercharge ($Y=0$) and a complex $SU(2)_L$ scalar singlet $S$. In this framework, we assume the neutral part of $\bm{T}$ as one of the two DM candidates, while the second DM candidate is a pseudo-Nambu Goldstone boson (pNGB) \cite{Nambu:1960tm,Nambu:1961tp,Goldstone:1961eq,Goldstone:1962es} DM \cite{Silveira:1985rk,Barger:2008jx,Chiang:2017nmu,Gross:2017dan,Huitu:2018gbc,Alanne:2018zjm,Karamitros:2019ewv,Cline:2019okt,Arina:2019tib} derived from $S$.
\subsection{Scalar potential} \label{subsec:scalar-potential}
To begin with, we assume that all the coefficients of the scalar potential are real, ensuring the $CP$ is conserved. Besides, there is a $\mathbb{Z}_2$ symmetry $\bm{T} \to -\bm{T}$, that forbids terms such as $H^{\dagger} {\bm{T}} H$, where $H$ represents the SM $SU(2)_L$ doublet Higgs. Under these assumptions, the admissible interaction terms in the scalar potential involving $H$ and $\bm{T}$ are given by
\bea
\label{eq:scalar-pot:pot-H-T}
V_1 &=& \mu_{H}^2 (H^\dagger H) + \l_{H} (H^\dagger H)^2 + \frac{\mu_{\bm T}^2}{2} {\rm Tr}[\bm{T}^\dagger \bm{T}] + \frac{\l_{\bm T}}{4} \Big({\rm Tr}[\bm{T}^\dagger \bm{T}]\Big)^2\nn \\
&& + \frac{\l_{HT}}{2} (H^{\dagger} H) {\rm Tr}[\bm{T}^{\dagger} \bm{T}].
\eea
Next, we write down the potential terms that involve $S$ and respect a global $U(1)$ symmetry $S \to e^{i \alpha} S$,
\bea
\label{eq:scalar-pot:pot-S}
V_2 &=& \mu_{S}^2 (S^\dagger S) + \l_{S} (S^\dagger S)^2 + \l_{SH} (S^{\dagger} S) (H^\dagger H)  + \frac{\l_{ST}}{2} (S^{\dagger} S) {\rm Tr}[\bm{T}^{\dagger} \bm{T}].
\eea
In addition, we introduce a cubic term in $S$ that explicitly, but softly\footnote{This soft-breaking term is introduced to provide a mass to the pNGB DM. Its ultraviolet (UV) origin is not considered here, however, some examples have been discussed in the literature, e.g., see Refs. \cite{Abe:2020iph,Okada:2020zxo}.}, breaks the aforementioned global $U(1)$ symmetry,
\bea
\label{eq:scalar-pot:soft-pot}
V_{\rm soft} = \frac{\mu_3}{2} \left(S^3 + S^{\dagger 3}\right).
\eea
Therefore, the whole tree-level scalar potential, including Eqs.~(\ref{eq:scalar-pot:pot-H-T}) - (\ref{eq:scalar-pot:soft-pot}), is now given as,
\bea
\label{eq:scalar-pot:tot-pot}
V_0(H, \bm{T}, S) = V_1 + V_2 + V_{\rm soft}.
\eea
Note that, the total tree-level scalar potential of Eq. (\ref{eq:scalar-pot:tot-pot}) also remains invariant under a discrete $\mathbb{Z}_3$ symmetry, where the singlet $S$ transforms as $S \to e^{\frac{2\pi i}{3}} S$, while $H$ and $\bm{T}$ transform as singlets. This symmetry, in fact, prevents any quartic couplings in Eq. (\ref{eq:scalar-pot:tot-pot}) that would lead to a {\it hard} \footnote{While soft breaking introduces symmetry-violating terms that preserve renormalisability and avoid additional UV divergences, hard breaking, in contrast, lead to UV divergences and can spoil renormalisability.} breaking of the $U(1)$, i.e., $S \to e^{i \alpha} S$ \cite{Kannike:2019mzk}. The singlet $S$, in principle, can develop a non-zero vacuum expectation value (VEV) that breaks the aforesaid $U(1)$ (or $\mathbb{Z}_3$) symmetry spontaneously.  After the EWSB, the scalar fields are represented as,
\beq\label{eq:scalar-pot:scalar-field-basis}
H = \frac{1}{\sqrt{2}}
\begin{pmatrix}
	~\sqrt{2} G^{+}~ \\
	v+ ~h + i G^0~
\end{pmatrix},\quad
{\bm{T}} =
\frac{1}{\sqrt{2}}\begin{pmatrix}
	~T^0~ & ~ -\sqrt{2}T^+ ~ \\
	~-\sqrt{2}T^- ~ & ~ -T^0~
\end{pmatrix},\quad
S = \frac{v_S + s + i \chi}{\sqrt{2}}.
\eeq
Here, $h, s, T^0$ are the CP-even scalars, $\chi$ is the CP-odd neutral scalar, $G^+, G^0$ are the charged and neutral Goldstone bosons, and $T^\pm$ are the charged scalars, respectively. The VEVs of $H$ and $S$ are given by $v=246~{\rm GeV}$ \cite{ParticleDataGroup:2024cfk} and $v_S$, respectively. Whereas, $T^0$ is a DM candidate, and so, we assign it a vanishing VEV \cite{Cirelli:2005uq,Cirelli:2007xd}. This, in addition to requiring $\l_{\bm{T}} > 0$ for the potential to remain bounded from below, also necessitates $\mu_{\bm{T}}^2 > 0$ (see Eq.~(\ref{eq:scalar-pot:pot-H-T})). We further assume $\mu_H^2 < 0$, $\lambda_H > 0$ and $\mu_S^2 < 0$, $\lambda_S > 0$ in Eqs.~(\ref{eq:scalar-pot:pot-H-T}) and (\ref{eq:scalar-pot:pot-S}), which ensure that both the $H$ and $S$ fields acquire non-zero VEVs, thereby spontaneously breaking the EW symmetry.

Note that, the potential of Eq. (\ref{eq:scalar-pot:tot-pot}) also exhibits a $\mathbb{Z}_2$-like symmetry $S \to S^\dagger$, just like the original $\mathbb{Z}_2$ pNGB DM model \cite{Gross:2017dan}. This symmetry eventually translates into $\chi \to -\chi$, ensuring the stability of the pNGB DM $\chi$ even when the $\mathbb{Z}_3$ symmetry gets spontaneously broken. However, spontaneously breaking of the $\mathbb{Z}_3$ symmetry can lead to undesirable domain wall problems \cite{Zeldovich:1974uw,Kibble:1976sj,Kibble:1980mv, Abel:1995wk,Friedland:2002qs}. To address this, one can introduce a small explicit $\mathbb{Z}_3$ breaking term, e.g., linear in $S$. Since such a term does not affect the DM phenomenology \cite{Kannike:2019mzk}, we exclude it from the potential in Eq. (\ref{eq:scalar-pot:tot-pot}) for a simplified analysis.
\subsection{Masses and mixing in the scalar sector} \label{subsec:masses-mixings}
Given that, in general, $v_S \neq 0$ (see Eq.~(\ref{eq:scalar-pot:scalar-field-basis})), the $\l_{SH}$ term in Eq.~(\ref{eq:scalar-pot:pot-S}) leads to a mixing between the CP-even interaction eigenstates $\{h, s\}$. By using the minimisation conditions of the scalar potential shown in Eq.~(\ref{eq:scalar-pot:tot-pot}), after the EWSB, one can write down the CP-even squared mass matrix in the $\{h, s\}$ basis as
\bea
\label{eq:masses:scalar-mass-HS-new}
\mathcal{M}^2 &=
\begin{pmatrix}
	~\mathcal{M}_{hh}^2 ~ & ~\mathcal{M}_{hs}^2 ~ \\
	~\mathcal{M}_{sh}^2 ~ & ~\mathcal{M}_{ss}^2
\end{pmatrix},~~\text{where,}
\quad
\begin{array}{l}
	\mathcal{M}_{hh}^2 = 2 \l_H v^2, \\[5pt]
	\mathcal{M}_{ss}^2 = 2 \l_S v_S^2 + \frac{3}{2\sqrt{2}} \mu_3 v_S, \\[5pt]
	\mathcal{M}_{hs}^2 = \mathcal{M}_{sh}^2 = \l_{SH} v v_S.
\end{array}
\eea
The matrix $\mathcal{M}^2$ is real and symmetric. It can be diagonalised by a unitary transformation with an orthogonal matrix to obtain the CP-even mass eigenstates $h_1, h_2$ having masses $m_{h_1}, m_{h_2}$, respectively, 
\bea\label{eq:masses:hsmixtheta}
\begin{pmatrix}
	~ h_1 ~ \\
	~ h_2 ~
\end{pmatrix}
&=
\begin{pmatrix}
	\cos\theta~ & -\sin\theta~ \\
	\sin\theta ~ & ~\cos\theta
\end{pmatrix}
\begin{pmatrix}
	~ h ~ \\
	~ s ~
\end{pmatrix},
\eea
where $\theta$ is the mixing angle. The CP-even physical mass eigenstates and mixing angle are given as,
\bea\label{eq:masses:mass-eigen-HS}
m_{h_1}^2 &=& \mathcal{M}_{hh}^2 \cos^2\th + \mathcal{M}_{ss}^2 \sin^2\th - \mathcal{M}_{hs}^2 \sin2\th, \nn\\
m_{h_2}^2 &=& \mathcal{M}_{hh}^2 \sin^2\th + \mathcal{M}_{ss}^2 \cos^2\th + \mathcal{M}_{hs}^2 \sin2\th, ~~\text{and}, \nn \\
\tan2\theta &=& \frac{\l_{SH} v v_S}{\l_{S} v_S^2 - \l_{H} v^2 + \frac{3}{4 \sqrt{2}} \mu_3 v_S}.
\eea
We consider $m^2_{h_1}$ to be the lighter eigenvalue and identify $h_1$ as the SM-like Higgs which gives $m_{h_1} = 125.20 \pm 0.11$ GeV \cite{CMS:2020xrn,ATLAS:2023oaq, ParticleDataGroup:2024cfk}. On the other hand, the mass of the pNGB DM $\chi$, taking into account the minimisation conditions, is given by
\bea
\label{eq:masses:pNGB-mass}
m^2_{\chi} = - \frac{9}{2 \sqrt{2}} \mu_3 v_S.
\eea
The pNGB mass, as given in Eq.~(\ref{eq:masses:pNGB-mass}), is proportional to $\mu_3$, which explicitly breaks the $U(1)$ symmetry (see Eq.~(\ref{eq:scalar-pot:soft-pot})). As we consider $v_S > 0$, therefore for a non-tachyonic $m_{\chi}$, one must get $\mu_3 < 0$. Using Eqs.~(\ref{eq:masses:scalar-mass-HS-new}), (\ref{eq:masses:mass-eigen-HS}), and (\ref{eq:masses:pNGB-mass}), the tree-level potential parameters involving $H$ and $S$ can be expressed in terms of physical masses ($m_{h_1}$, $m_{h_2}$, $m_{\chi}$), mixing angle $\theta$, and VEVs ($v$, $v_S$) as,
\bea
\label{eq:dependent-param}
\l_H &&= \frac{m_{h_1}^2 \cos^2\th + m_{h_2}^2 \sin^2\th}{2 v^2},\quad
\l_{SH} =  \frac{\big( m_{h_2}^2 - m_{h_1}^2 \big)\sin2\theta}{2 v v_S}, \nn \\
\l_S &&= \frac{3 m_{h_1}^2 \sin^2\theta + 3 m_{h_2}^2 \cos^2\theta + m_{\chi}^2}{6 v_S^2}, \quad
\mu _3 = -\frac{2 \sqrt{2}}{9} \frac{m_{\chi}^2}{v_S}.
\eea
Finally, the squared mass elements for the triplet scalar are,
\bea\label{eq:mass-triplet}
m_{T^0, T^{\pm}}^2 &= \mu_{\bm{T}}^2 + \frac{1}{2} \left(\l_{HT} v^2 + \l_{ST} v_S^2 \right).
\eea
The tree-level degeneracy between $m_{T^0}$ and $m_{T^\pm}$ is lifted at the one-loop level \cite{Cirelli:2005uq, Cirelli:2009uv}, resulting a mass splitting $\Delta m = m_{T^\pm} - m_{T^0} \approx 166 \, \text{MeV}$ \cite{Cirelli:2005uq} for $m_{T^0} \gg m_W, m_Z$. Here, $m_{W(Z)}$ represents mass of the SM $W^{\pm}(Z)$ boson. This small mass hierarchy\footnote{The value $166 \, \text{MeV}$ shifts slightly upwards with the second-order corrections \cite{Ibe:2012sx}.} ensures $T^0$ remains the lightest triplet state, justifying its role as one of the two DM candidates.

Before continuing to the next section, we outline the independent parameters of the chosen model. The tree-level scalar potential $V(H,{\bm{T}},S)$ (see Eq.~(\ref{eq:scalar-pot:tot-pot})) includes ten bare parameters: $\mu_H$, $\l_H$, $\mu_S$, $\mu_3$, $\l_S$, $\l_{SH}$, $\mu_{\bm{T}}$, $\l_{\bm{T}}$, $\l_{ST}$, and $\l_{HT}$, as outlined in Eqs. (\ref{eq:scalar-pot:pot-H-T}), (\ref{eq:scalar-pot:pot-S}), and (\ref{eq:scalar-pot:soft-pot}). By accounting for Higgs sector mixing between $H$ and $S$, and using Eqs. (\ref{eq:dependent-param}) and (\ref{eq:mass-triplet}), we reframe these ten input parameters into $\{m_{h_1}, m_{h_2}, m_{\chi}, m_{T^0},$ $v, v_S, \sin\theta, \lambda_{\bm{T}}, \lambda_{ST}, \lambda_{HT}\}$. In the SM, $v = 246$ GeV \cite{ParticleDataGroup:2024cfk}, and for this study, as previously mentioned, we identify $h_1$ as the SM-like Higgs fixing  $m_{h_1} = 125.20 \pm 0.11$ GeV \cite{CMS:2020xrn,ATLAS:2023oaq, ParticleDataGroup:2024cfk}. Consequently, our framework contains eight independent free parameters:
\beq\label{eq:8freeparam}
m_{h_2}, m_{\chi}, m_{T^0}, v_S, \sin\theta, \lambda_{\bm{T}}, \lambda_{ST}, {\rm{~and~}} \lambda_{HT}.
\eeq
However, as will be discussed in the upcoming sections, not all eight parameters significantly influence the DM phenomenology or the PT dynamics.
\section{Observables and constraints} \label{sec:obs-cons}
In this section, we briefly discuss relevant constraints applied for the forthcoming analyses and the observables that we shall address in the subsequent sections.
\subsection{Theoretical bounds} \label{subsec:theoretical-bounds}
In our analysis, we require the model parameters to satisfy, primarily, the two theoretical bounds: (i) {\it vacuum stability}, and (ii) {\it tree-level perturbative unitarity}. The vacuum stability requires the tree-level potential of Eq.~(\ref{eq:scalar-pot:tot-pot}) to remain bounded from below, and this translates into the following lower bounds \cite{Kannike:2012pe,Kannike:2016fmd}:
\bea
\label{eq:constraints:vaccum-stability}
\l_{H}, \l_S, \l_{\bm{T}} \geq 0, \quad \l_{SH} \geq -2 \sqrt{\l_H \l_S}, \quad \l_{HT} \geq -2 \sqrt{\l_H \l_{\bm{T}}}, \quad \l_{ST} \geq -2 \sqrt{\l_S \l_{\bm{T}}}\,.
\eea
The scalar quartic couplings are further constrained by the
perturbative unitarity of the tree-level $2 \to 2$ scattering amplitudes and it imposes the following bounds on the relevant parameters \cite{Pruna:2013bma,Kanemura:2015fra},
\bea
\label{eq:constraints:perturbativity}
|\l_H|,\,|\l_S|,\, |\l_{\bm{T}}|,\, |\l_{HT}|,\, |\l_{ST}| \leq 4\pi, \quad |\l_{SH}| \leq 8 \pi, \nn \\
|3 \l_{H} + 2 \l_{S} \pm \sqrt{(3 \l_{H} - 2 \l_{S})^2 + 2 \l_{SH}^2}| \leq 8\pi.
\eea
We further verify the unitarity constraints at a finite energy \cite{Goodsell:2018tti} with the help of the {\tt SARAH-4.15.2} package \cite{Staub:2008uz,Staub:2011dp,Staub:2012pb,Staub:2013tta,Staub:2015kfa,Goodsell:2018tti}, which we also use to write down the model Lagrangian.

Additionally, we require $ \mathcal{HS} \equiv (\langle H \rangle, \langle S \rangle, \langle \bm{T} \rangle) = (v, v_S, 0)$ to be the global minimum at zero temperature. In the singlet-doublet Higgs sector, this requirement implies \cite{Kannike:2019mzk}:
\bea
\label{eq:constraints:HS-global-vac}
m_{\chi}^2 < \frac{9~ m_{h_1}^2 m_{h_2}^2}{m_{h_1}^2 \cos^2\theta + m_{h_2}^2 \sin^2\theta}.
\eea
In the limit of a negligible mixing between $SU(2)_L$ singlet and doublet Higgs, i.e., $\sin\theta \to 0$, this simplifies to $m_{\chi} \lesssim 3 m_{h_2}$. This bound appears to be stronger than the constraint on the cubic coupling $\mu_3$ (see Eq.~(\ref{eq:dependent-param})) derived from minimisation conditions\footnote{The global vacuum condition will be further verified using {\tt cosmoTransitions} \cite{Wainwright:2011kj} while discussing the PT dynamics.}. However, the bound on $\mu_3$ can be relaxed if the SM vacuum is a metastable local minimum with a lifetime exceeding the age of the Universe, as advocated in Ref.~\cite{Belanger:2012zr}.
\subsection{Experimental constraints} \label{subsec:exp-constraints}
In addition to the theoretical bounds, we must account for the existing limits from various experimental observations in order to thoroughly examine the parameter space of the chosen model. Below, we outline the key experimental constraints considered in this study.

$\bullet$ {\it Electroweak precision observables (EWPO):}
The presence of additional scalars in our setup contributes to the gauge boson self-energies, affecting the EWPOs, which can be parameterised by the oblique parameters $S$, $T$, and $U$ \cite{Peskin:1991sw}. The new physics contributions to these parameters can be written as:
\bea
\Delta \mathcal{X} = \Delta \mathcal{X}_{\rm IT} + \Delta \mathcal{X}_{\rm xS}.
\eea
Here $\Delta \mathcal{X} \equiv \mathcal{X} - \mathcal{X}^{\rm{SM}}$ for $\mathcal{X} = {S, T, U}$, where the subscripts ${\rm IT}$ and ${\rm xS}$ denote contributions from the ``inert triplet'' and the ``complex scalar singlet'', respectively. Given that, only the real component of the complex scalar acquires a non-zero VEV, the pNGB DM $\chi$ does not contribute to the gauge boson self energies. Thus, the oblique parameters follow the same dependence as observed in the extended triplet-real singlet scenarios (e.g., see Ref. \cite{Borah:2024emz} and references therein). The deviations in the EWPO parameters can be expressed as \cite{Borah:2024emz},
\bea\label{eq:STU-inertTriplet}
\Delta S_{\textrm{IT}} &&= 0, \nn \\
\Delta T_{\textrm{IT}} && \simeq \frac{(\Delta m)^2}{24 \pi s^2_W m^2_W},\quad
\Delta U_{\textrm{IT}} \simeq \frac{\Delta m}{3 \pi m_{T^{\pm}}}, \quad \mathrm{for}\, \Delta m 
= m_{T^\pm}-m_{T^0} \ll m_{T^0}, ~~\text{and,} \nn \\
\Delta \mathcal{X}_{\rm xS} &&= \sin^2\theta \left[f_2(m_{h_2}) - f_1(m_{h_1}) \right].
\eea
Here, $s_W = \sin\theta_W$, with $\theta_W$ denoting Weinberg angle \cite{ParticleDataGroup:2024cfk}, $\Delta m$ is defined following Eq.~(\ref{eq:mass-triplet}), and $m_W = 80.3692$ GeV \cite{ParticleDataGroup:2024cfk}. Neglecting, for the moment, the updated $W$-boson mass measurement by CDF \cite{CDF:2022hxs}, the global electroweak fit \cite{Lu:2022bgw}, and fixing $\Delta U = 0$, the central values of $\Delta S$ and $\Delta T$ are given as \cite{Lu:2022bgw,ParticleDataGroup:2024cfk}:
\bea
\label{eq:constraints:ST-param}
\Delta S = 0.05 \pm 0.08, \quad \Delta T = 0.09 \pm 0.07,\quad {\rm with~ correlation}~~ \rho_{ST} = 0.92.
\eea
We consider a $\chi^2$-test with the two remaining degrees of freedom ({\it d.o.f.}) on our parameter space and exclude points having a $\chi^2$ larger than $5.99$ to remain consistent with the EWPOs constraints.

$\bullet$ {\it Constraints from Higgs boson properties:}
Precise measurements of Higgs properties at the LHC impose significant constraints on the parameter space. Key constraints from the SM and BSM Higgs searches at colliders must be considered for a comprehensive analysis.

(i) {\underline{Invisible Higgs decays:}}
For a light DM below ${m_{h_1}}/{2}$ threshold, the SM-like Higgs ($h_1$) can decay invisibly into two DM particles, $h_1 \to T^0 T^0, \chi \chi$. Visible decays like $h_1 \to h_2 h_2$ or $T^{\pm} T^{\mp}$ are also possible if kinematically allowed. However, in our setup, $h_2$ is heavier than $h_1$, and considering lower bounds on $T^0$ (discussed later in this subsection), decays such as $h_1 \to T^0 T^0$ and $h_1 \to T^{\pm} T^{\mp}$ are kinematically forbidden. Thus, the only allowed invisible $h_1$ decay is $h_1 \to \chi \chi$. Under such circumstances, we need to employ the bound on the invisible Higgs decay branching ratio ($Br$) of the SM Higgs boson as \cite{ParticleDataGroup:2024cfk,ATLAS:2023tkt,CMS:2023sdw}
\bea
\label{eq:constrains:Higgs-invisible-psi}
Br (h_1\to \chi \chi)= \frac{\Gamma(h_1 \rightarrow \chi \chi)}{\Gamma^{\rm{tot}}_{h_1} + \Gamma(h_1 \rightarrow \chi \chi)} < 0.107,
\eea
where $\Gamma^{\rm {tot}}_{h_1}$ and $\Gamma(h_1 \rightarrow \chi \chi)$ represent the total decay width of $h_1$ into \textit{all the SM} modes and the decay width into $\chi \chi$ decay channel, respectively. Apart from Eq. (\ref{eq:constrains:Higgs-invisible-psi}), one also needs to check whether 
$\Gamma^{\rm{tot}}_{h_1} + \Gamma(h_1 \rightarrow \chi \chi)< 3.7^{+1.9}_{-1.4}$ MeV \cite{ParticleDataGroup:2024cfk,ATLAS:2023dnm,CMS:2022ley}.

(ii) {\underline{Higgs signal strength measurements:}}
Even if new Higgs decays are kinematically forbidden, the BSM states can still affect Higgs signal strengths, defined as $\mu_{ij} = (\sigma_i \times Br_j)^{\rm obs}/(\sigma_i \times Br_j)^{\rm SM}$, where $\sigma_i$ is the production cross-section for $i$-th channel (e.g., gluon fusion) and $Br_j$ is the branching ratio for the $j$-th decay mode (e.g., $\bar{b}b$), relative to the SM predictions. With $h_1$ as the $\approx 125$ GeV SM-like Higgs, $\mu_{ij}$ values for various production and decay modes like $\gamma\gamma$, $WW^*$, and $ZZ^*$ must comply with the experimental constraints \cite{ParticleDataGroup:2024cfk}, limiting the singlet-doublet mixing parameter to $|\sin\theta| < 0.2$ across most of the parameter space. However, future projections of Higgs boson signal strength measurements at HL-LHC \cite{Cepeda:2019klc} and ILC \cite{ILC:2013jhg}/GigsZ \cite{ILC:2001igx,Heinemeyer:2010yp} provide a more stringent constraint on $|\sin\theta|$, which almost excludes $\sin\theta \gtrsim 0.16$ (see, for e.g., Ref.~\cite{Papaefstathiou:2020iag}). Additionally, new charged states such as $T^\pm$ can modify $h_1 \to \gamma\gamma$ decays through loops \cite{Gunion:1989we,YaserAyazi:2014jby}. To ensure a viable parameter space or propose benchmark points, $\mu_{\gamma \gamma}$ must remain within the measured range of $1.10 \pm 0.06$, as per the latest LHC observations \cite{ATLAS:2022tnm, CMS:2022dwd}.

(iii) {\underline{Heavy Higgs searches:}}
The singlet-doublet mixing introduces a second CP-even Higgs state, $h_2$, in addition to the SM-like Higgs $h_1$. While $h_1$ is predominantly doublet-like and identified as the 125 GeV Higgs, $h_2$ represents a true BSM {\it heavy} Higgs (see subsection \ref{subsec:masses-mixings}). A small but non-vanishing singlet-doublet mixing allows $h_2$ to decay into the SM states, though these decays are tightly constrained by collider searches. Heavy Higgs searches at the LHC can probe $h_2$ via final states like $b\bar{b}\tau^+\tau^-$, $b\bar{b}b\bar{b}$, and $b\bar{b}\gamma\gamma$ (see Refs~\cite{ATLAS:2021jki,AEDGE:2019nxb,ATLAS:2016paq,CMS:2016jvt,CMS:2016cma,
ATLAS:2015sxd,CMS:2015hra,ATLAS:2015oxt,ATLAS:2015pre,CMS:2013vyt} for more details). Additionally, $h_2$ can decay into BSM states such as the inert triplet scalars $T^0$ and $T^\pm$, as well as the pNGB DM candidate $\chi$, whenever allowed kinematically. The decay widths of $h_2$ into the SM and BSM states depend on the mixing angle $\theta$ and couplings, making $h_2$ a key probe for the extended scalar sector.

In this study, we use {\tt SARAH-4.15.2}\cite{Staub:2008uz,Staub:2011dp,Staub:2012pb,Staub:2013tta,Staub:2015kfa,Goodsell:2018tti} - {\tt SPheno-4.0.5}\cite{Porod:2003um,Porod:2011nf,Staub:2011dp,Porod:2014xia,Goodsell:2014bna,Goodsell:2015ira,Goodsell:2016udb,Braathen:2017izn,Goodsell:2018tti,Goodsell:2020rfu} pipeline to calculate the mass spectrum and observables such as decay widths, $Br$s, etc. The SLHA2 \cite{Allanach:2008qq} output is then interfaced with {\tt HiggsTools v.1.2} \cite{Bahl:2022igd}, which incorporates the latest experimental constraints on Higgs properties. By utilizing its sub-packages, {\tt HiggsSignals v.3} (database {\tt v.1.1}) \cite{Bechtle:2013xfa,Bechtle:2014ewa,Bechtle:2020uwn} and {\tt HiggsBounds v.6} (repository {\tt v.1.6}) \cite{Bechtle:2008jh, Bechtle:2011sb, Bechtle:2013wla, Bechtle:2020pkv}, we compare our model predictions with the most up-to-date Higgs data and experimental limits.

$\bullet$ {\it Disappearing charged track:}
For the $SU(2)_L$ inert triplet ${\bm{T}}$, the small mass splitting between its charged and neutral components complicates traditional search strategies for the singly charged scalar $T^{\pm}$. The decay of the unstable $T^\pm$ into $T^0 \pi^\pm$ \cite{Cirelli:2005uq} (with $\Delta m = m_{T^\pm} - m_{T^0} \approx 166$ MeV) produces a soft pion that is difficult to detect, alongside the neutral component $T^0$, a DM candidate. This leads to a distinctive collider signature: a disappearing charged track, which has garnered significant interest. Recent studies \cite{Chiang:2020rcv} shows that 13 TeV LHC excludes real triplets below $275~(248)$ GeV for $\mathcal{L} = 36 \, \text{fb}^{-1}$ with $\Delta m = 166~(172)$ MeV. Projections for $\mathcal{L} = 300 \, \text{fb}^{-1}$ and $3000 \, \text{fb}^{-1}$ extend limits to $590~(535)$ GeV and $745~(666)$ GeV, respectively, though 30\% systematic uncertainty reduces these to $382~(348)$ GeV and $520~(496)$ GeV, respectively.

In this study, we take a conservative approach, adopting a lower mass limit of $m_{T^0} > 300$ GeV based on the current LHC constraints\footnote{This partly excludes the ``desert region'' \cite{Cirelli:2005uq, Fischer:2011zz,Araki:2011hm, Khan:2016sxm} below $300$ GeV.}.
\subsection{Dark matter observations} \label{subsec:DM-observables}
The parameter space of the present model, besides the bounds and constraints stated in subsections \ref{subsec:theoretical-bounds} and \ref{subsec:exp-constraints}, needs to be constrained by the measured value of the DM relic density. The total relic density contribution from both the DMs, i.e., $\Omega_{\rm{tot}} h^2 = \Omega_{T^0} h^2 + \Omega_\chi h^2$, must satisfy the DM relic abundance $\Omega_{\rm DM}^{\rm exp} h^2$ as measured by the PLANCK collaboration \cite{Planck:2018vyg},
\bea
\label{eq:constraints:DM-relic-Planck}
\Omega_{\rm DM}^{\rm exp} h^2 = 0.1198 \pm 0.0012.
\eea

In addition to this, the observed and the projected sensitivity reaches of the existing and the upcoming DM direct search experiments, e.g., XENON1T \cite{XENON:2019gfn}, LZ-2022 \cite{LZ:2022lsv}, recently updated as LZ-2024 \cite{LZCollaboration:2024lux}, and DARWIN \cite{DARWIN:2016hyl} further restrict the model parameter space. To assess these limits, we need to obtain the DM-nucleon ($N$) scattering cross-section, which in this model arises from $t$-channel exchanges of $h_1$ or $h_2$. The spin-independent (SI) $N$-DM cross-section, in the non-relativistic limit, is given by
\bea
\label{eq:DD-SI-appx}
\sigma^{\rm SI}_{\text{DM-}N}~ =~ \frac{f^2_N m^2_N \mu^2_{\text{DM-}N}}{4 \pi m^2_{i} v^2} \left[ \frac{\l_{h_1 i i} \cos\th}{m^2_{h_1} } + \frac{\l_{h_2 i i} \sin\th}{m^2_{h_2}} \right]^2,
\eea
where $i \equiv {\rm{DM}} = \{T^0, \chi$\} and $\mu_{\text{DM-}N}$ is the reduced mass of the DM-$N$ system, defined as $\mu_{\text{DM-}N} = \frac{m_i m_N}{m_i + m_N}$ with $m_N = 0.946$ GeV, $m_{i} = \{m_{T^0}, m_{\chi} \}$. The nucleon form factor, which depends on the hadronic matrix elements and, approximately, is given as $f_N =0.28$ \cite{Alarcon:2012nr}. The trilinear couplings $\l_{h_1 i i}$ and $\l_{h_2 i i}$, using Eqs. (\ref{eq:scalar-pot:pot-H-T})-(\ref{eq:scalar-pot:scalar-field-basis}) and (\ref{eq:masses:hsmixtheta}), for the corresponding DMs are given as,
\bea
\label{eq:trilinear-couplings-DMDD}
\begin{array}{l}
	\l_{h_1 T^0 T^0} = \l_{HT} v c_{\theta} - \l_{ST} v_S s_{\theta}, \\
	\l_{h_2 T^0 T^0} =  \l_{HT} v s_{\theta} + \l_{ST} v_S c_{\theta},
\end{array}
\,\, \text{and,}\,\,
\begin{array}{l}
	\l_{h_1 \chi \chi} = \l_{SH} v c_{\theta} - 2 \l_{S} v_S s_{\theta} + \frac{3 \mu_3}{\sqrt{2}} s_{\theta}, \\
	\l_{h_2 \chi \chi} = \l_{SH} v s_{\theta} + 2 \l_{S} v_S c_{\theta} - \frac{3 \mu_3}{\sqrt{2}} c_{\theta},
\end{array}
\eea
where $s_{\theta} (c_{\theta})$ is the shorthand notation used for $\sin\theta(\cos\theta)$, as defined in Eq.~(\ref{eq:masses:mass-eigen-HS}). We note that, unlike the $\mathbb{Z}_2$ pNGB DM model (see e.g., Ref.~\cite{Gross:2017dan}), the tree-level DD amplitude includes a term that survives at the zero momentum transfer. This can be seen using the expressions of Eq.~(\ref{eq:dependent-param}) and the trilinear couplings for the $\chi$-DM, given in Eq.~(\ref{eq:trilinear-couplings-DMDD}), becomes
\bea
\begin{array}{l}
\sigma^{\rm SI}_{\chi-N} \propto \l^2_{\chi-\text{SI}}
\end{array}
~~\text{with,}~~
\begin{array}{l}
	\l_{\chi-\text{SI}} \propto \frac{c_{\theta} s_{\theta}}{v_S m^2_{h_1} m^2_{h_2}} m^2_{\chi} \left(m^2_{h_2} - m^2_{h_1}\right),
\end{array}
\eea
where $\l^2_{\chi-\text{SI}}$ corresponds to the term in square bracket in Eq.~(\ref{eq:DD-SI-appx}). For most of the parameter space, this term remains sufficiently small \cite{Kannike:2019mzk,Alanne:2019bsm,Coito:2021fgo}, explaining the negative results in the DD experiments across a wide range of $m_\chi$.

The DD limits are often supplemented by constraints from the ID of the DM searches, which can be observed at the Earth-based facilities like H.E.S.S. \cite{Abdalla:2016olq,Abdallah:2016ygi,HESS:2015cda,Abramowski:2014tra,Abramowski:2011hc} and CTA \cite{CTA:2015yxo,CTAConsortium:2010umy}, as well as satellite-based experiments such as FERMI-LAT \cite{Fermi-LAT:2016uux,Fermi-LAT:2015kyq,Fermi-LAT:2015att,Ackermann:2015tah,Ackermann:2013yva,Ackermann:2013uma,Abramowski:2012au,Ackermann:2012nb,Fermi-LAT:2011vow,Abdo:2010ex} and AMS \cite{AMS:2002yni, AMS:2016oqu, AMS:2021nhj}. Fermi-LAT observations of photons from dwarf spheroidal galaxies impose strong constraints on the sub-electroweak scale DM masses annihilating into the SM fermion pairs (e.g., $b \bar{b}$, etc.). The current Fermi-LAT limits for the ID through photon range from $\langle \sigma v \rangle \approx 3 \times 10^{-26} cm^3 s^{-1}$ to $\approx 2.5 \times 10^{-25} cm^3 s^{-1}$, for DM masses in the window of 100-1000 GeV\footnote{Here, $\langle \sigma v \rangle$ represents the thermal average of the annihilation (co-annihilation) cross-section.}. In our model, as will be discussed later, the DM annihilation cross-sections are, in general, largest for the SM gauge boson ($VV$) final states. Therefore we require\footnote{In order to constrain our model parameter space with the ID limits, we employ a method opted in Ref. \cite{Belanger:2021lwd}.}, $\langle \sigma v \rangle_{VV}$ to lie below the 95\% confidence level (CL) given by Fermi-LAT \cite{Fermi-LAT:2015att}. Further, the AMS-02 \cite{AMS:2002yni, AMS:2016oqu, AMS:2021nhj} anti-proton searches also provide strong constraints on WIMP DM masses. Again, we consider $\langle \sigma v \rangle_{VV}$ to limit our model parameter space and use the bounds obtained in Ref. \cite{Reinert:2017aga}.

At this juncture, one should note that the DD and ID limits, in the context of a model with multi-component DM, should be rescaled by the relative relic abundance parameters $f_{i}$ and $f_{i}^2$ \cite{Cao:2007fy,Aoki:2012ub} respectively, where $f_i$ are defined as
\bea
\label{eq:constraints:DM-relic-rescaled}
f_i = \frac{\Omega_i h^2}{\Omega^{\rm exp}_{\rm{DM}} h^2}, \quad~~\text{for}~~ i =\{T^0, \chi\}.
\eea
Also, the rescaled upper limit on the total SI DM-nucleon scattering cross-section is written as \cite{Cao:2007fy},
\bea
\label{eq:DM-pheno:rescaled-DD}
\sigma^{\rm SI}_{\rm tot} = f_{T^0} \sigma^{\rm SI}_{T^0} + f_{\chi} \sigma^{\rm SI}_{\chi} < \sigma^{\rm SI}_{\rm exp},
\eea
where $ \sigma^{\rm SI}_{i=T^0,\,\chi}$ represents the SI DM-$N$ scattering cross-section for the individual DM species and  $\sigma^{\rm SI}_{\rm exp}$ is the experimental limit from various experiments, searching for the DD of the DM. In this work, our results for the relic density, scattering amplitudes and annihilation cross-section are obtained using the publicly available code {\tt micrOMEGAs-6.0.5} \cite{Belanger:2001fz,Alguero:2023zol}.
\section{DM phenomenology} \label{sec:DM-pheno}
In this section, we consider both the DM components to be WIMP-like, making them testable in the ongoing and the upcoming DD experiments. Initially, in the thermal equilibrium, WIMPs undergo freeze-out when their interaction rate drops below the Hubble expansion rate. Various possible DM-DM annihilation (co-annihilation) and conversion processes, possible in the chosen model, are presented in Table \ref{tab:DM_processes}. The DM abundance of each component, after freeze-out, is determined by solving the coupled Boltzmann equations (cBEQs), which in their simplest forms, for the chosen setup, are given by:
\begingroup
\allowdisplaybreaks
\bea
\frac{d Y_{T^0}}{d x} &=& -\frac{s(x)}{x ~\mathcal{H}(x)} \Big[ \langle \sigma v \rangle_{T^0 T^0 \rightarrow X X } \left( Y^2_{T^0} - Y^2_{T^0, eq} \right)  
+ \langle \sigma v \rangle_{T^0 T^{\pm} \rightarrow X X^\prime} \left( Y_{T^0} Y_{T^{\pm}} - Y_{T^0, eq} Y_{T^{\pm}, eq} \right) \nn \\
&&+ \langle \sigma v \rangle_{T^0 T^0 \rightarrow \chi \chi} \left( Y^2_{T^0} - \frac{
	Y^2_{T^0, eq}}{Y^2_{\chi, eq}} Y^2_{\chi} \right)
\Big], \nn \\
\label{eq:DM-pheno-BEQs}
\frac{d Y_{\chi}}{d x} &=& -\frac{s(x)}{x ~\mathcal{H}(x)} \Big[ \langle \sigma v \rangle_{\chi \chi \rightarrow X X} \left( Y^2_{\chi} - Y^2_{\chi, eq} \right) 
+ \langle \sigma v \rangle_{ \chi \chi \rightarrow T^0 T^0 } \left( Y^2_{\chi} - \frac{
	Y^2_{\chi, eq}}{Y^2_{T^0, eq}} Y^2_{T^0} \right)
\Big]. 
\eea
\endgroup
In Eq.~(\ref{eq:DM-pheno-BEQs}), we define $x = \frac{\mu_{\rm red}}{T}$, where $T$ is the thermal bath temperature, $\mu_{\rm red} = \frac{m_{T^0} m_{\chi}}{m_{T^0} + m_{\chi}}$ is the reduced mass of the two DM components and $X$ corresponds to various annihilation (co-annihilation) states. Parameters $Y_{i, eq}, Y_{T^\pm, eq}$ represent the co-moving yields, at equilibrium, of the DM candidates, with $i = {T^0, \chi}$, and $T^\pm$, respectively. Further, $s(x) = \frac{2 \pi^2}{45} g_s \frac{\mu_{\rm red}^3}{x^3}$ denotes the comoving entropy density, and $\mathcal{H}(x) = \sqrt{\frac{\pi^2 g_{\rho}}{90}} \frac{\mu_{\rm red}^2}{x^2 M_{\rm Pl}}$ is the Hubble parameter. Here, $g_s$ and $g_{\rho}$ are the relativistic {\it d.o.f.} for entropy and matter, respectively, and $M_{\rm Pl} = 2.4 \times 10^{18}$ GeV is the reduced Planck mass. Although $g_s$ and $g_{\rho}$ vary slightly during the Universe's evolution, we treat them as constants, equal to the effective {\it d.o.f.} $g_*(T)^{1/2} = \frac{g_s}{\sqrt{g_{\rho}}} \left(1 + \frac{1}{3} \frac{T}{g_s} \frac{d g_s}{d T}\right)$ \cite{Gondolo:1990dk}. In our notation, annihilation (co-annihilation) final states $X$ includes $\{h_1, h_2, W^{\pm}, Z, \text{leptons, quarks}\}$, while $X^\prime$ also includes $\gamma$ in the final states (see Table \ref{tab:DM_processes}). Individual relic densities can be derived from solutions of the cBEQs, and it can be expressed as \cite{Kolb:1990vq}
\beq
\Omega_i h^2 \simeq 2.752\times 10^{8}\left(\frac{m_i}{\rm GeV}\right)Y_i({x \to \infty}).
\eeq
The total relic density $\Omega_{\rm{tot}} h^2$, defined in subsection \ref{subsec:DM-observables}, is the sum of individual relic densities and must satisfy the PLANCK limit of Eq.~(\ref{eq:constraints:DM-relic-Planck}). In our analysis, we allow for a $3\sigma$ variation of this limit, corresponding to the range $0.1162 - 0.1234$.
\begin{table}[!htpb]
    \hspace{1.5cm}
	\renewcommand{\arraystretch}{1.3} 
	\arrayrulecolor{black} 
	\begin{tabular}{|l|l|}
        \hhline{|= =|}
		\rowcolor{lightgray!35} 
		\multicolumn{1}{|c|}{\textbf{$T^0$-annihilation}} & \multicolumn{1}{c|}{\textbf{$T^0$-co-annihilation}} \\ \hline
		$T^0 T^0 \to f\bar{f},\, ZZ,\, W^\pm W^\mp, h_i h_j$ & $T^0 T^\pm \to f f',\, W^\pm Z,\, W^\pm h_i, W^\pm \gamma$ \\ 
		& $T^\mp T^\pm \to f \bar{f},\, ZZ,\, W^\pm W^\mp, Z h_i,\, h_i h_j$ \\
		\hhline{|= =|}
		\rowcolor{lightgray!35} 
		\multicolumn{2}{|c|}{\textbf{$\chi$-annihilation}} \\ \hline
		\multicolumn{2}{|l|}{$\chi \chi \to f\bar{f},\, ZZ,\, W^\pm W^\mp,\, h_i h_j$} \\
		\hhline{|= =|}
		\rowcolor{lightgray!35}
		\multicolumn{2}{|c|}{\textbf{DM conversion}} \\
		\hline
		$\chi \chi \to T^0 T^0 ~~\textnormal{when}  ~m_{\chi} > m_{T^0}$ & $T^0 T^0 \to \chi \chi ~~\text{when}  ~m_{\chi} < m_{T^0}$ \\ \hline
	\end{tabular}
	\caption{Annihilation, co-annihilation, and DM conversion processes. Here, $i,j=1,2$ and $f,\,f'$ represent any SM fermion, ensuring charge conservation of course.}
	\label{tab:DM_processes}
\end{table}

As already mentioned, we use {\tt microMEGAS-6.0.5} to numerically solve the cBEQs, obtaining values for the DM relic density, SI DD cross-sections, annihilation rates $\langle \sigma v \rangle$, and other relevant observables. In this respect, for a detailed numerical scan, we have varied the independent model parameters, as outlined in subsection \ref{subsec:masses-mixings}, within the following range,
\bea \label{eq:scan-param}
	50 \leq m_{\chi} [\text{GeV}] \leq 1500, ~~300 \leq m_{T^0} [\text{GeV}] \leq 1500,\nn\\
	130 \leq m_{h_2} [\text{GeV}] \leq 2000,  ~~50 \leq v_S [\text{GeV}] \leq 2000, \nn \\
	10^{-3} \leq |\sin\theta| \leq 0.15,  ~~0.01 \leq \lambda_{HT} \leq 0.3,
	~~0.01 \leq \lambda_{ST} \leq 0.3.
\eea
It is worth mentioning that, the upper DM mass bounds are motivated by current limits from mono-X searches at colliders \cite{CMS:2018ffd,ATLAS:2018nda,CMS:2019ykj,CMS:2020ulv,ATLAS:2020fgc,ATLAS:2020uiq,CMS:2021far,CMS:2021mjq,ATLAS:2021gcn,ATLAS:2021jbf,ATLAS:2021kxv,ATLAS:2021shl,MonoJetSearch2021} while the lower bound for $m_{T^0}$ stems from the disappearing charged track constraints (see subsection \ref{subsec:exp-constraints}). We treat $m_{h_2}$ as the heavier Higgs, with $|\sin\theta|$ chosen to satisfy Higgs search and the EWPO constraints. The lower limit of $m_{h_2}$ is set from $m_{h_2} > m_{h_1}$ requirements. For the upper bound on $|\sin\theta|$, we adopt a conservative approach by taking into account the future projections of Higgs signal strength measurements at HL-LHC and ILC/GigaZ \cite{Papaefstathiou:2020iag} (see subsection \ref{subsec:exp-constraints}). The triplet scalar quartic couplings are set to ensure perturbativity and to avoid DD limits. As $\lambda_{\bm{T}}$ has negligible impact on the DM phenomenology and the PT dynamics, we fix it at 0.01.
\subsection{Parameter scan and results} \label{subsec:DM-results}
We performed a random scan over $2 \times 10^6$ model points, as specified in Eq.~(\ref{eq:scan-param}), and identified $\sim 27,000$ model points that meet all constraints outlined in Sec.~\ref{sec:obs-cons}, including the DM relic density limit within the $3\sigma$ range (see Eq.~(\ref{eq:constraints:DM-relic-Planck})). To refine the analysis, we incorporated also the DM SI-DD and ID constraints, classifying the results into two DM mass hierarchies: (i) {\it Region-I:} $m_{\chi} > m_{T^0}$, and (ii) {\it Region-II:} $m_\chi < m_{T^0}$.
\subsubsection{Region-I: the $m_{\chi} > m_{T^0}$ regime} \label{subsubsec:DM-results:R-I}
\begin{figure*}[!htpb]
	\centering
	\subfigure[\label{fig:R-I:param-dependence-a}]{\includegraphics[height=5.8cm,width=7.5cm]{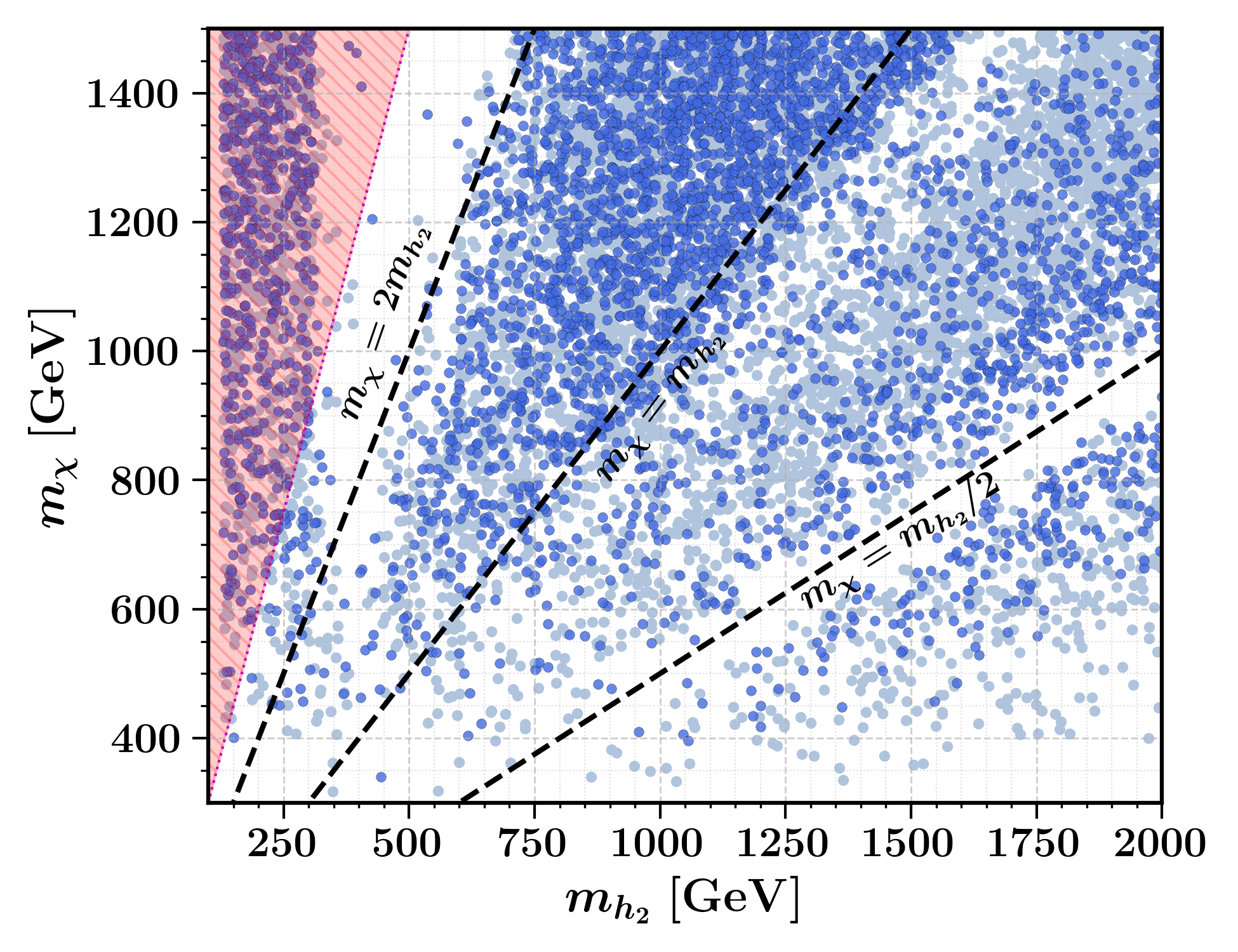}}
	\subfigure[\label{fig:R-I:param-dependence-b}]{\includegraphics[height=5.8cm,width=7.5cm]{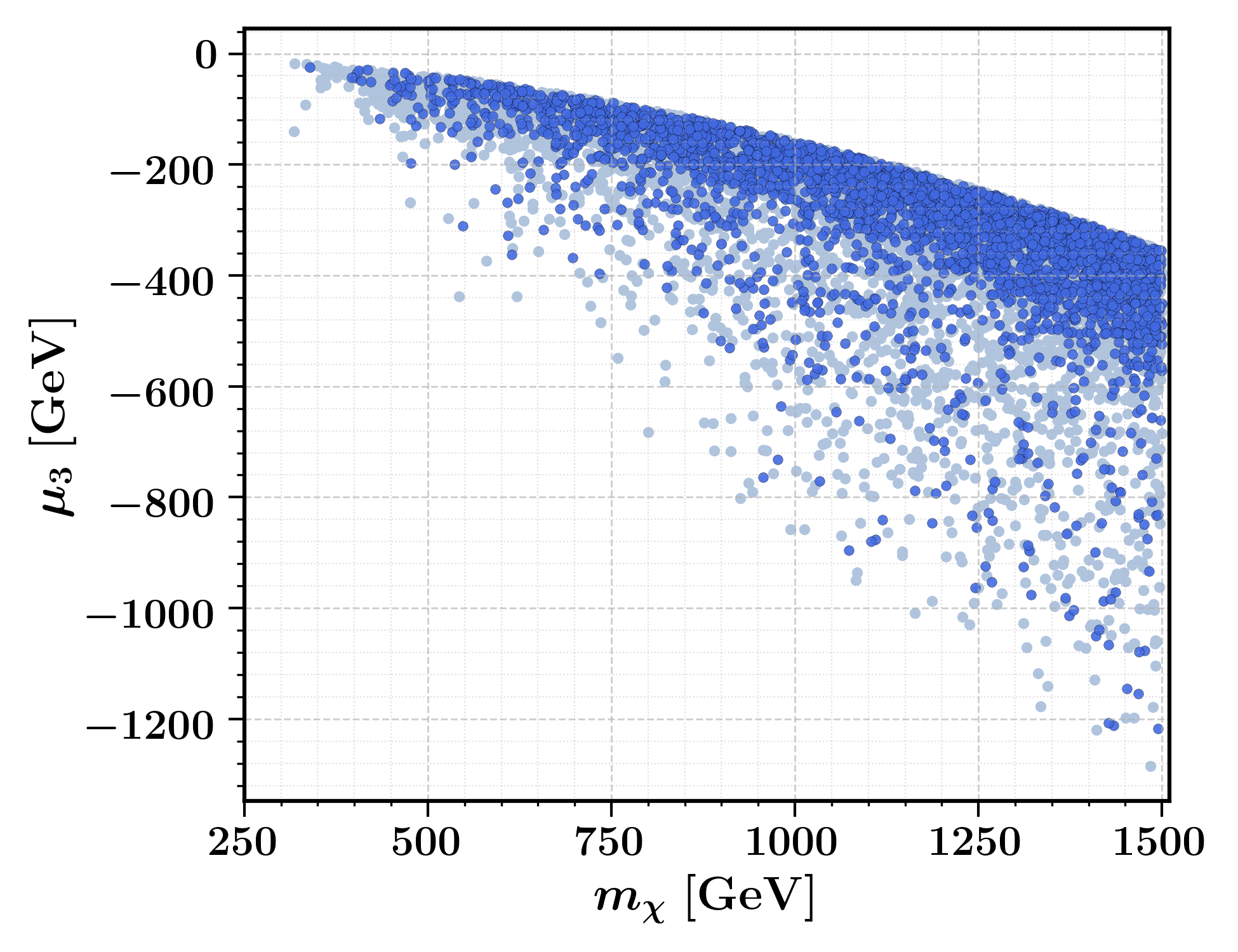}}
	\subfigure[\label{fig:R-I:param-dependence-c}]{\includegraphics[height=5.8cm,width=7.8cm]{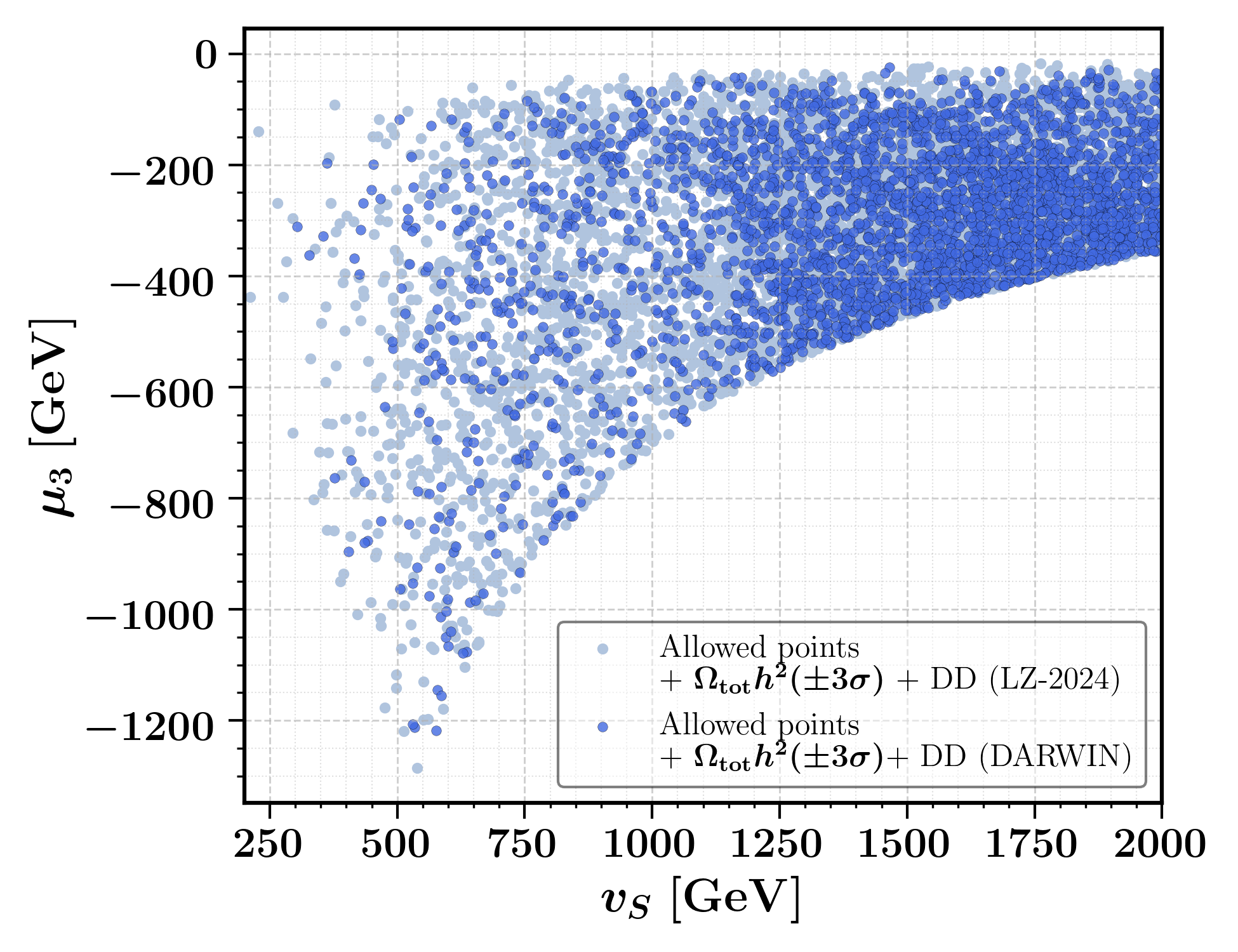}}

\caption{The plots display projections of model points onto various planes: (a) $m_{h_2}$-$m_\chi$, (b) $m_\chi$-$\mu_3$ and (c) $v_S$-$\mu_3$, illustrating their inter-dependence. The light blue coloured points satisfy various constraints as outlined in Sec.~\ref{sec:obs-cons}, align with the $3\sigma$ relic density limit and also pass the LZ-2024 limit. The dark blue coloured points represent the model samples that are further allowed by the stricter DD limits projected by DARWIN. In (a), the light red coloured shaded region is excluded by the requirement of an $\mathcal{HS}$ global minimum (see Eq.~(\ref{eq:constraints:HS-global-vac})), and the differently styled black coloured lines correspond to various threshold limits, i.e., $m_\chi = m_{h_2}$, $m_\chi = m_{h_2}/2$ and $m_\chi = 2 m_{h_2}$. See text for more details.}
	\label{fig:R-I:param-dependence}
\end{figure*}

We begin our discussion by illustrating our scan results for the $m_{\chi} > m_{T^0}$ regime, region-I (\textbf{R-I}). Fig. \ref{fig:R-I:param-dependence} shows a sample of viable model points projected onto different planes. We emphasise that, rather than examining the parameter dependence of the entire model samples for the concerned DM phenomenology, our analysis focuses exclusively on points that satisfy the $3\sigma$ relic density constraint and evade all other constraints outlined in Sec.~\ref{sec:obs-cons}. In Fig.~\ref{fig:R-I:param-dependence}, ``light blue'' coloured points denote the model samples that further pass DD bounds from the recently updated LZ-2024, while ``dark blue'' coloured points are consistent with the DD limits projected by DARWIN.

Fig.~\ref{fig:R-I:param-dependence-a} displays the correlation between the pNGB DM mass $m_\chi$ and the heavy Higgs mass $m_{h_2}$. The black coloured solid, dashed and dot-dashed lines indicate the degenerate regime ($m_\chi = m_{h_2}$), the $h_2$-resonance condition ($m_\chi = m_{h_2}/2$), and the $m_{\chi} = 2 m_{h_2}$ threshold, respectively. The shaded light-red coloured region represents the exclusion zone defined by Eq.~(\ref{eq:constraints:HS-global-vac}), for $\sin\theta \to 0$. Any model points falling within this region are excluded based on the requirement of the $\mathcal{HS}$ global vacuum condition. However, in Fig. \ref{fig:R-I:param-dependence-a}, we retain these points to highlight the stringency of the $\mathcal{HS}$ global minimum constraint in our model. For subsequent analyses, including Figs. \ref{fig:R-I:param-dependence-b}, \ref{fig:R-I:param-dependence-c}, and beyond, we omit all points that fail to satisfy the condition given in Eq.~(\ref{eq:constraints:HS-global-vac}). In \textbf{R-I}, viable points span from $m_\chi > 300$ GeV\footnote{This lower bound arises from the requirement as $m_\chi > m_{T^0} > 300$ GeV.}, to the maximum scan value shown in Eq.~(\ref{eq:scan-param}). Notably, $m_\chi$ increases with $m_{h_2}$, particularly in the range $m_{h_2}/2 \lesssim m_\chi \lesssim 2 m_{h_2}$. In contrast, for $m_{h_1} \lesssim m_{h_2} \lesssim 2 m_{h_1}$, $m_{\chi}$ can take any value provided $m_{\chi} \gtrsim 2 m_{h_2}$, however, majority of model samples in this region are ruled out due to the $\mathcal{HS}$ global minimum requirement at the zero temperature.  While LZ-2024 does not significantly constrain the parameter space, most points sensitive to DARWIN are concentrated in the $m_{h_2} \lesssim m_\chi \lesssim 2 m_{h_2}$ region. Additionally, a subset of points near the $h_2$-resonance ($m_\chi \approx m_{h_2}/2$) can also evade the DARWIN bound.

Fig.~\ref{fig:R-I:param-dependence}(b) illustrates the dependence between $m_\chi$ and the soft $U(1)$ breaking parameter $\mu_3$ (see Eq.~(\ref{eq:scalar-pot:soft-pot})). We observe that $|\mu_3|$ increases with $m_\chi$, reaching up to $\sim 1.2$ TeV near the maximum chosen mass range for pNGB DM, as of Eq.~(\ref{eq:scan-param}). While LZ-2024 permits $|\mu_3|$ as large as $\sim 1.2$ TeV, most model points testable by DARWIN are concentrated in the region $|\mu_3| \lesssim 600$ GeV. Finally, Fig.~\ref{fig:R-I:param-dependence}(c) largely demonstrates an anti-correlation between the singlet VEV $v_S$ and the parameter $|\mu_3|$, as can be seen from Eq.~(\ref{eq:dependent-param}). As $v_S$ increases, smaller values of $|\mu_3|$ are required to satisfy the relic density constraints from PLANCK. In \textbf{R-I}, the region with, mostly, $v_S \gtrsim 350$ GeV is favoured to achieve the correct relic density while evading DD bounds from LZ-2024 or DARWIN. The highest $v_S$ value in our sample, $\sim 2.0$ TeV, corresponds to $|\mu_3| \lesssim 350$ GeV. Furthermore, it is observed that increasing $v_S$ further necessitates an even smaller $|\mu_3|$ to satisfy the DM relic density requirement. Additionally, the parameter region with $v_S \gtrsim 1.0$ TeV have higher chances of being probed by DARWIN.

As already mentioned, this study investigates the ``desert region'' of the $Y=0$ scalar triplet DM model \cite{Cirelli:2005uq, Fischer:2011zz,Araki:2011hm, Khan:2016sxm}. While $T^0$ alone remains underabundant in this region, the DM conversion process $\chi \chi \rightarrow T^0 T^0$ in $\textbf{R-I}$ of our model framework can crucially enhance $T^0$'s relic density—unlike the pure $Y=0$ scalar triplet scenario. To estimate the relevance of DM conversion that contributes to the relic density of $T^0$, it is convenient to define the following parameter \cite{Belanger:2020hyh},
\bea
\label{eq:DM-conversion-R-I}
\zeta^{\chi}_{\rm conv.} = \frac{\sigma^{\chi}_v (\chi \chi \rightarrow T^0 T^0)}{\sum \sigma^{\chi}_v (\chi \chi \rightarrow all)}.
\eea
Here, the numerator represents the thermal averaged annihilation cross-section of $\chi$-DM conversion ($\sigma^{\chi}_v$), while the denominator sums over $\sigma^{\chi}_v$ for all $\chi$-DM processes as detailed in Table \ref{tab:DM_processes}, including the DM conversion. The parameter $\zeta^{\chi}_{\rm conv.}$ varies between 0 and 1, with $\zeta^{\chi}_{\rm conv.} \approx 1$ indicating that $T^0$ relic has the predominant contribution due to $\chi$-DM-conversion.
\begin{figure*}[!htpb]
	\hspace*{0.35cm} 
	\subfigure[\label{fig:R-I:DM-conversion-a}]{\includegraphics[height=5.9cm,width=7.8cm]{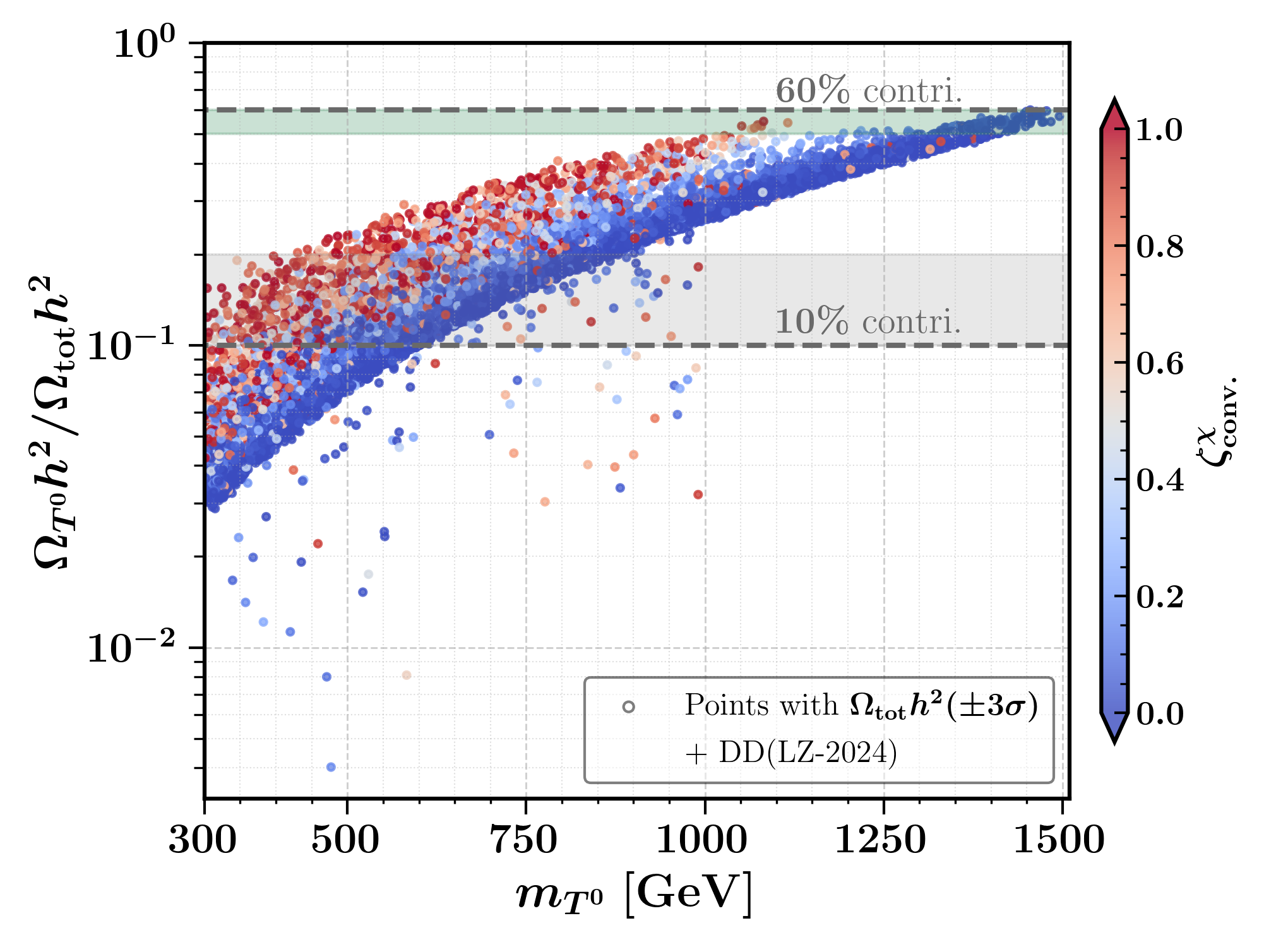}}
	\subfigure[\label{fig:R-I:DM-conversion-b}]{\includegraphics[height=5.8cm,width=6.8cm]{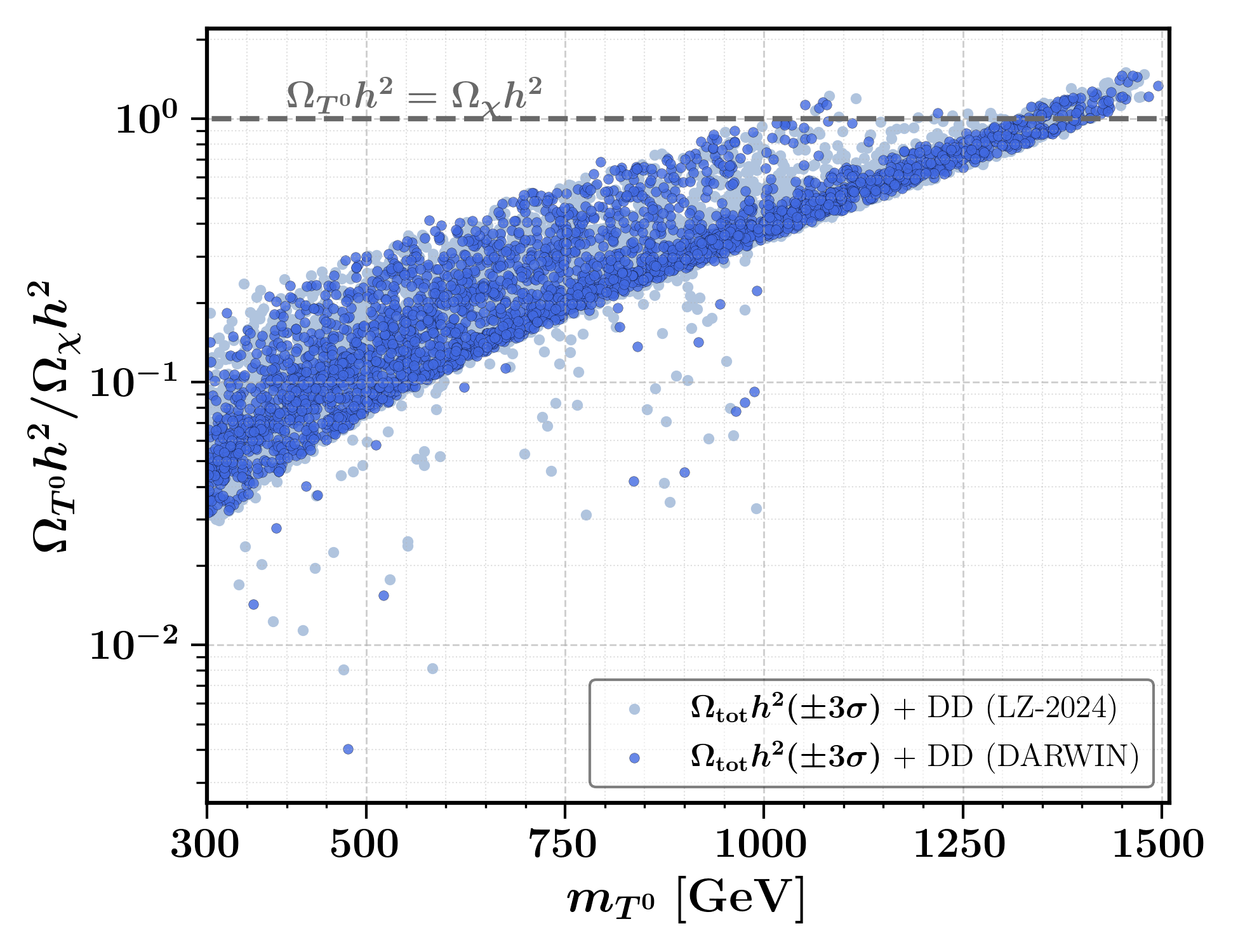}}
	\caption{Contribution of $T^0$-DM relic density $\Omega_{T^0} h^2$ to the total relic density, $\Omega_{\rm{tot}} h^2$, in \textbf{R-I}: (a) Projection on the $m_{T^0}$-plane with different values of $\zeta^{\chi}_{\rm conv.}$ (see Eq.~(\ref{eq:DM-conversion-R-I})) depicted by the colour bar, showing the impact of the DM conversion. The green coloured shaded band corresponds to $50-60\%$ contribution of $\Omega_{T^0} h^2$ to $\Omega_{\rm{tot}} h^2$. (b) Relative contributions of $T^0$ and $\chi$ relic densities to $\Omega_{\rm tot} h^2$. The model points shown in both plots satisfy the constraints outlined in Sec.~\ref{sec:obs-cons} and also accommodate the $3\sigma$ relic density. In (a), samples consistent with the LZ-2024 DD bounds are displayed, while (b) illustrates points passing both the LZ-2024 (light blue coloured points) and DARWIN (dark blue coloured points) limits. }
	\label{fig:R-I:DM-conversion}
\end{figure*}

Fig.~\ref{fig:R-I:DM-conversion-a} shows contribution of the $T^0$-DM to the total relic density and its projection on the $m_{T^0}$-plane with different values of $\zeta^{\chi}_{\rm conv.}$, as depicted through the colour bar. In our scan, we find that $\chi$-DM-conversion plays a crucial role in boosting the $T^0$-relic in the sub-TeV mass region of the triplet DM. Particularly for $m_{T^0} \gtrsim 500$ GeV, it over powers the predictions of the $Y=0$ scalar triplet DM model \cite{Cirelli:2005uq, Fischer:2011zz,Araki:2011hm, Khan:2016sxm}. The darker red coloured points in Fig.~\ref{fig:R-I:DM-conversion-a} indicates the model points where $\chi \chi \rightarrow T^0 T^0$ process yields a leading $\Omega_{T^0} h^2$ contributions to $\Omega_{\rm{tot}} h^2$. In fact, we find an enhancement up to $\sim 50-60\%$ (depicted by the green coloured band in Fig.~\ref{fig:R-I:DM-conversion-a}) in the $\Omega_{T^0} h^2$ near $m_{T^0} \approx 1.0$ TeV due to the DM-conversion, which is a significant improvement over a pure $Y=0$ scalar triplet DM model, where $T^0$ contribution to the relic density in the sub-TeV region remains only up to $\sim 10-20\%$ (represented by the light grey coloured band in Fig.~\ref{fig:R-I:DM-conversion-a}) \cite{Cirelli:2005uq,Araki:2011hm, Khan:2016sxm}. However, for $m_{T^0} \gtrsim 1.0$ TeV, the impact of the DM conversion diminishes, with contributions becoming less significant within the scanned mass range. Regarding the relative contributions of the two DM particles to the total relic density in $\textbf{R-I}$, Fig.~\ref{fig:R-I:DM-conversion-b} shows that the pNGB DM $\chi$ leads in most of the parameter space. However, the contribution of $T^0$-DM increases with $m_{T^0}$, becoming dominant ($\gtrsim 50\%$) near $m_{T^0} \approx 1.0$ TeV and beyond. Unlike the pure $Y=0$ scalar triplet DM model, this enhancement in our model setup is driven by the DM conversion process, as illustrated in Fig.~\ref{fig:R-I:DM-conversion-a}, particularly in the sub-TeV mass regime, i.e., $m_{T^0} \gtrsim 500$ GeV.
For $m_{T^0} \gtrsim 1.0$ TeV, the contribution of pNGB DM to the total relic density becomes subdominant. This behaviour arises because, in $\textbf{R-I}$ where $T^0$ is the lighter DM component, the relic density of the heavier $\chi$-DM is thermally suppressed relative to the $T^0$-DM as its mass increases.
Thus, while DM conversion enhances $T^0$-DM contributions in the sub-TeV region, the hierarchy in DM masses ($m_\chi > m_{T^0}$) naturally leads to $T^0$-DM dominance at higher masses, i.e., $m_{T^0} \gtrsim 1.0$ TeV.
\begin{figure}[htpb!]
	\centering
	\hspace{-0.2cm}
	\subfigure[\label{fig:R-I:param-dependence-conv-a}]{\includegraphics[height=4.5cm,width=5.1cm]{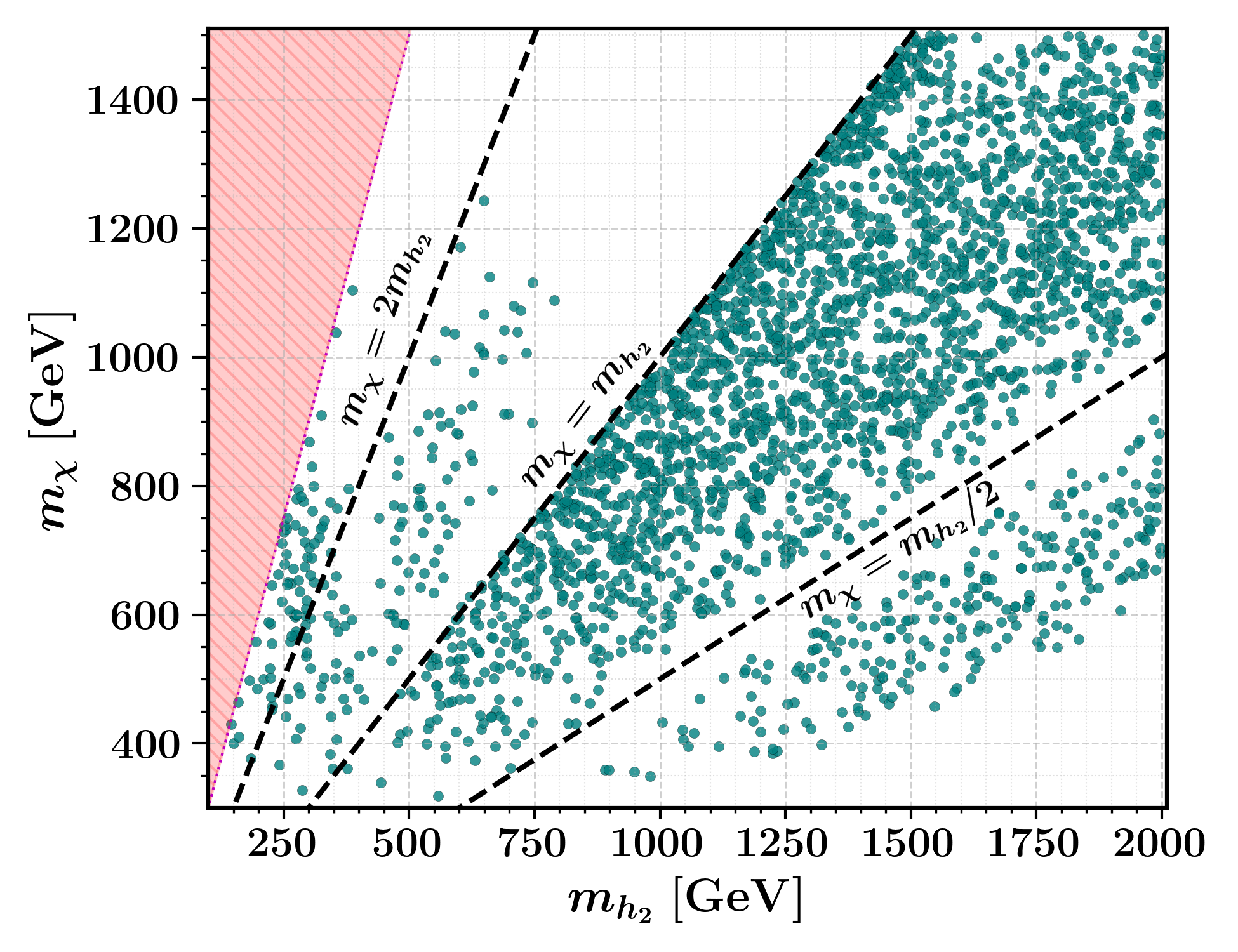}}
	\subfigure[\label{fig:R-I:param-dependence-conv-b}]{\includegraphics[height=4.5cm,width=5.1cm]{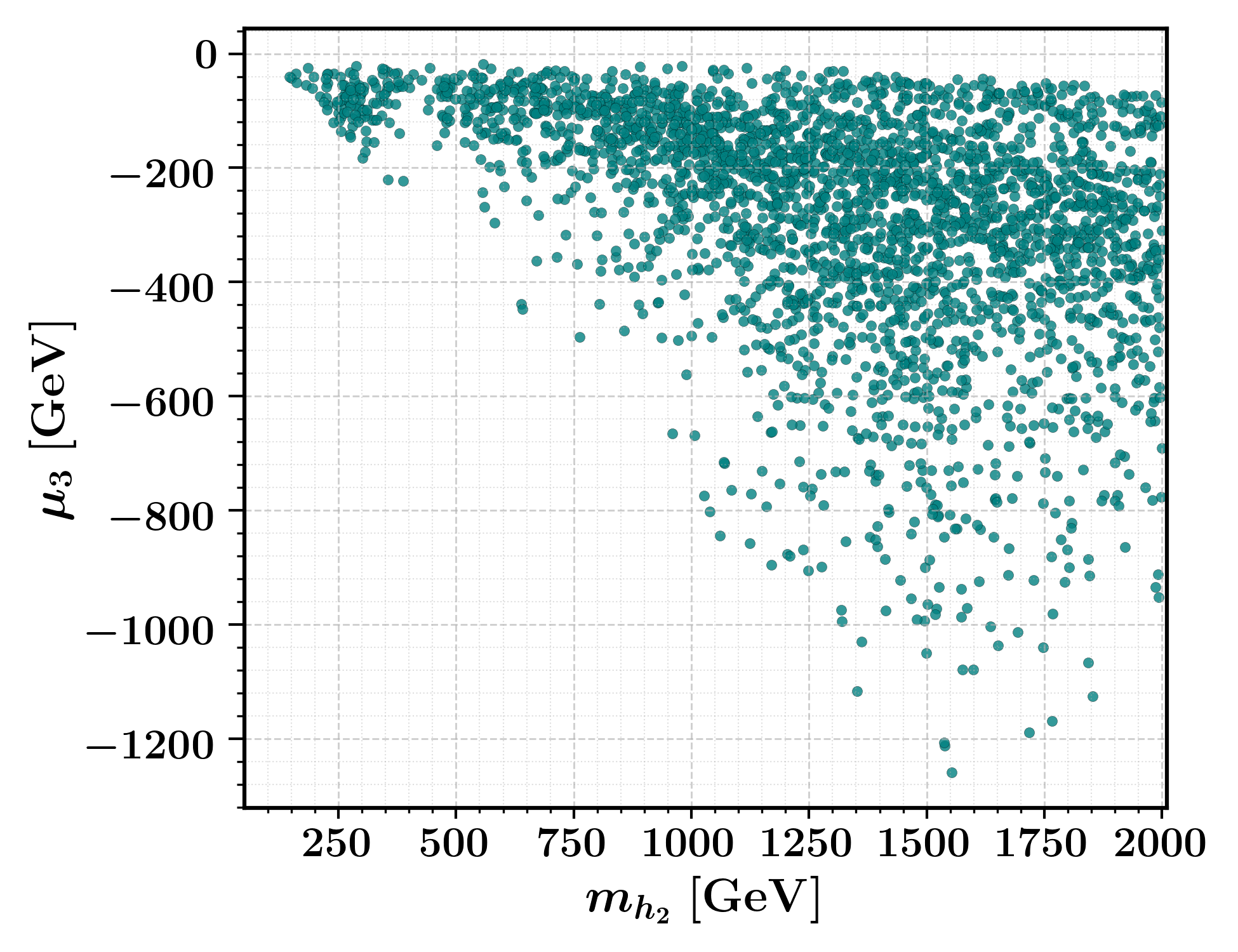}}
	\subfigure[\label{fig:R-I:param-dependence-conv-c}]{\includegraphics[height=4.5cm,width=5.1cm]{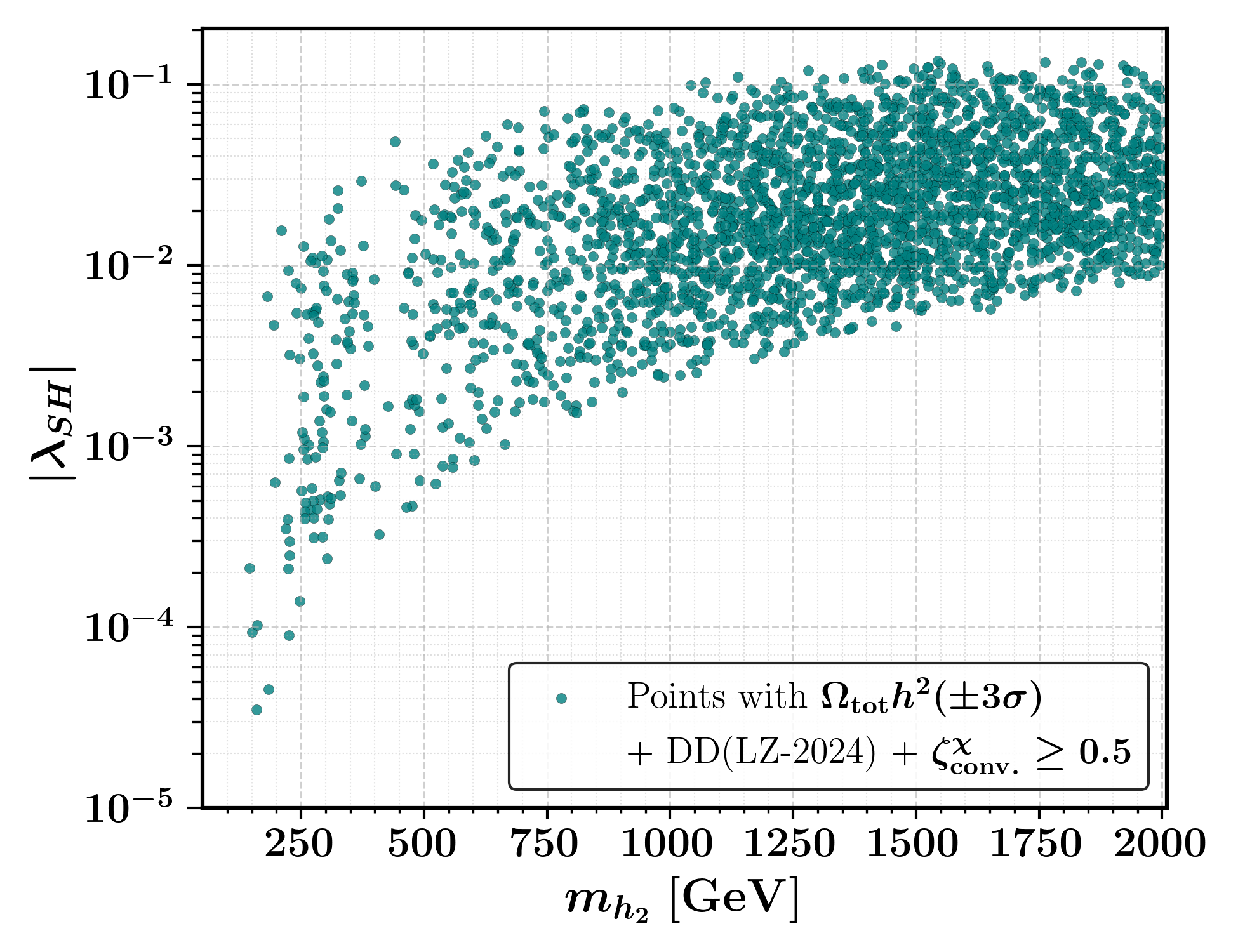}}
	%
	\caption{Parameter space of model points with a leading DM conversion, i.e., $\zeta^{\chi}_{\rm conv.} \geq 0.5$, in \textbf{R-I}. (a) Distribution of $m_\chi$ in the $m_{h_2}$-$m_\chi$ plane. (b) Correlation between $m_{h_2}$ and $\mu_3$. (c) Variation of $|\lambda_{SH}|$ with $m_{h_2}$. In (a), the light red coloured shaded region and distinctively styled black coloured lines have the same meaning as in Fig.~\ref{fig:R-I:param-dependence}(a). All model samples, apart from being consistent with the constraints discussed in Sec. \ref{sec:obs-cons} and maintaining $\zeta^{\chi}_{\rm conv.} \geq 0.5$, obeys the $3\sigma$ relic density limit and evades the DD bounds reported by LZ-2024, as detailed in the legend of plot (c).}
	\label{fig:R-I:param-dependence-conv}
\end{figure}

Given the significant role of the DM conversion in $\textbf{R-I}$, we analyse the parameter space of model points contributing dominantly to this process. We select points with $\zeta^{\chi}_{\rm conv.} \geq 0.5$ and present our findings in Fig.~\ref{fig:R-I:param-dependence-conv}. Here, Fig.~\ref{fig:R-I:param-dependence-conv}(a) reveals that pNGB DM masses yielding $\gtrsim 50\%$ enhancement to the $T^0$-relic via conversion are concentrated in the range $m_{h_2}/2 \lesssim m_\chi \lesssim m_{h_2}$, spanning the $h_2$-resonance to the degenerate mass case with $m_{h_2}$. While the remaining $m_\chi$ regions can also enhance $T^0$-relic, the density of such points is rather sparser. Across most of the parameter space with a significant DM conversion, $m_{h_2}$ exhibits a positive correlation with $|\mu_3|$, as shown in Fig.~\ref{fig:R-I:param-dependence-conv}(b). With increasing $m_{h_2}$, it prefers larger $|\mu_3|$ to have significant impact on the DM conversion with $\zeta^{\chi}_{\rm conv.} \geq 0.5$.
Next, we examine the role of the Higgs portal coupling $|\lambda_{SH}|$\footnote{$\l_{SH}$, as shown in Eq.~(\ref{eq:dependent-param}), is a derived parameter. Hence, values of $\l_{SH}$ are obtained using $m_{h_1} \approx 125$ GeV, and $m_{h_2}, v_S, \sin \theta$ ranges as shown in Eq.~(\ref{eq:scan-param}).} in the DM conversion, and is shown in the $\lambda_{SH}-m_{h_2}$ plane in Fig.~\ref{fig:R-I:param-dependence-conv}(c). Notably, $|\lambda_{SH}|$ increases with $m_{h_2}$, saturating near the maximum allowed $m_{h_2}$ in our scan. For the singlet-like Higgs mass, i.e., $m_{h_2} \lesssim 700$ GeV, $|\lambda_{SH}|$ as small as $10^{-5}$ to $10^{-3}$ can achieve $\zeta^{\chi}_{\rm conv.} \geq 0.5$. However, this region is less favoured as the number of model points are less populated in this region. In contrast, for the singlet-like Higgs mass $\gtrsim 700$ GeV, $|\lambda_{SH}|$ is predominantly concentrated between $10^{-3}$ and $10^{-1}$.
\subsubsection{Region-II: the $m_{T^0} > m_{\chi}$ regime} \label{subsubsec:DM-results:R-II}
\noindent We now examine region-II ($\textbf{R-II}$), characterised by the other possible DM mass hierarchy, i.e., $m_{T^0} > m_{\chi}$, and present our findings. Similar to Fig.~\ref{fig:R-I:param-dependence}, we project viable model points on the various planes, with the colour code and dashed lines retaining their meanings. 
\begin{figure}[!h]
	\centering
	\subfigure[\label{fig:R-II:param-dependence-a}]{\includegraphics[height=5.8cm,width=7.5cm]{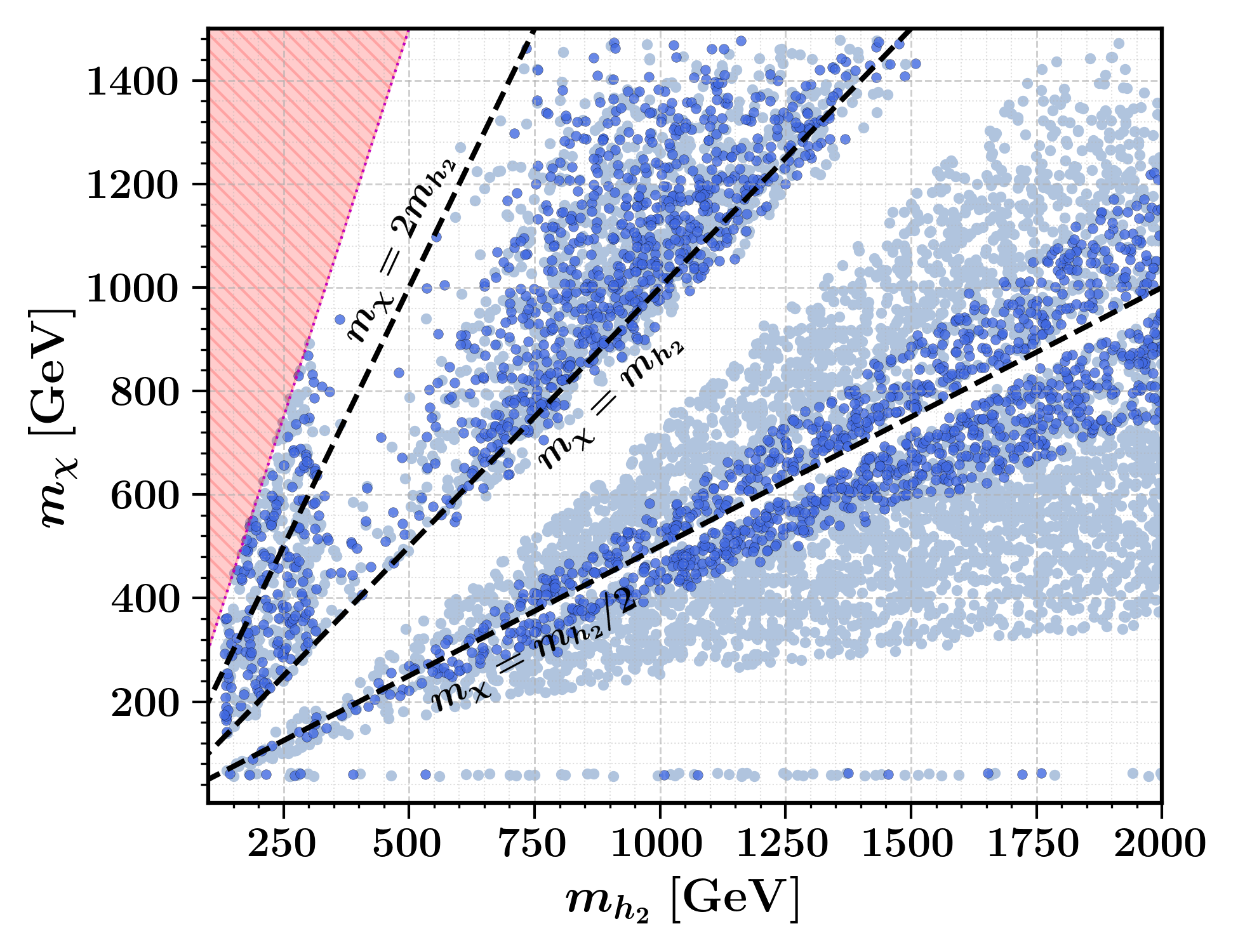}}
	\subfigure[\label{fig:R-II:param-dependence-b}]{\includegraphics[height=5.8cm,width=7.5cm]{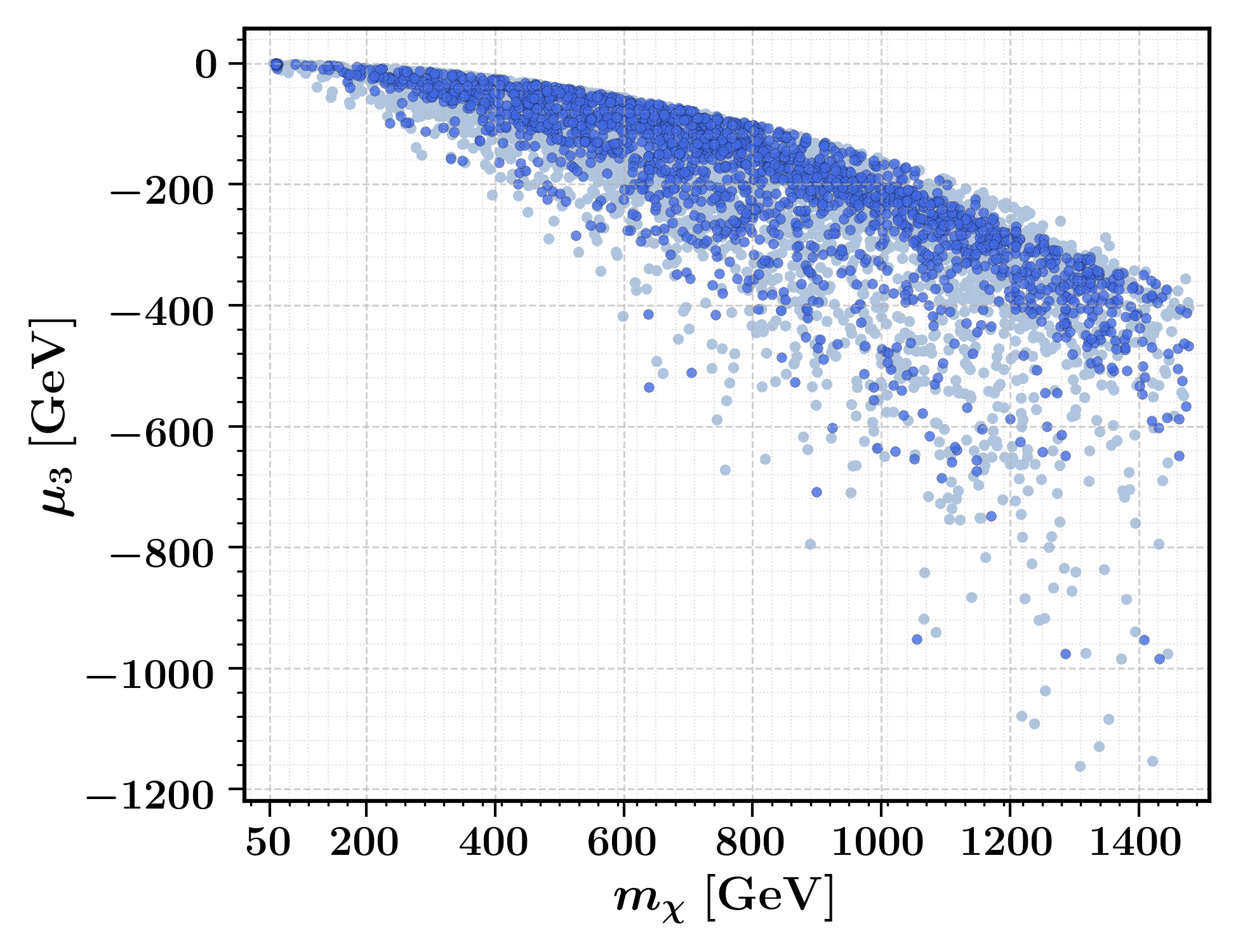}}
	\subfigure[\label{fig:R-II:param-dependence-c}]{\includegraphics[height=5.8cm,width=7.8cm]{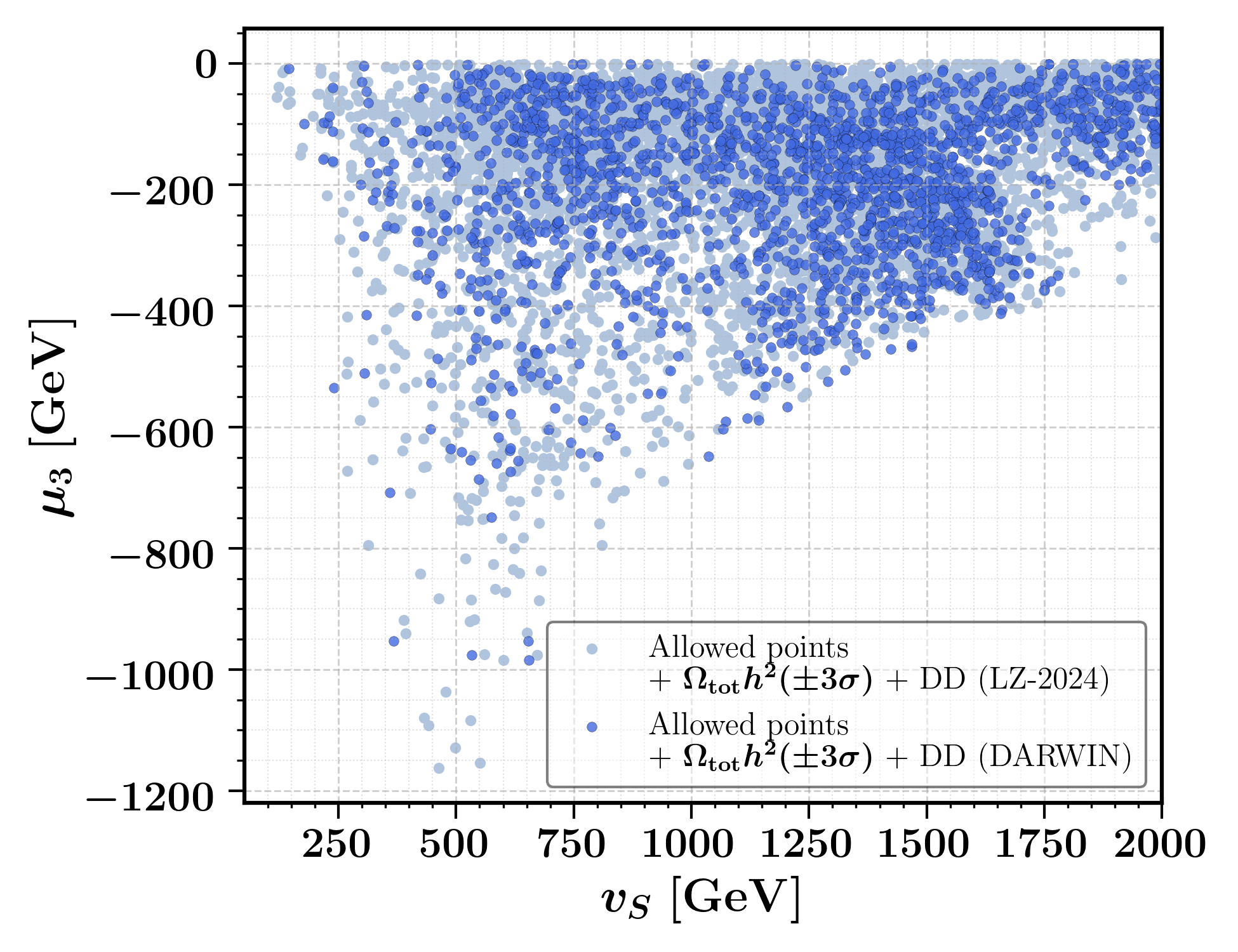}}
	
\caption{Similar to Fig.~\ref{fig:R-I:param-dependence}, but depicting $\textbf{R-II}$ with $m_{T^0} > m_{\chi}$.} \label{fig:R-II:param-dependence}
\end{figure}
In $\textbf{R-II}$, $m_{\chi}$ is the lighter DM, thereby, allowing for lower pNGB DM masses near the $h_1$-resonance ($m_\chi \approx m_{h_1}/2$) to bypass various experimental constraints, as indicated by the horizontal model points parallel to $m_{h_2}$-axis in Fig.~\ref{fig:R-II:param-dependence-a}.
Unlike $\textbf{R-I}$, the parameter points evading DD bounds from LZ-2024 shows a notable concentrations near the $h_2$-resonance ($m_\chi = m_{h_2}/2$), the degenerate regime ($m_\chi \approx m_{h_2}$), and extend up to the $2 m_{h_2}$ threshold, but with greater sparsity beyond $m_\chi > m_{h_2}$ and fewer points near the upper threshold. When DARWIN limits are applied, the allowed parameter space shrinks around the $h_2$-resonance (dark blue coloured points), while the remaining mass regime exhibits a distribution similar to that allowed by the LZ-2024 bound. The $U(1)$ soft breaking parameter $|\mu_3|$ exhibits a similar behaviour with the pNGB DM mass $m_\chi$, as of $\textbf{R-I}$, which can be seen in Fig.~\ref{fig:R-II:param-dependence-b}. Finally, Fig.~\ref{fig:R-II:param-dependence-c} illustrates the correlation between $|\mu_3|$ and the singlet VEV $v_S$, retaining the general features similar to those in $\textbf{R-I}$. However, unlike $\textbf{R-I}$, the lower limit for $v_S$ shifts to $v_S \gtrsim 100$ GeV to satisfy relic density constraints and evade DD bounds from LZ-2024 and DARWIN. This shift arises because $m_\chi$ can now approach the lower limit of the scan range, owing to $m_\chi < m_{T^0}$, relaxing the earlier lower bound on $v_S$. Additionally, in $\textbf{R-II}$, the allowed parameter points can approach very small $|\mu_3|$, i.e., $|\mu_3| \lesssim 10$ GeV, covering the entire allowed $v_S$ parameter space. Whereas in $\textbf{R-I}$, such a trend appears near $v_S \sim 1500$ GeV or beyond.
\begin{figure*}[!h]
	\centering
	\subfigure[\label{fig:R-II:DM-conversion-a}]{\includegraphics[height=5.8cm,width=6.8cm]{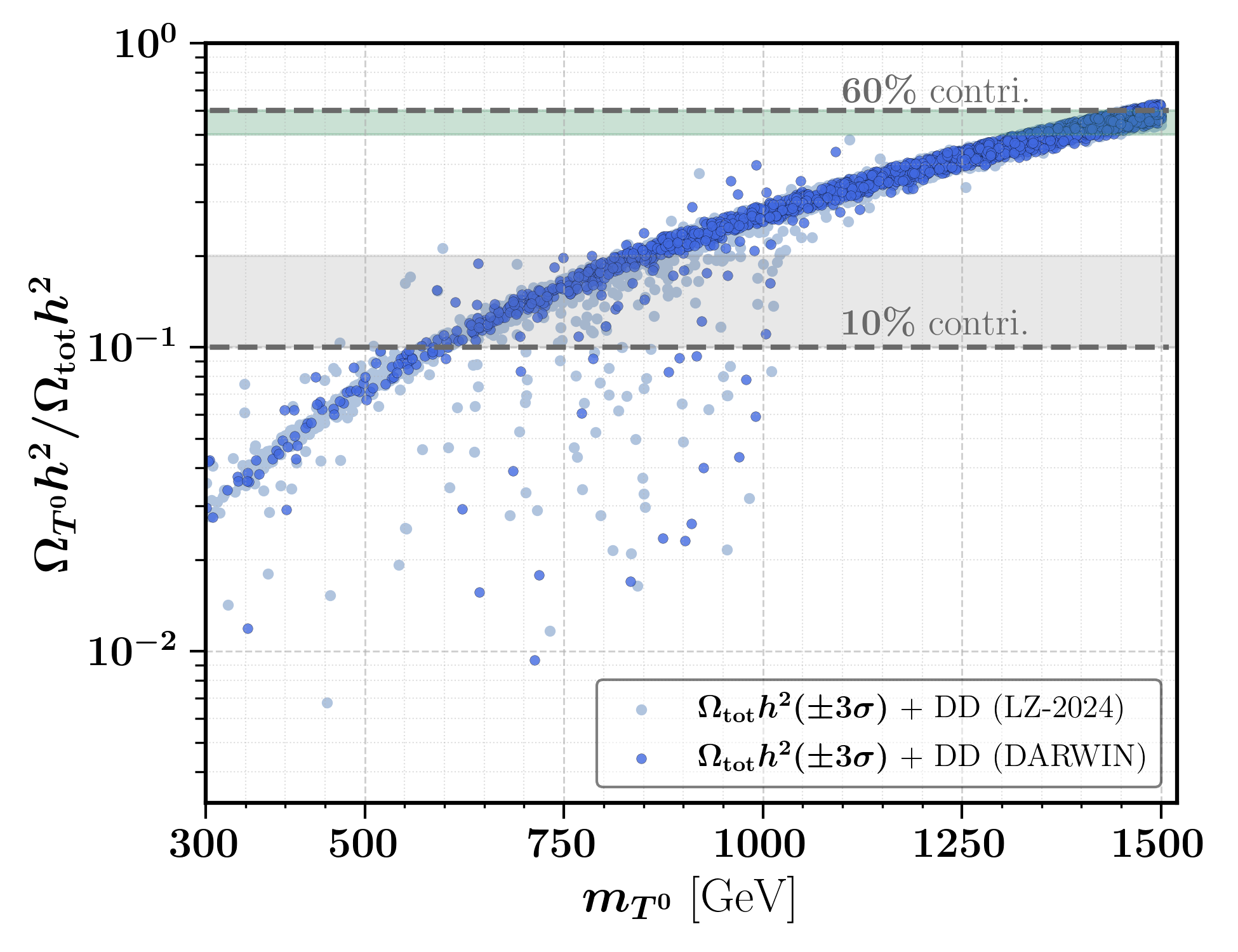}}
	\subfigure[\label{fig:R-II:DM-conversion-b}]{\includegraphics[height=5.8cm,width=6.8cm]{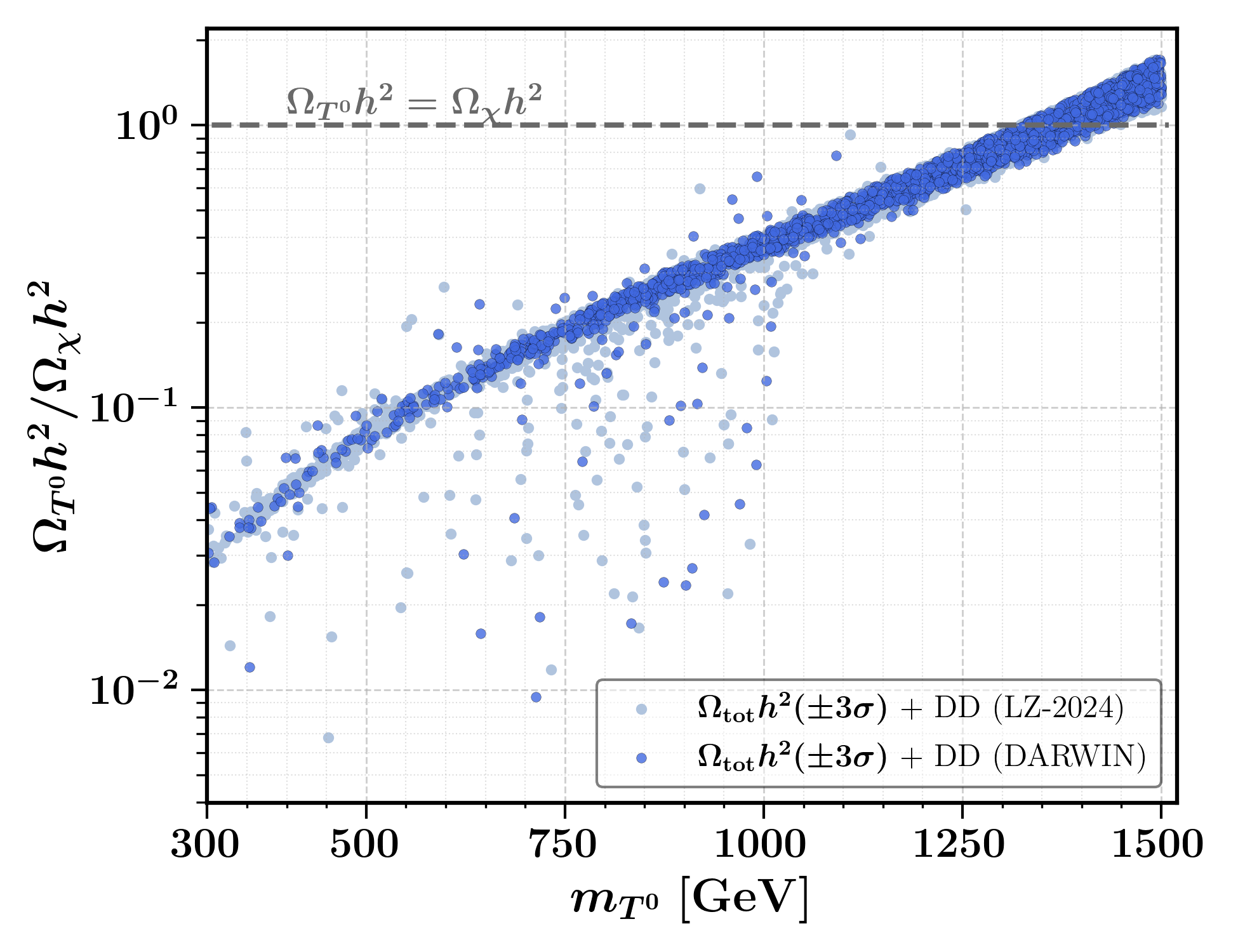}}
\caption{Contribution of the $T^0$-DM relic density to $\Omega_{\rm{tot}} h^2$ in \textbf{R-II}. (a) Projection on the $m_{T^0}$-plane, showing that $T^0$-DM remains subdominant throughout the sub-TeV region and only reaches $\sim 50\%$ contribution near $m_{T^0} \sim 1.35$ TeV. (b) Relative contributions of $T^0$ and $\chi$ relic densities to $\Omega_{\rm tot} h^2$, where $\chi$-DM dominates in most of the parameter space, with $T^0$-DM surpassing $\chi$ only near $m_{T^0} \sim 1.35$ TeV and beyond.}
	\label{fig:R-II:DM-conversion}
\end{figure*}

In $\textbf{R-II}$, $\chi$ is the lighter DM, making the process $T^0 T^0 \rightarrow \chi \chi$ kinematically allowed. However, the heavier DM, $T^0$, remains underabundant in our mass range of interest, $300 \lesssim m_{T^0} ~[{\rm GeV}] \lesssim 1500$, thereby rendering its contribution to $\Omega_\chi h^2$ via DM conversion negligible. In fact, in $\textbf{R-II}$, the triplet scalar DM behaves similarly to the pure $Y=0$ scalar triplet DM model. Our results for the $T^0$'s contribution to $\Omega_{\rm{tot}} h^2$ and its relative contribution compared to $\Omega_\chi h^2$ are presented in Fig.~\ref{fig:R-II:DM-conversion}. We find that it is the pNGB DM that plays the dominant role in this mass regime in contributing to the total relic density that satisfies the $3 \sigma$ PLANCK limit. Unlike $\textbf{R-I}$, the contribution of $T^0$-DM to the total relic density never reaches $~50\%$ or higher within the sub-TeV region. This is, however, observed near the upper mass limit of $\sim 1.5$ TeV, as shown in Fig.~\ref{fig:R-II:DM-conversion-a}. Further, the relative contribution of the $T^0$-DM, compared to the pNGB DM, only becomes dominant ($\gtrsim 50\%$) near $m_{T^0} \sim 1.35$ TeV and beyond (see Fig.~\ref{fig:R-II:DM-conversion-b}), in contrast to $\textbf{R-I}$, where this transition occurs near $m_{T^0} \sim 1.0$ TeV. 
\subsubsection{Direct and indirect detections of the Dark Matter} \label{subsubsec:DM-results:indirect-detection}
We now discuss the DD and ID prospects of the DM detection in our model. We emphasise that the DD constraints from the LZ-2024 and DARWIN have already been incorporated in our analysis of the parameter dependence in $\textbf{R-I}$ and $\textbf{R-II}$, as detailed in \ref{subsubsec:DM-results:R-I} and \ref{subsubsec:DM-results:R-II}. Here, we first present the DD prospects for the two DM mass regimes, $\textbf{R-I}$ and $\textbf{R-II}$, followed by a discussion of the ID prospects for the general case.
\begin{figure*}[!htbp]
	\hspace*{0.10cm} 
	\subfigure{\includegraphics[height=5.9cm,width=7.9cm]{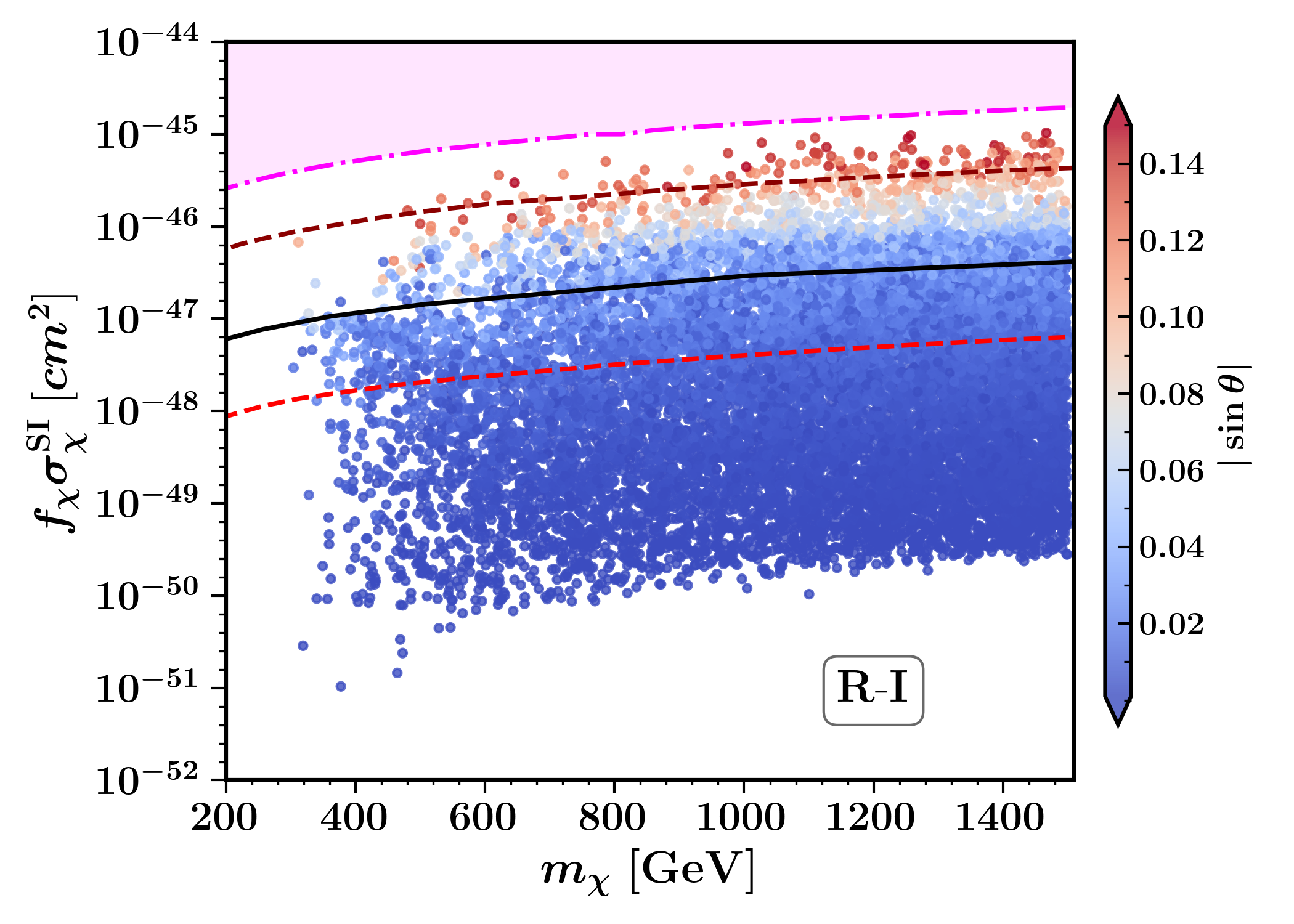}}
	\subfigure{\includegraphics[height=5.9cm,width=7.8cm]{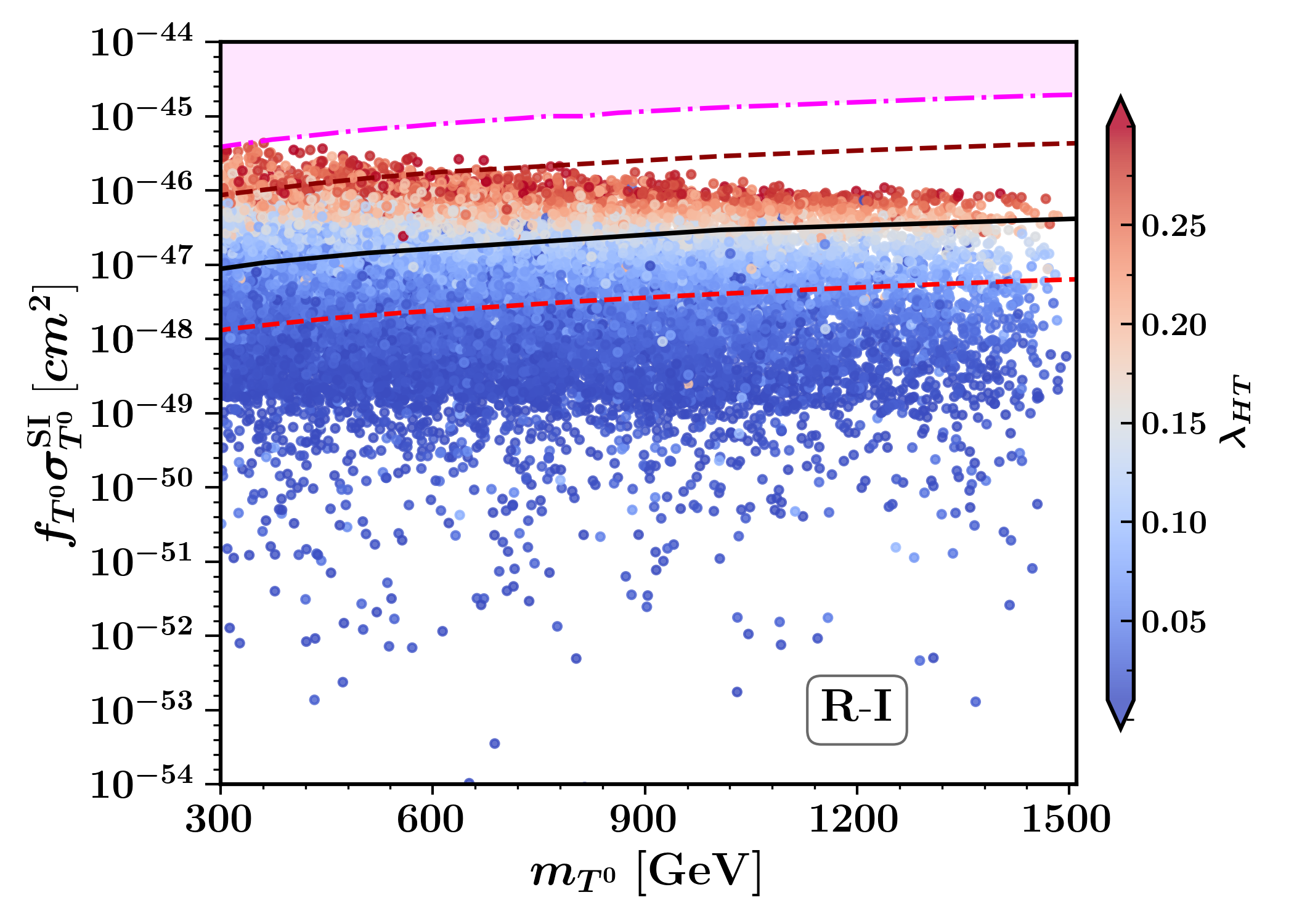}}
	\hspace*{0.10cm}
	\subfigure{\includegraphics[height=5.9cm,width=7.9cm]{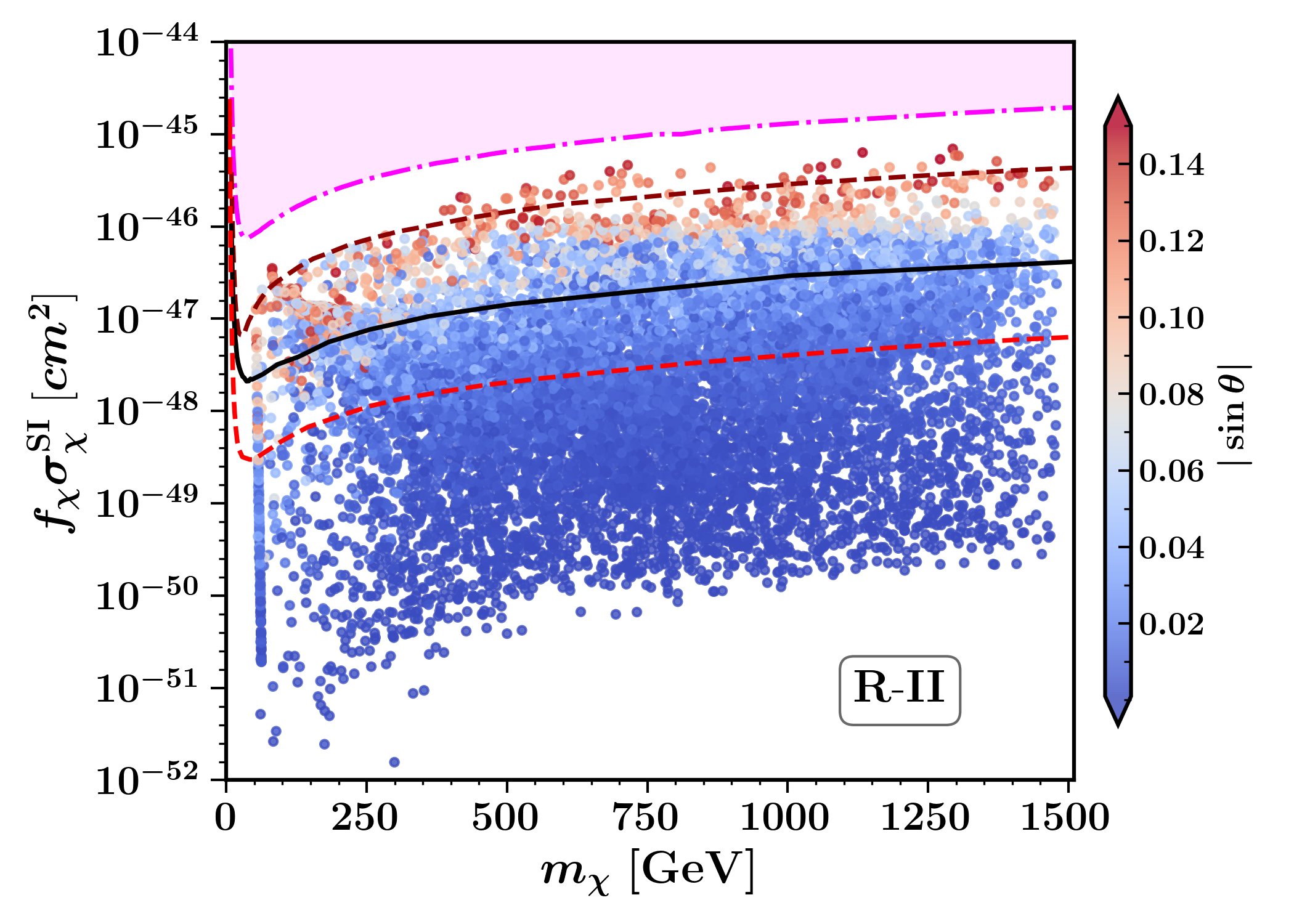}}
	\subfigure{\includegraphics[height=5.9cm,width=7.8cm]{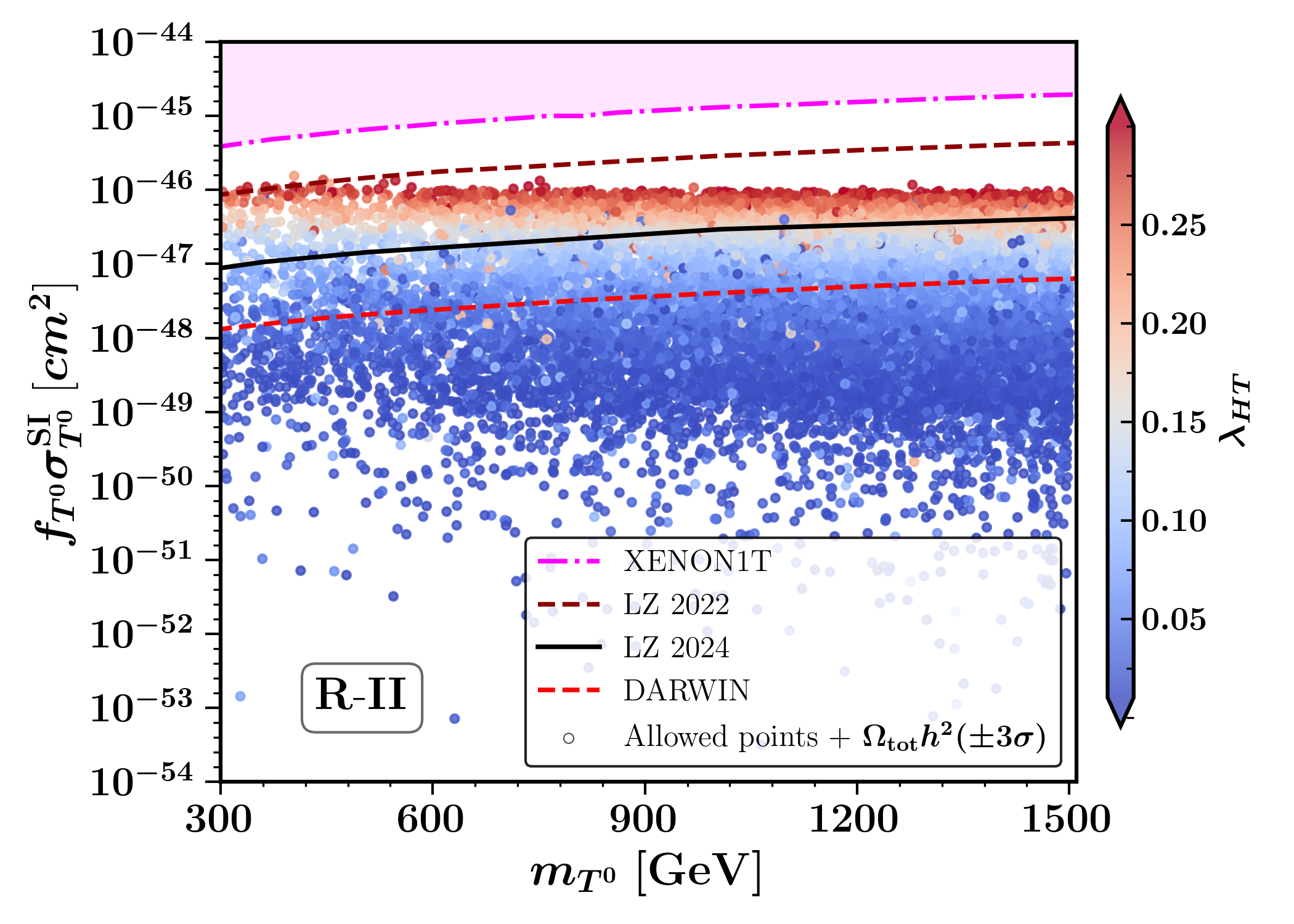}}
\caption{Distributions of the SI cross-sections for the elastic scattering of $\chi$ and $T^0$ with nuclei, scaled by their respective fractions, $f_\chi$ and $f_{T^0}$, and projected on the $m_\chi$ and $m_{T^0}$ mass planes. The upper (lower) panel depicts the $\textbf{R-I}$ ($\textbf{R-II}$) regime. The colour bars in the upper-left (upper-right) panel and lower-left (lower-right) panel reflect the variations of $|\sin\theta|$ ($\l_{HT}$). Various DD limits from different experiments are represented by differently styled and coloured lines, as detailed in the legend (bottom right panel). The allowed points, for all these four plots, satisfy the constraints discussed in Sec.~\ref{sec:obs-cons}.}
	\label{fig:DD}
\end{figure*}

Fig.~\ref{fig:DD} shows the DD prospects of the viable model points that pass all the constraints outlined in Sec.~\ref{sec:obs-cons}, including the $3 \sigma$ DM relic density limit. The top panel compares the SI DM-nucleon scattering cross-sections, for the pNGB DM $\chi$ (left) and scalar triplet DM $T^0$ (right), scaled to their relative abundances, i.e., $f_{i=\{\chi, T^0\}}$, against the current limit set by XENON1T (pink coloured dashed-dotted line), LZ-2022 (brown coloured dashed line), recently updated LZ-2024 (black coloured solid line) and the expected sensitivity reach of the DARWIN (dashed red coloured line) for $\textbf{R-I}$. The lower panel provides a similar comparison for $\textbf{R-II}$. The shaded pink region represents the exclusion limit from the XENON1T\footnote{For the subsequent analyses in this work, we will consider XENON1T pass model points for illustrative purposes. Stronger DD limits, e.g., from DARWIN, however, impose tighter constraints and would significantly reduce the number of viable parameter points.} experiment. For the pNGB DM case, we further show variation of the mixing angle $|\sin\theta|$ in the colour bar, while the variation of the Higgs-triplet parameter $\l_{HT}$ is shown for the triplet scalar DM case. We observe that, for the pNGB DM, larger values of $|\sin\theta|$ correspond to stronger DD signals in both regimes. Similarly, for the triplet DM, larger $\lambda_{HT}$ values lead to enhanced DD signals in both regimes. Notably, the maximum allowed values of $|\sin\theta|$ and $\lambda_{HT}$ within our scan range (see Eq.~(\ref{eq:scan-param})) remain consistent with DD bounds for both regimes.

In $\textbf{R-I}$ (top panel), the entire pNGB DM mass range, from $m_\chi > 300$ GeV (top left panel) up to the maximum allowed mass in our scan, remains testable in the current and the proposed DD experiments. Similarly, for the $T^0$-DM case in $\textbf{R-I}$ (top right panel), the chosen full triplet DM mass range remains unaffected by the DD constraints. Note that, these model points, as already mentioned, further pass the relic limits imposed by PLANCK. Thus, extending the $Y=0$ scalar triplet DM model with a pNGB DM offers a significant amount of allowed parameter space that accommodates the correct relic density while remaining accessible at the current and future DD experiments. A similar conclusion also holds for the regime $m_{T^0} > m_\chi$, $\textbf{R-II}$ (bottom panel). This demonstrates that our framework, consistent with constraints outlined in Sec.~\ref{sec:obs-cons}, successfully revives the ``desert region'' of the pure $Y=0$ scalar triplet DM, a feat unachievable in a single-component DM scenario \cite{Cirelli:2005uq,Araki:2011hm, Khan:2016sxm}.
 
The key difference regarding the DD prospects between the two regimes lies in the lower allowed mass limits for the pNGB DM $\chi$, which stem from the distinct mass hierarchies in each case. For example, in $\textbf{R-II}$, where $\chi$ is the lighter DM, it can satisfy both DD and relic density constraints ranging from the $h_1$-resonance region up to $\sim 1.5$ TeV. In contrast, the lower bound for $\chi$ in $\textbf{R-I}$ is around $\sim 300$ GeV, a limit enforced by the mass hierarchy $m_\chi > m_{T^0} > 300$ GeV, which shrinks the viable parameter space for $\chi$ compared to $\textbf{R-II}$. Regarding the $T^0$-DM, the DD constraints are slightly more stringent in $\textbf{R-I}$, particularly for $m_{T^0} \lesssim 500$ GeV, due to a larger Higgs-triplet coupling $\lambda_{HT} \gtrsim 0.20$. In $\textbf{R-II}$, where the pNGB DM is the dominant contributor to the relic density (see discussion in \ref{subsubsec:DM-results:R-II}), the relative abundance of $T^0$-DM, denoted as $f_{T^0}$ (see Eq.~(\ref{eq:constraints:DM-relic-rescaled})), is significantly suppressed compared to $\textbf{R-I}$. Consequently, $T^0$ experiences a weaker DD constraints, even for $m_{T^0} \lesssim 500$ GeV and $\lambda_{HT} \gtrsim 0.20$, as shown in Fig. \ref{fig:DD} (bottom right panel).

Following the discussion on the DD prospects, Fig.~\ref{fig:DD} demonstrates that our model points, which adhere to the PLANCK relic density constraint, lie within the projected sensitivity of DARWIN and a significant fraction of them feature SI DD cross-sections just below the current limit. Thus, both the regimes, i.e., $\textbf{R-I}$ and $\textbf{R-II}$, offer excellent prospects to be tested in the current and the planned DD experiments.
\begin{figure}[htbp!]
    \hspace*{2.5cm}
	\begin{minipage}{0.75\textwidth}
		\subfigure[\label{fig:indirect-a}]{\includegraphics[height=6.0cm,width=8.0cm]{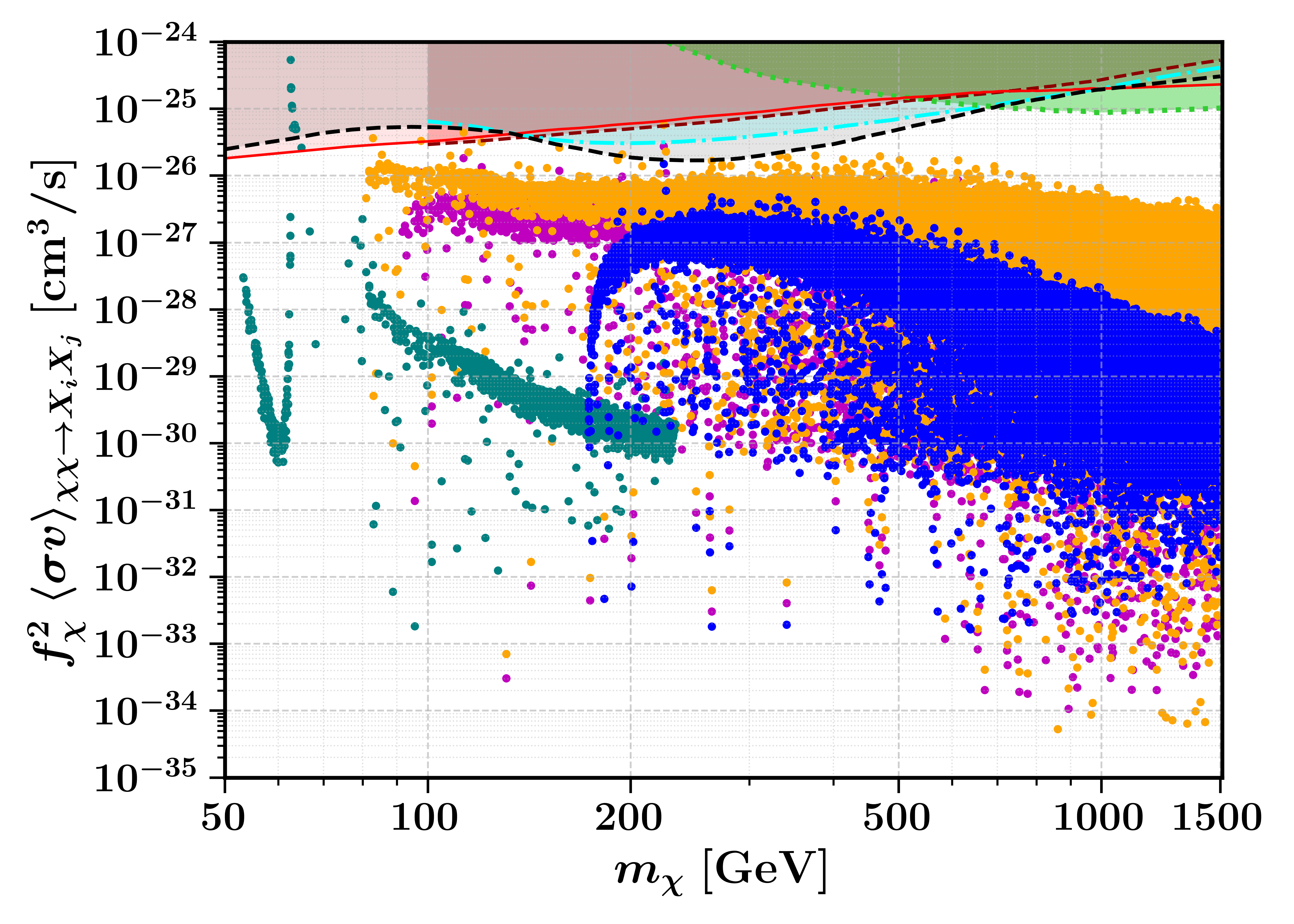}}
		\vspace{0.3cm} 
		\subfigure[\label{fig:indirect-b}]{\includegraphics[height=6.0cm,width=8.0cm]{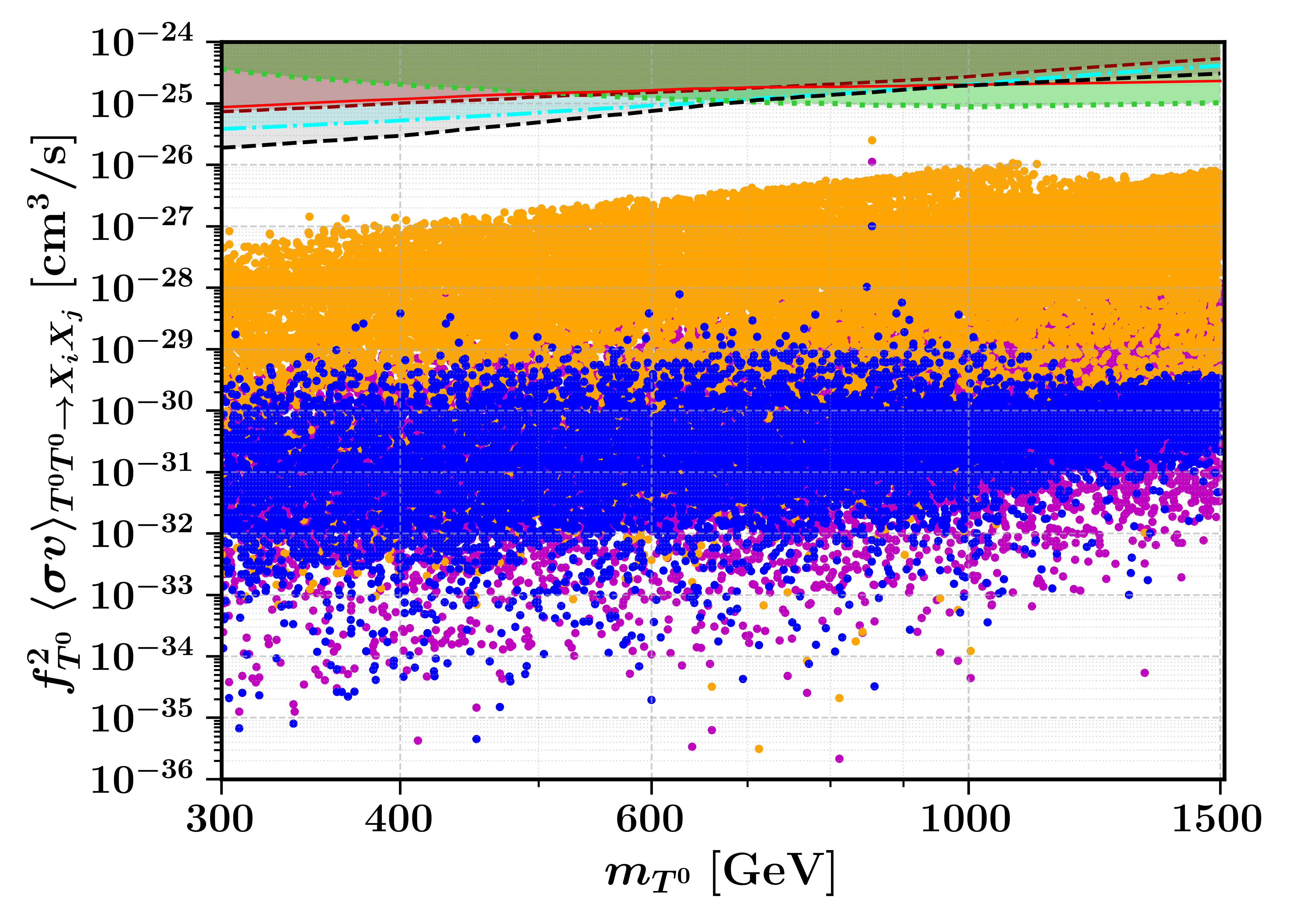}}
	\end{minipage}%
	\hspace{-3.0cm} 
	\begin{minipage}{0.20\textwidth}
		\raisebox{1.5cm}{
		\subfigure{\includegraphics[height=4cm,width=\linewidth]{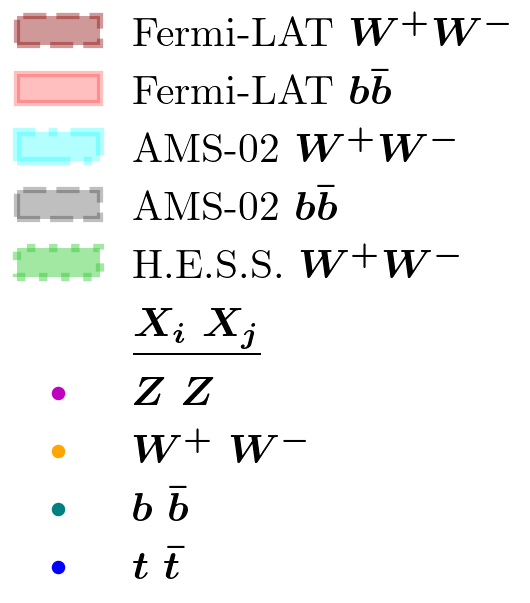}}
	    }
	\end{minipage}
\caption{Summary of important ID constraints on the model parameter space. (a) Dominant DM annihilation channels and the scaled cross-sections for pNGB DM $\chi$ and (b) corresponding results for the scalar triplet DM ($T^0$), satisfying the $3\sigma$ relic density and DD bounds (at least from the XENON1T). The different final states $X_i X_j$ ($X_{i(j)} = Z, W^+ (W^-), b ( \bar{b}), t (\bar{t})$) are differently colour-coded, with exclusion limits from Fermi-LAT, AMS-02, and H.E.S.S. are indicated by shaded regions (see the legend).
	}
	\label{fig:indirect}
\end{figure}

For ID, the most relevant DM annihilation channels are shown in Fig. \ref{fig:indirect} for the model points that accommodate the correct relic density and evade the DD bounds (at least from XENON1T), along with their corresponding scaled cross-sections. We adopt the strategy as discussed earlier in subsection~\ref{subsec:DM-observables}, and examine whether $\langle \sigma v \rangle_{VV}$ annihilations dominate across most of the parameter space. To test our model predictions, we compare these results with experimental limits obtained from gamma-ray searches for the DM \cite{Fermi-LAT:2015kyq, Fermi-LAT:2015att, AMS:2016oqu, HESS:2022ygk}, using the exclusion lines provided in Ref.~\cite{Reinert:2017aga}. Note that an exception arises for the low ($\lesssim 100$ GeV) pNGB DM masses, where we instead apply the $\langle \sigma v \rangle_{b\bar{b}}$ limits derived from Refs.~\cite{Reinert:2017aga, Fermi-LAT:2015att}. 
Fig.~\ref{fig:indirect-a} shows the dominant annihilation channels and their scaled cross-sections for the pNGB DM $\chi$, while Fig.~\ref{fig:indirect-b} presents the corresponding results for the scalar triplet DM $T^0$. The annihilation cross-sections into various $X_i X_j$ final states, with $X_i X_j = ZZ, W^+W^-,b\bar{b},t\bar{t}$, are distinguished by different colour codes, with experimental limits from and Fermi-LAT \cite{Fermi-LAT:2016uux,Fermi-LAT:2015kyq,Fermi-LAT:2015att,Ackermann:2015tah,Ackermann:2013yva,Ackermann:2013uma,Abramowski:2012au,Ackermann:2012nb,Fermi-LAT:2011vow,Abdo:2010ex} and AMS-02 \cite{AMS:2002yni, AMS:2016oqu, AMS:2021nhj} indicated by differently coloured shaded regions, as detailed in the legend of Fig.~\ref{fig:indirect}. Note that, we also include a limit coming from H.E.S.S. \cite{HESS:2022ygk}, which is important for DM masses above $\sim 500$ GeV. We find that most of the parameter space, that already passed the $3\sigma$ relic density bound, evaded DD limits (at least from XENON1T\footnote{As we mentioned earlier, XENON1T is chosen solely for demonstration purposes. Imposing other DD limits, such as DARWIN, would significantly reduce the phenomenological viable points.}) and other constraints outlined in Sec.~\ref{sec:obs-cons}, remains unconstrained by the current ID experiments, except for a few points with pNGB DM masses in the range of $\approx 60-65$ GeV, when the dominant contribution arises from $\langle \sigma v \rangle_{b \bar{b}}$. Meanwhile, for the scalar triplet DM, the entire mass range under consideration, i.e., $300-1500$ GeV, remains unconstrained, as illustrated in Fig.~\ref{fig:indirect-b}. As a result, we can conclude that ID experiments remain ineffectual on the viable parameter space of the chosen model, which are consistent with the correct relic density, obey DD bounds and other constraints as addressed in Sec.~\ref{sec:obs-cons}.

In concluding this section, where we have characterised the viable parameter space of our model and determined its direct and indirect detection prospects, we review our key findings as below:
\begin{itemize}
	\item \textbf{Revival of the Desert Region:} Our model, consistent with the collider searches, satisfies relic density and the current DD/ID limits from various experiments across a broad range of DM masses, as depicted via Figs. \ref{fig:DD} and \ref{fig:indirect}. Notably, the \emph{``desert region''}, which is excluded in the $Y=0$ scalar triplet DM model, becomes viable due to the presence of a pNGB DM and the mechanism of DM conversion.
	
	\item \textbf{Enhanced Relic Contribution via DM Conversion:} In the regime where $T^0$ is the lighter DM particle (\textbf{R-I}), the DM conversion process $\chi \chi \rightarrow T^0 T^0$ significantly enhances the relic density contribution of $T^0$, reaching $\sim 50$--$60\%$ within the sub-TeV mass range, as shown in Fig. \ref{fig:R-I:DM-conversion}. This feature is unachievable in a pure $Y=0$ scalar triplet DM scenario. Conversely, in \textbf{R-II}, where the pNGB DM is the heavier one, the triplet scalar DM behaves like the conventional $Y=0$ scalar triplet model, with $\chi$ dominantly contributing to the total relic density, as presented in Fig. \ref{fig:R-II:DM-conversion}.
\end{itemize}

Therefore, besides being well-motivated, our model provides a consistent and verifiable framework for the two-component DM. We now turn to another distinctive aspect of the chosen framework: the dynamics of PTs and their associated cosmological signatures.
\section{Phase transitions and gravitational waves} \label{sec:PT-GW}
In the framework of EWBG, the EWPT (see Ref.~\cite{Athron:2023xlk} for a recent review) plays a key role in generating the observed baryon asymmetry of the Universe \cite{Steigman:1976ev,Cohen:1997ac,Planck:2018vyg} by providing an out-of-equilibrium environment -- one of the three Sakharov conditions \cite{Sakharov:1967dj}. Any EWPT analysis begins with the thermal effective scalar potential. In this section, we introduce the one-loop effective potential first at zero temperature and successively at finite temperature, followed by a discussion on the emergence of the FOPT in the chosen model. Subsequently, we will discuss the associated generation and detection prospects of GWs.
\subsection{Thermal effective potential} \label{subsec:PT-GW:thermal-pot}
In our investigation of the EWPT dynamics, we choose the three CP-even scalar fields $\varphi \equiv h,s, T^0$ as the dynamical fields. The zero temperature tree-level potential can be derived from Eq.~(\ref{eq:scalar-pot:tot-pot}), and in terms of the dynamical fields it can be written as,
\bea
\label{eq:pot:tree-level}
V_0(h,s,T^0) \equiv V_0(\varphi) &=& \frac{1}{2} \mu_{H}^2 h^2 + \frac{1}{4} \l_H h^4 + \frac{1}{2} \mu_S^2 s^2 + \frac{1}{4} \l_S s^4 + \frac{1}{2} \mu_{\bm{T}}^2 {T^0}^2 + \frac{1}{4} \l_{\bm{T}} {T^0}^{4}\nn\\
&&  + \frac{1}{4} \l_{SH}\, h^2 s^2 + \frac{1}{2 \sqrt{2}} \mu_3 s^3 + \frac{1}{4} \l_{HT}\, h^2 {T^0}^2 + \frac{1}{4} \l_{ST}\, s^2 {T^0}^2.
\eea
The one-loop correction to the tree-level potential at zero temperature is given by the Coleman-Weinberg (CW) potential \cite{Coleman:1973jx,Weinberg:1973am}. In the on-shell renormalisation scheme \cite{Delaunay:2007wb, Curtin:2014jma}, it can be written as
\bea
\label{eq:pot:V_CW}
V_{\rm CW} (\varphi, T=0) = \sum_{i} (-1)^{F_i} \frac{d_i}{64 \pi^2} \left[ m_i^4 (\varphi) \left(\log\frac{m_i^2(\varphi)}{m^2_{0i}} - \frac{3}{2}\right) + 2m_i^2(\varphi) m^2_{0i} \right],
\eea
where the index $i$ runs over all the particles contributing to the potential with $F_i = 0\,(1)$ for bosons (fermions), $d_i$ is the number of {\it d.o.f.} of the particle species, $m_i(\varphi)$ is the field dependent mass (see Appendix~\ref{appx:A1} for details) and $m_{0i}$ denotes its value at the EW vacuum. In the on-shell renormalisation scheme, the one-loop contributions preserve the tree-level minimisation conditions, eliminating the need for counter terms to maintain physical minima and masses at the EW vacuum, as encountered in the $\ovr{\rm MS}$ renormalisation scheme \cite{Coleman:1973jx}. The contribution of Goldstone bosons ($G^0, G^\pm$) to the CW potential requires special treatment, as their vanishing masses at the physical EW minimum lead to infrared (IR) divergences. While this issue is an artifact of perturbation theory \cite{Martin:2014bca,Elias-Miro:2014pca}, it can be mitigated by resumming their contributions or introducing an infrared regulator, say, $\mu_{\rm IR}^2 \simeq 1~{\rm GeV^2}$ \cite{Baum:2020vfl,Borah:2023zsb,Borah:2024emz}. However, since the numerical impact of this resummation is minimal \cite{Martin:2014bca}, we drop Goldstone boson contributions in our study.

Finally, the finite temperature one-loop contribution is encapsulated as \cite{Dolan:1973qd}
\bea
\label{eq:pot:thermal}
V_{\rm T} (\varphi, T) = \frac{T^4}{2 \pi} \left[ \sum_{i=B} n_i^B J_B\left(\frac{m_i^2(\varphi, T)}{T^2} \right) + \sum_{i=F} n_i^F J_F\left(\frac{m_i^2(\varphi, T)}{T^2} \right)\right], 
\eea
where the two sums are over all the possible bosonic ($n_i^B$) and fermionic ($n_i^F$) {\it d.o.f}. respectively. $m^2_i(\varphi,T)$ is defined in Eq.~(\ref{eq:field_mass:daisy-coeff}) and the corresponding thermal functions \cite{Anderson:1991zb} are,
\bea
\label{eq:pot:thermal-functions}
J_{B}(x) =  \int_{0}^{\infty} dy ~y^2 \ln \left(1 - e^{-\sqrt{y^2+x^2}}\right)\&~~
J_{F}(x) = - \int_{0}^{\infty} dy ~y^2 \ln \left(1 + e^{-\sqrt{y^2+x^2}}\right).
\eea 
Here, $B\,(F)$ denotes bosons (fermions), and $x^2 \equiv \frac{m_i^2(\varphi, T)}{T^2}$. The resummation of the leading self-energy daisy diagrams \cite{Carrington:1991hz} is required for a consistent treatment of the thermal corrections, which shifts the field-dependent masses \cite{Curtin:2016urg}, see Appendix \ref{appen:A2} for details.

With the above pieces of information, the finite temperature one-loop effective potential can be expressed as a combination of Eqs.~(\ref{eq:pot:tree-level}), (\ref{eq:pot:V_CW}) and (\ref{eq:pot:thermal}), as
\bea
\label{eq:pot:Eff-pot}
V_{\rm eff}(\varphi,T) \equiv V^T_{\rm eff} = V_0(\varphi) + V_{\rm CW} (\varphi, T=0) + V_{\rm T}(\varphi,T).
\eea
\subsection{Nucleation and percolation} \label{subsec:PT-GW:perco-nuc}
The cosmological FOPT occur via nucleating true vacuum bubbles which expand in a space filled with the false vacuum. The time of nucleation at which the probability of a true vacuum bubble forming within
a horizon radius becomes significant, can be obtained from \cite{Enqvist:1991xw}
\bea
\label{eq:nucleation:volume}
N(T_n) = \int_{T_n}^{T_c} \frac{dT}{T} \frac{\Gamma(T)}{H(T)^4} = 1.
\eea
Here, $T_c$ is the critical temperature where the minima of $V_{\rm eff} (\varphi, T)$ (see Eq.~(\ref{eq:pot:Eff-pot})) become degenerate, and $T_n$ is the nucleation temperature. The Hubble expansion rate is given by \cite{Lewicki:2024xan}
\bea
\label{eq:nucleation:tota-Hubble-rate}
H(T)^2 = \frac{\rho_{\rm rad}}{3 M^2_{\rm Pl}} + \frac{\Delta V(\varphi,T)}{3 M^2_{\rm Pl}}, ~~\text{with}~~ \rho_{\rm rad} = \frac{\pi^2}{30} g_{\ast}(T) T^4,
\eea
where we have also included the vacuum contribution from $\Delta V(\varphi,T)$\footnote{This is essential \cite{Lewicki:2021pgr,Ellis:2022lft} when the model can undergo a sufficiently strong transitions, $\alpha \gtrsim \mathcal{O}(1.0)$ \cite{Athron:2024xrh} (defined later).}, representing the potential difference between the false and true vacua. The radiation energy density of the relativistic particles is denoted by $\rho_{\rm rad}$, with $M_{\rm Pl}$ being the reduced Planck mass. The temperature-dependent effective number of {\it d.o.f}, $g_{\ast}(T)$, is derived from the tabulated data provided in Ref.~\cite{Saikawa:2018rcs}. Next, the nucleation probability per unit time per unit volume is given as,
\bea
\label{eq:nucleation:probability}
\Gamma(T) = \left(\frac{S_E}{2\pi T}\right)^{3/2} T^4 e^{-S_E/T}.
\eea
Here, $S_E$ represents the three-dimensional Euclidean action of the theory, which is obtained by integrating the equation of motion for the scalar fields $\varphi$. The condition for nucleation can be determined either from Eq.~(\ref{eq:nucleation:volume}) or by finding a solution that satisfies the following equality at $T_n$,
\bea
\label{eq:nucleation:possible-nucleation}
\Gamma(T_n) = H (T_n)^4.
\eea

After confirming the nucleation, it is crucial to check whether it completes properly \cite{Ellis:2018mja}. This involves finding the temperature where the probability of a point remaining in the false vacuum falls below 0.71 \cite{Lorenz:2001lor} and ensuring the false vacuum volume decreases at that temperature \cite{Lewicki:2021pgr, Ellis:2022lft}
\bea
\label{eq:percolation:percolation-condition}
I(T) = \frac{4 \pi}{3} \int_{T}^{T_c} \frac{dT^{\prime}}{H(T^{\prime})} \Gamma(T^{\prime}) \frac{r(T,T^{\prime})^3}{{T^{\prime}}^4} = 0.34, \quad T\frac{dI(T)}{dT} < -3,
\eea
where $I(T)$ represents the fragment of the space that has already been converted to the broken phase, and $r(t,t^{\prime}) = \int_{t^{\prime}}^{t} \frac{v_w(\tilde{t}) d\tilde{t}}{a(\tilde{t})}$ is the comoving radius of a bubble nucleated at time $t^{\prime}$ and propagated until $t$ with a bubble wall velocity $v_w$. The temperature at which both conditions stated in Eq.~(\ref{eq:percolation:percolation-condition}) are satisfied is called the percolation temperature, $T_p$.

It is often assumed that the FOPTs are instantaneous and complete at $T \simeq T_n$, however, the percolation temperature $T_p$ provides a more precise measure of when the transition completes. Recent studies, such as Ref.~\cite{Athron:2022mmm}, suggest that $T_n$ may not be suitable for strong transitions, whereas $T_p$ can capture the dynamics of the process more efficiently. Therefore, for accurately estimating parameters relevant to the GW spectra, it is preferable to evaluate them at $T_p$. The necessary thermodynamic parameters to estimate the GW signal will be defined in the relevant subsection.

Before moving to the next subsection, we define the criterion for determining whether a FOPT is strong or not. For EWBG to succeed, a strong FOPT (SFOPT) is necessary to suppress the baryon number violation after the EWPT\footnote{An SFOPT in the context of the EW sector, which is also the case of the present analyses, is also called an SFOEWPT.}. In general, it is known as the baryon washout condition \cite{Farrar:1993hn} and translates into
\bea
\label{eq:PT-strength}
\xi_c = \frac{v_c}{T_c} \gtrsim 1.0, \quad \xi_n = \frac{v_n}{T_n} = \frac{\sqrt{\left(\langle \varphi^{lT}_i\rangle - \langle \varphi^{hT}_i\rangle \right)^2} }{T_n} \gtrsim 1.0,
\eea
where $\xi_c$ ($\xi_n$) is defined at the critical (nucleation) temperature $T_c~(T_p)$. In Eq.~(\ref{eq:PT-strength}), $v_c$ is the VEV of the SM-like Higgs at $T = T_c$, and $\langle\varphi^{lT}_i \rangle$ ($\langle\varphi^{hT}_i \rangle$)  denotes the low (high) temperature minimum at true (false) vacuum, 
with $\varphi_i \equiv \{h,s,T^0\}$.
\subsection{Remarks on the gauge dependency} \label{subsec:PT-GW:gauge-inv}
Determination of quantities that are crucial to study the PT dynamics in a gauge theory from the full $V_{\rm eff} (\varphi, T) \equiv V^T_{\rm eff}$ of Eq.~(\ref{eq:pot:Eff-pot}) depends on the particular gauge choice ($\xi$) one undertakes. The quantity is only gauge-invariant at its global minima \cite{Nielsen:1975fs,Fukuda:1975di}. Therefore, estimation of $v_c/T_c$ or tunnelling calculations
are in general gauge dependent. However, gauge-invariant treatments of the effective potential are proposed in the literature, and interested readers can follow, for e.g., Refs.~\cite{Fukuda:1975di,Laine:1994zq,Kripfganz:1995jx,Patel:2011th,DiLuzio:2014bua} for more detail.

A straightforward way to ensure a gauge-invariant treatment of $V^T_{\rm eff}$ is possible through the high-temperature (HT) expansion approximation of $V^T_{\rm eff}$. This approach assumes that the EWPT occurs at a sufficiently high temperature (i.e., $T \gg$ involved mass scale), where one-loop thermal corrections dominate over the $V_{\rm CW}(\varphi, T=0)$ (see Eq.~(\ref{eq:pot:V_CW})), allowing an expansion up to $\mathcal{O}(T^2)$ \cite{Profumo:2007wc,Profumo:2014opa,Cho:2021itv}. Since gauge dependence in the PT calculations arises from the leading $\mathcal{O}(g^3)$ terms in $V^T_{\rm eff}$, truncating the expansion at $\mathcal{O}(g^2)~(\text{or}~\mathcal{O}(T^2))$ makes the HT expansion free from gauge dependence \cite{Patel:2011th,Kozaczuk:2019pet}. Therefore, ignoring $T=0$ contributions from CW corrections, the finite temperature cubic and tadpole terms from the $V^T_{\rm eff}$ (see Eq.~(\ref{eq:pot:Eff-pot})), the overall potential in the HT expansion can be expressed as\footnote{A tadpole term, $\propto \mu_3 s T^2$, is allowed in Eq.(\ref{eq:pot:HT}). However, as noted earlier, such a term can introduce gauge dependence at higher perturbative orders \cite{Profumo:2014opa}, which is why we have omitted it in Eq.~(\ref{eq:pot:HT}). Further, previous studies \cite{Profumo:2007wc, Profumo:2014opa} with similar potential structures considered for this work have found that the numerical impact of these terms in the PT dynamics is negligible, reinforcing our choice to exclude them.}
\bea
\label{eq:pot:HT}
V^{\rm HT}_{\rm eff} = V_0(h,s,T^0) + \frac{1}{2} \left( \Sigma_h h^2 + \Sigma_s s^2\right) T^2,
\eea
where $V_0(h,s,T^0)$ is given by Eq.~(\ref{eq:pot:tree-level}) and the thermal mass contributions are
\bea
\label{eq:pot:HT-thermal-masses}
	\Sigma_h = \frac{g_1^2}{16} + \frac{3g_2^2}{16} + \frac{y_t^2}{4} + \frac{\l_{H}}{2} + \frac{\l_{SH}}{12} + \frac{\l_{HT}}{8}, ~\text{and},~ 
	\Sigma_s = \frac{\l_{S}}{3} + \frac{\l_{SH}}{6} + \frac{\l_{ST}}{8}.
\eea
Here, $y_t$ being the top Yukawa coupling and $g_1, g_2$ are the $U(1)_Y, SU(2)_L$ gauge couplings of the SM, respectively.

Even though the HT potential is simple and gauge-independent, the validity of the HT expansion of $J_{B(F)}$ of Eq.~(\ref{eq:pot:thermal-functions}) is not guaranteed as the temperature decreases after a transition. Moreover, neglecting the one-loop $T=0$ contributions and higher-order thermal corrections may lead to an incomplete description of the model in quantitative studies. Ref.~\cite{Patel:2011th} introduces the Patel-Ramsey-Musolf (PRM) scheme, which incorporates higher-order corrections in a gauge-invariant manner. In this approach, $v_c$ is determined using the HT potential, while the gauge-invariant $T_c$ is obtained via the Nielsen-Fukuda-Kugo (NFK) identity \cite{Nielsen:1975fs, Fukuda:1975di}. Interested readers can refer to Ref.~\cite{Patel:2011th} for further details. However, the PRM scheme introduces explicit renormalisation scale dependence through the $V_{\rm CW}(\varphi, T=0)$ in the $\ovr{\rm MS}$ scheme, which can be significant in some cases \cite{Chiang:2017nmu, Chiang:2018gsn}. Nevertheless, this scale dependence can be mitigated, as discussed in Ref.~\cite{Chiang:2017nmu}.

In this study, instead of the PRM scheme, we adopt the HT potential for a gauge-invariant exploration of our model space. Whereas, we employ the full effective potential $V^T_{\rm eff}$ (see Eq.~(\ref{eq:pot:Eff-pot})) for detailed numerical analysis to capture the effects of higher-order corrections. Further, $V_{\rm CW} (\varphi,T=0)$ (see Eq.~(\ref{eq:pot:V_CW})) is treated in the on-shell renormalisation scheme, and we work in Landau gauge. Notably, previous studies (e.g., Refs.~\cite{Kozaczuk:2019pet, Chiang:2019oms, Croon:2020cgk, Arunasalam:2021zrs, Borah:2023zsb, Borah:2024emz}) showed that when a potential barrier is formed at the tree-level, the PT dynamics in Landau gauge align well with the gauge-independent treatments\footnote{Moreover, in Landau gauge, its dependence on the PT observable, such as the baryon washout criterion (Eq.~(\ref{eq:PT-strength})), is expected to be sub-dominant compared to other sources of theory uncertainty \cite{Garny:2012cg}, such as renormalisation scale dependence, missing higher-order corrections, and ambiguities in the thermal resummation procedure \cite{Bittar:2025lcr}.}. In our model, cubic terms, e.g., $\sim \mu_3 s^3$, in Eq.~(\ref{eq:pot:tree-level}), ensure the presence of a tree-level barrier. Thus, incorporating higher-order corrections via $V^T_{\rm eff}$ would yield results consistent with the PRM scheme, justifying its use.

Before concluding this subsection, we highlight that refinements to thermodynamic predictions beyond the one-loop effective potential have been achieved through two-loop calculations \cite{Niemi:2021qvp, Schicho:2021gca, Schicho:2022wty, Ekstedt:2022bff} and non-perturbative methods \cite{Gould:2022ran}. A dimensionally reduced 3-dimensional effective field theory (3D EFT) framework is also being developed for gauge-independent evaluations with thermal resummation \cite{Niemi:2021qvp, Schicho:2021gca, Schicho:2022wty, Ekstedt:2022bff,Gould:2022ran}. We leave an in-depth analysis beyond the one-loop or using 3D EFT for future work.
\subsection{Gravitational waves} \label{subsec:PT-GW:GWs}
In this subsection, we estimate the GW spectrum from the bubble dynamics during a FOPT. The relevant thermodynamic parameters, derived from the PT dynamics, required to determine the GW spectra are: $T_{\ast}, \alpha$ and $\beta$. Here, $T_{\ast}$ is the temperature at which FOPT completes, $\alpha$ relates to the energy budget of the FOPT and quantifies the strength of a cosmological PT, which is given as \cite{Espinosa:2010hh,Caprini:2019egz,Lewicki:2021pgr}
\bea
\label{eq:PT-GW:alpha}
\alpha = \frac{\Delta V_{\rm eff} (\varphi, T) - \frac{T}{4} \Delta \frac{\partial V_{\rm eff} (\varphi, T)}{\partial T}}{\rho_{\rm rad}}\Bigg|_{T = T_*}.
\eea
Here, $\Delta V_{\rm eff} (\varphi, T)$ represents the difference of $V^T_{\rm eff}$ between the false ($\varphi_{\rm{false}}$) and true ($\varphi_{\rm{true}}$) vacua, with a similar definition for $\Delta \frac{\partial V_{\rm eff}(\varphi, T)}{\partial T}$ and $\rho_{\rm rad}$ is defined in Eq.~(\ref{eq:nucleation:tota-Hubble-rate}). The parameter $\beta$, which indicates the inverse duration time of a PT, is usually normalised to the Hubble rate at the transition, $H(T_*) \equiv H_*$, and can be expressed as \cite{Grojean:2006bp}
\bea
\label{eq:PT-GW:beta}
\frac{\beta}{H_*} = T_{*} \frac{d}{dT} \left(\frac{S_E}{T}\right)\Bigg|_{T = T_*}.
\eea
The parameters in Eqs.~(\ref{eq:PT-GW:alpha}) and (\ref{eq:PT-GW:beta}) are evaluated at the time of GW production, which, in our case, corresponds to the temperature at which bubbles of the new phase percolate. Thus, we estimate the GW spectrum at $T_{\ast} \approx T_p$.

An SFOPT can give rise to stochastic GW background mainly through three different mechanisms: (i) bubble wall collisions \cite{Turner:1990rc,Kosowsky:1991ua,Kosowsky:1992rz}, (ii) sound waves \cite{Hindmarsh:2013xza,Hindmarsh:2015qta,Hindmarsh:2016lnk}, and (iii) magneto-hydrodynamic (MHD) turbulence in the plasma \cite{Kamionkowski:1993fg,Kosowsky:2001xp,Dolgov:2002ra,Gogoberidze:2007an,Caprini:2009yp}. In the absence of significant supercooling \cite{Witten:1980ez,Ellis:2018mja}, which is the case of the present model, as we will realise later, the dominant contribution would be from the sound wave source \cite{Caprini:2019egz}. However, percolation may also cause turbulence in the plasma and would have a non-negligible impact on the GW spectrum. Therefore, ignoring the contribution\footnote{In fact, for a PT proceeding in a thermal plasma, contribution from bubble collisions to the total GW energy density is assumed to be negligible \cite{Bodeker:2017cim}.} from bubble collisions, the full GW signal can be approximately written as
\bea
\label{eq:GW:total-contribution}
\Omega_{\rm GW} h^2 \simeq \Omega_{\rm sw} h^2 + \Omega_{\rm t} h^2,
\eea
where, $h=H_0/(100~{\rm km} \cdot {\rm sec}^{-1} \cdot {\rm Mpc}^{-1})$ with $H_0$ corresponding to Hubble’s constant at the present epoch, and $\Omega_{\rm sw} h^2(\Omega_{\rm t} h^2)$ corresponds to contribution from sound waves (MHD turbulence).

The contribution from sound waves to the total GW spectrum can be parameterised as a function of the frequency `$f$' as \cite{Hindmarsh:2013xza,Hindmarsh:2016lnk,Hindmarsh:2017gnf,Caprini:2019egz,Guo:2020grp},
\bea
\label{eq:GW:soundwave-amp}
\Omega_{\rm sw} h^2 &=& 4.13 \times 10^{-7} \left(1 - \frac{1}{\sqrt{1+ 2 \tau_{\rm sw} H_*}}\right) \left(\frac{100}{g_*(T)}\right)^{\frac{1}{3}} \left(\frac{f}{f_{\rm sw}}\right)^3 \left[\frac{4}{7} + \frac{3}{7} \left(\frac{f}{f_{\rm sw}}\right)^2\right]^{-\frac{7}{2}}\nn\\
&& \times 
\begin{dcases}
	\left(\frac{\kappa_{\rm sw} \alpha}{1 + \alpha}\right)^2 (R_* H_*), ~~~~{\rm for} ~~~\left(R_* H_* \lesssim \sqrt{\frac{3}{4} \kappa_{\rm sw} \alpha/(1+\alpha)}\right)\\
	\left(\frac{\kappa_{\rm sw} \alpha}{1 + \alpha}\right)^{\frac{3}{2}} (R_* H_*)^2, ~~~{\rm for} ~~~\left(R_* H_* > \sqrt{\frac{3}{4} \kappa_{\rm sw} \alpha/(1+\alpha)}\right),
\end{dcases} 
\eea
where, $R_* = (8\pi)^{1/3} v_w/\beta$ denotes the mean bubble separation with $v_w$ as the bubble wall velocity. The red-shifted peak frequency at the present day, using the results of Ref. \cite{Saikawa:2018rcs}, is expressed as
\bea
\label{eq:GW:soundwave-peak-freq}
f_{\rm sw} = 2.6 \times 10^{-5} {\rm Hz}\, (R_* H_*)^{-1} \left(\frac{T_*}{100\,{\rm GeV}}\right) \left(\frac{g_*(T)}{100}\right)^{\frac{1}{6}}.
\eea
For the sound wave lifetime, $\tau_{\rm sw}$, normalised to the Hubble rate, we adopt the approximation from Refs.~\cite{Hindmarsh:2017gnf, Ellis:2018mja, Ellis:2019oqb}:  
	\begin{equation}
		\label{eq:GW:soundwave-lifetime}
		\tau_{\rm sw} H_* = \frac{H_* R_*}{U_f}, \quad U_f \simeq \sqrt{\frac{3}{4} \frac{\alpha}{1 + \alpha} \kappa_{\rm sw}},
	\end{equation}
where $U_f$ represents the root-mean-squared fluid velocity. The sound wave efficiency factor, $\kappa_{\rm sw}$, can be determined using the fluid velocity ($v$) and temperature profiles~\cite{Lewicki:2021pgr}, following Ref.~\cite{Espinosa:2010hh}:  
	\begin{equation}
		\label{eq:GW:soundwave-kappa_sw}
		\kappa_{\rm sw} = \frac{3}{\alpha \rho_{\rm rad} v_w^3} \int w ~(\widetilde{\xi})^2 \frac{v^2}{1-v^2} d\widetilde{\xi}, 
	\end{equation}
where $w$ is the plasma enthalpy, and $\widetilde{\xi}$ has the dimension of velocity~\cite{Lewicki:2021pgr}. In practice, numerical fits can be used to obtain $\kappa_{\rm sw}$. The full expressions for $\kappa_{\rm sw}$ which we have utilised in our work can be found in Ref.~\cite{Espinosa:2010hh} \footnote{Note that $\kappa_{\rm sw}$ is denoted as $\kappa$ in Ref.~\cite{Espinosa:2010hh}.}.

During the time of a FOPT, the plasma becomes fully ionised. This results in MHD turbulence in the plasma, leading to a stochastic GW background. Its contribution to 
$\Omega_{\rm GW} h^2$ a function of the frequency `$f$' can be modelled as \cite{Caprini:2009yp}
\beq
\label{eq:GW:turb-amp}
\Omega_{\rm t} h^2 = 3.35 \times 10^{-4} \left( \frac{H_{\ast}}{\beta} \right) v_w \left( \frac{\kappa_{\rm t} \alpha}{1 + \alpha} \right)^{\frac{3}{2}} \left( \frac{100}{g_{\ast}(T)}\right)^{\frac{1}{3}} \left[ \frac{\left( f/f_{\rm t} \right)^3}{\left[ 1 + \left( f/f_{\rm t} \right) \right]^{11/3} \left( 1 + \frac{8 \pi f}{h_{\ast}} \right)} \right],
\eeq
with $\kappa_{\rm t}$ denoting the fraction of latent heat that is transformed into the MHD turbulence, and the frequency parameter $h_{\ast}$ is
\bea
\label{eq:GW:turb-hstar}
h_{\ast} = 16.5 \times 10^{-6} \left( \frac{T_*}{100 {\rm ~GeV}} \right) \left( \frac{g_{\ast}(T)}{100} \right)^{1/6} {\rm Hz}.
\eea
The red-shifted peak frequency at the present day, associated with the MHD turbulence, is given by
\bea
\label{eq:GW:turb-peak-freq}
f_{\rm t} = 2.7 \times 10^{-5} \frac{1}{v_w} \left( \frac{\beta}{H_{\ast}} \right) \left( \frac{T_*}{100 {\rm~GeV}} \right) \left( \frac{g_{\ast}(T)}{100} \right)^{1/6} {\rm Hz.}
\eea
In our analysis, we set $\kappa_{\rm t}=\epsilon \kappa_{\rm sw}$ with $\epsilon = 0.1$, which is motivated from simulations \cite{Caprini:2009yp}, and $\epsilon$ stands for the fraction of the bulk motion which is turbulent. Moreover, in our study, we consider $v_w$ to be a free parameter. In general, it is not free, and one should obtain it by solving the necessary Boltzmann equations from the first principle of the PT dynamics, leading to $v_w$ as a solution. However, taking $v_w$ as a free parameter with $v_w \approx 1$ \cite{Caprini:2015zlo} is an optimistic choice which enhances the possibility of GW detection. In fact, this is a conventional choice in literature, and one can take further inspiration from this approximation from the recent NANOGrav report \cite{NANOGrav:2023hvm}, where this value was adopted.
\subsection{Results} \label{subsec:PT-GW:results}
In this subsection, we analyse the PT dynamics of the chosen model, investigate the correlations between the key model parameters in the light of an SFOPT, and explore their potential implications for the DM observables discussed in Sec. \ref{sec:DM-pheno}. We then assess the detectability of GWs from such an SFOPT. To compute various thermodynamic quantities, we use publicly available code \texttt{cosmoTransitions-2.0.6} \cite{Wainwright:2011kj}. Additionally, our numerical analyses include only model points that satisfy all constraints outlined in Sec. \ref{sec:obs-cons}, accommodate the correct relic density within $3\sigma$ (see Eq.~(\ref{eq:constraints:DM-relic-Planck})), and remain allowed by both the DD (at least from the XENON1T) and ID limits.
\subsubsection{Phase transition along the $s$-direction} \label{subsubsec:PT-GW:PT-s}
The presence of three dynamical fields, i.e., $h^0,s,T^0$, allows for a rich variety of PT patterns. In principle, the transition can proceed from the symmetric phase, $\mathscr{O} \equiv \{0,0,0\}$, at very high temperatures ($T \gg T_c$) to a single field direction or along a mixed field trajectory. However, a PT along the $T^0$ direction is less favourable in our case. For a FOPT in this direction and to significantly influence transitions along the $h$-field, the triplet-Higgs quartic coupling $\lambda_{HT}$ must be large, i.e., $\gtrsim \mathcal{O}(1.0)$, with $m_{T^0} \lesssim 250$ GeV \cite{Niemi:2018asa,Niemi:2020hto}. However, such parameter choices are disfavoured by the DM constraints. To ensure that $T^0$ evades the DD (at least from the XENON1T) bounds (see Sec. \ref{sec:DM-pheno}), we restrict $\lambda_{HT}$ and $\lambda_{ST}$ to $\sim 0.3$. Additionally, $m_{T^0} < 300$ GeV is already excluded from the disappearing charged track constraints, as discussed in subsection \ref{subsec:exp-constraints}. Thus, the PT trajectory in our study is restricted to the $\{h,s\}$ field space. Even though $T^0$ does not directly influence the PT dynamics, it can still modify the $V^T_{\rm eff}$ through loop effects.

In the $\{h,s\}$-field space, with the presence of a cubic term in $s$ in the tree-level potential (see Eq.~(\ref{eq:pot:tree-level})), the symmetry in the $s$-direction is not necessarily restored. Hence, a transition can occur from high-temperature vacuum with $\langle S \rangle \neq 0$, i.e., $\mathscr{O}^{\prime} \equiv \{0,s_0\}$, to a low-temperature vacuum configuration. As the temperature drops, the system can undergo a transition from $\{0, s_0\} \xrightarrow{1^{\rm st}} \{0, s_1\}$ in the first step. A second transition, $\{0, s_1\} \xrightarrow{2^{\rm nd}} \{h_2, s_2\}$, may follow in the next step, eventually settling in the true EW global minimum $\{v, v_S\}$ at $T = 0$. In the phenomenologically viable DM parameter space, the second transition is typically weaker ($\xi_n \ll 1.0$, see Eq.~(\ref{eq:PT-strength})) and does not generate detectable GWs, whereas the former transition is strongly first-order and can produce a testable GW spectrum. Therefore, in the following, we focus on PTs only along the $s$-direction.
\begin{figure*}[!htpb]
	\centering
	{\includegraphics[height=7.3cm,width=14.0cm]{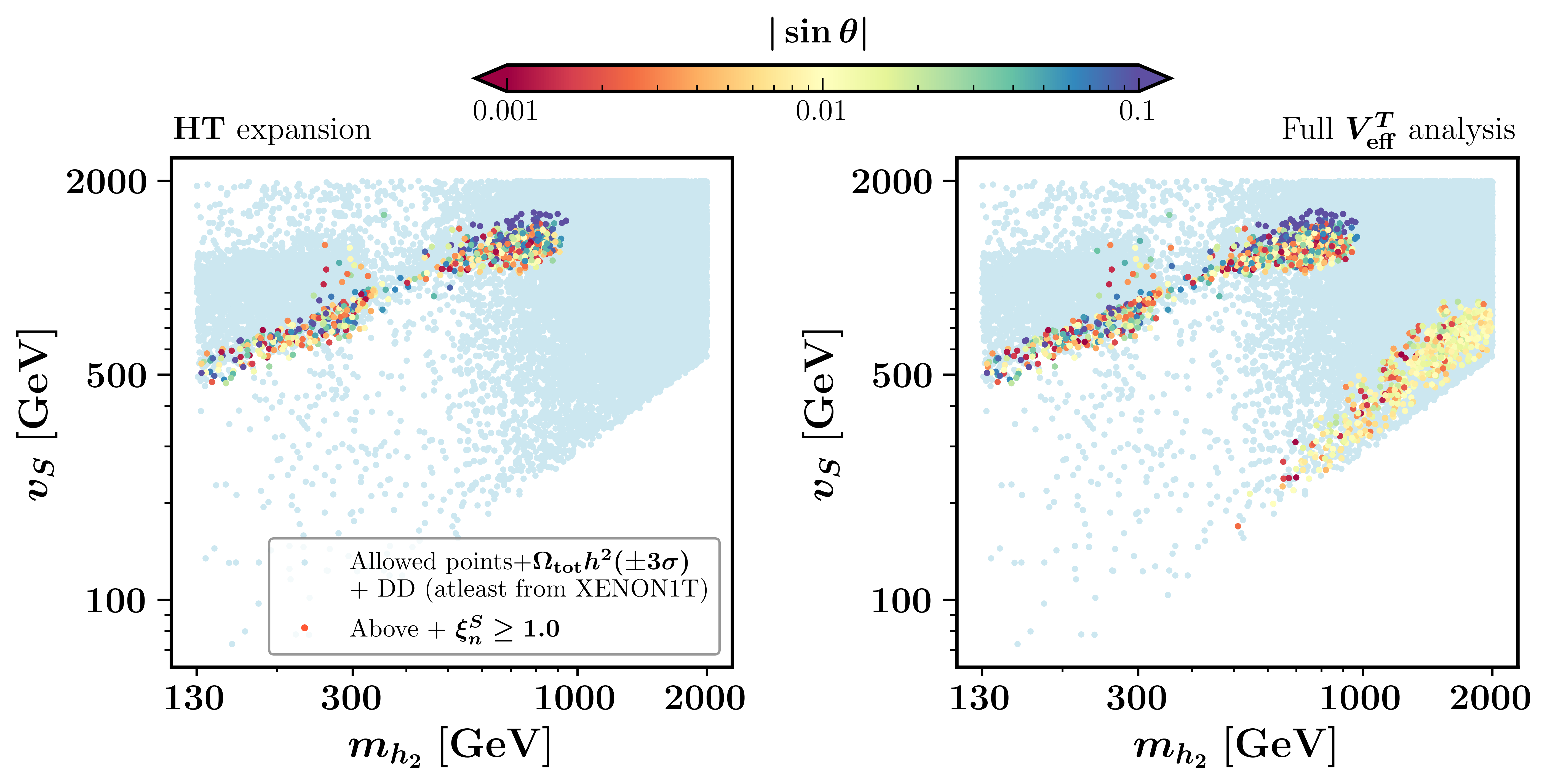}}
\caption{Comparison between two methods of calculating the thermal potential for the PTs: (Left panel) HT expansion approximation, as given by Eq.~(\ref{eq:pot:HT}), and (right panel) the full $V^T_{\rm eff}$, as shown by Eq.~(\ref{eq:pot:Eff-pot}). The light-blue coloured points represent model points that are consistent with the $3 \sigma$ relic density, bypass constraints outlined in Sec.~\ref{sec:obs-cons}, and survive both ID and DD (at least from XENON1T) bounds. The graded coloured points further show an SFOPT along the $s$-direction, and their dependence on the singlet-doublet mixing angle ($|\sin\theta|$) is shown above. See the text for more details. 
	}
	\label{fig:result:HT-VeffFull}
\end{figure*}

In Fig.~\ref{fig:result:HT-VeffFull}, we compare the results obtained after evaluating the $V^T_{\rm eff}$ using two approaches: the HT expansion approximation given in Eq.~(\ref{eq:pot:HT}) (left panel) and the $V^T_{\rm eff}$ of Eq.~(\ref{eq:pot:Eff-pot}) (right panel). The results are projected on the singlet-like Higgs mass ($m_{h_2}$) and its VEV ($v_S$) plane, with the Higgs mixing angle $|\sin\theta|$ represented in the colour bar. The ``sky-blue'' coloured points indicate viable model parameter space and the graded coloured points correspond to an SFOPT in the $s$-direction, i.e, $\xi^S_n \geq 1.0$, for some of those ``sky-blue'' coloured points as shown in Fig. \ref{fig:result:HT-VeffFull}. 

We observe in Fig.~\ref{fig:result:HT-VeffFull} that the HT expansion (left panel) imposes an upper bound on the singlet-like Higgs mass, restricting it to $m_{h_2} \lesssim 950$ GeV, beyond which an SFOPT along the $s$-field is not realised. In contrast, the full $V^T_{\rm eff}$ (right panel) reveals two distinct mass regions supporting an SFOPT: (i) a range similar to the HT expansion and (ii) an extended region where singlet Higgs masses remain viable from, $m_{h_2} \gtrsim 500$ GeV, up to the maximum scanned limit. The upper bound from the HT expansion aligns with the generic constraint of $\sim 700$ GeV discussed in Ref.~\cite{Ramsey-Musolf:2019lsf}. However, in our model, this bound is slightly relaxed since the singlet-like Higgs mass receives additional contributions from the singlet VEV $v_S$ (see Eqs.~(\ref{eq:masses:scalar-mass-HS-new}) and (\ref{eq:masses:mass-eigen-HS})), unlike the scenario in Ref.~\cite{Ramsey-Musolf:2019lsf} where it depends solely on the SM Higgs VEV $v$. In the full $V^T_{\rm eff}$ case, the inclusion of a complete one-loop thermal corrections significantly enhances the possibility of an SFOPT, allowing even larger values of $m_{h_2}$ to support an SFOPT. Additionally, to achieve $\xi^s_n > 1.0$ in the HT expansion requires a larger singlet VEV, in the range $500~\text{GeV} \lesssim v_S \lesssim 1800 ~\text{GeV}$, while the entire scanned range of $|\sin\theta|$ remains viable. 
However, with the full $V^T_{\rm eff}$ analysis, although an SFOPT can still occur for very small mixing angles, i.e., $|\sin\theta| \lesssim 0.01$, the upper limit on $v_S$ drops to $\lesssim 900$ GeV. Thus, as evident from the right panel of Fig.~\ref{fig:result:HT-VeffFull}, a singlet VEV of $v_S \lesssim 500$ GeV can still induce an SFOPT, unlike in the HT expansion case. Our scan also indicates a lower bound on the singlet Higgs VEV, $v_S \gtrsim 150$ GeV for the case with full $V^T_{\rm eff}$. To ensure a comprehensive quantitative assessment, our subsequent discussion of the model parameter space will be based on results obtained from the full $V^T_{\rm eff}$ calculation.
\begin{figure*}[!htpb]
	\centering
	{\includegraphics[height=7.3cm,width=14.0cm]{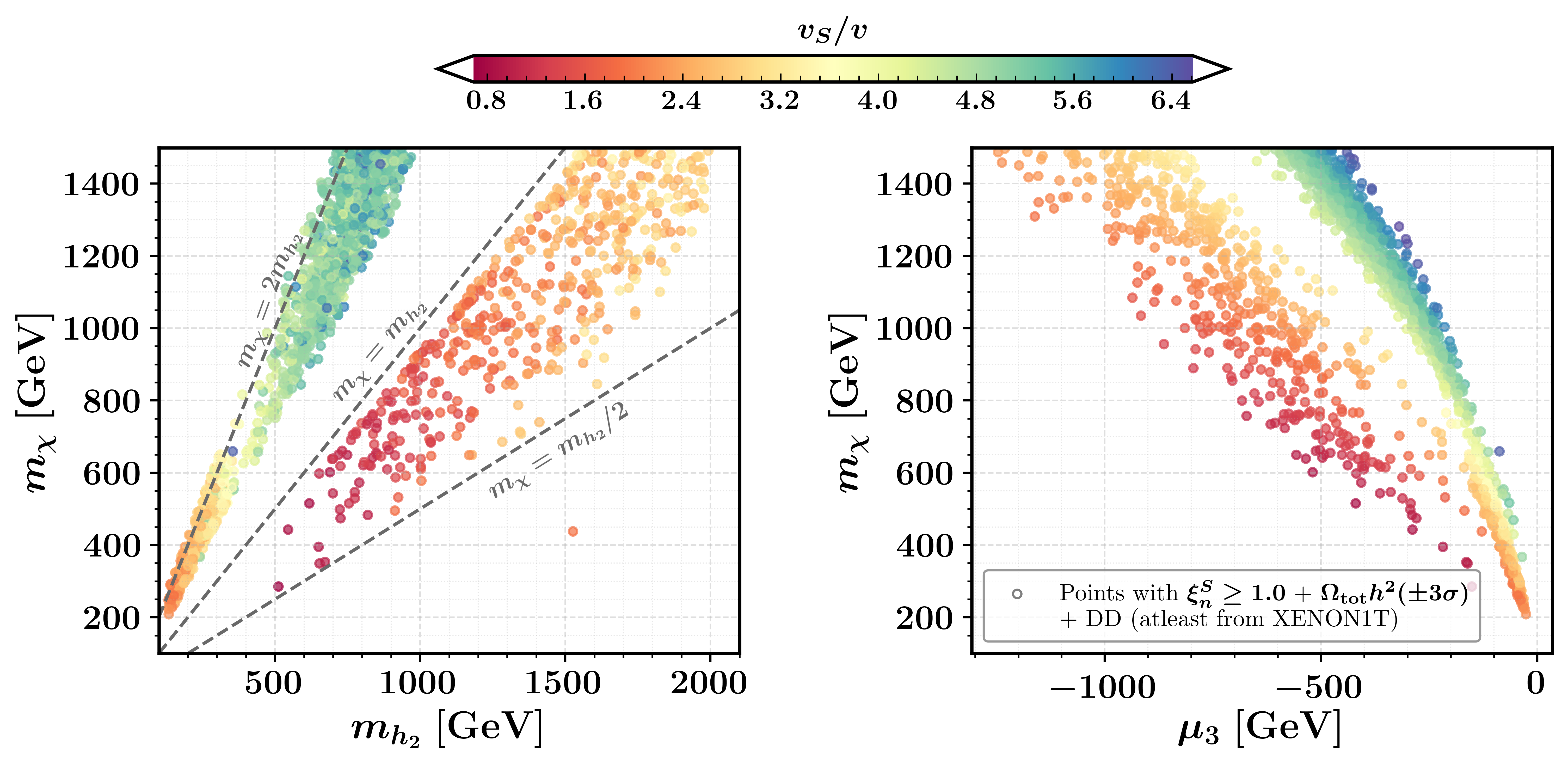}}
	\caption{Plots depicting the allowed mass regions of the pNGB DM $\chi$, associated with an SFOPT in the $s$-direction, in the $m_{h_2}$-$m_\chi$ (left) and $\mu_3$-$m_\chi$ (right) planes. The colour bar on the top indicates the variation of the singlet VEV $v_S$, normalised to the SM Higgs VEV $v$. The sample points are obtained by evaluating the full $V^T_{\rm eff}$ (Eq.~(\ref{eq:pot:Eff-pot})). The differently styled lines in the left plot carry the same meaning as Fig.~\ref{fig:R-I:param-dependence}(a).}
	\label{fig:result:DM-model-param}
\end{figure*}

Fig.~\ref{fig:result:DM-model-param} illustrates the correlation between the pNGB DM mass $m_{\chi}$ with the heavy singlet-like Higgs mass $m_{h_2}$ (left panel) and the $U(1)$ soft-breaking parameter $\mu_3$ (right panel). The colour bar shows variation of the singlet VEV $v_S$, normalised to the SM Higgs VEV $v$. The differently styled black-coloured lines of Fig.~\ref{fig:result:DM-model-param} (left panel) carry the same meaning as defined for Fig.~\ref{fig:R-I:param-dependence-a}. As revealed in the left panel of Fig.~\ref{fig:result:DM-model-param}, for $\xi^S_n \geq 1.0$, the pNGB DM mass is allowed from $m_{\chi} \gtrsim 200$ GeV up to the maximum scanned value of $\sim 2.0$ TeV. The lower restriction on $m_{\chi}$ arises due to the lower bound on $v_S$\footnote{It also can be noticed from Eq.~(\ref{eq:dependent-param}).}, as already depicted in Fig.~\ref{fig:result:HT-VeffFull} (right panel). Although, the viable pNGB DM mass region is found to be entirely covered in the $m_{h_2}$–$m_{\chi}$ plane for the DM mass regimes \textbf{R-I} and \textbf{R-II}, as depicted by Fig.~\ref{fig:R-I:param-dependence-a} and Fig.~\ref{fig:R-II:param-dependence-a}. However, demanding an SFOPT along the $s$-direction limits the distribution density of the allowed points. Notably, most sample points are concentrated either in the range $m_{h_2}/2 \lesssim m_\chi \lesssim m_{h_2}$ or cluster near the threshold $m_\chi = 2m_{h_2}$. In both regions, we observe a positive correlation with the singlet VEV $v_S$. The latter region, however, favours larger $v_S$ as the pNGB DM mass increases, while in the former, a relatively smaller $v_S$ with $v_S/v \lesssim 3.5$ is sufficient to ensure an SFOPT in the $s$-direction. In addition, $m_\chi < m_{h_2}/2$ is almost disfavoured in our scan. However, as evident from the left panel of Fig. \ref{fig:result:DM-model-param}, the entire mass range for the singlet Higgs, $h_2$, remain available for an SFOPT in the $s$-direction. The generic trend of the explicit $U(1)$ soft breaking parameter $\mu_3$ and $m_\chi$, shown in Fig.~\ref{fig:result:DM-model-param} (right panel), has already been discussed in subsection \ref{subsec:DM-results}. In this plane, we do not observe a strong correlation with $v_S$. The variation of the graded coloured points reflects two distinct pNGB mass regions relative to $m_{h_2}$, as seen in the left panel of Fig.~\ref{fig:result:DM-model-param}, and their dependence on $v_S$ remains consistent in this plane as well.
\begin{figure*}[!htbp]
	\centering
	\subfigure[\label{fig:result:PT-obs-model-param-a}]{\includegraphics[height=5.8cm,width=6.8cm]{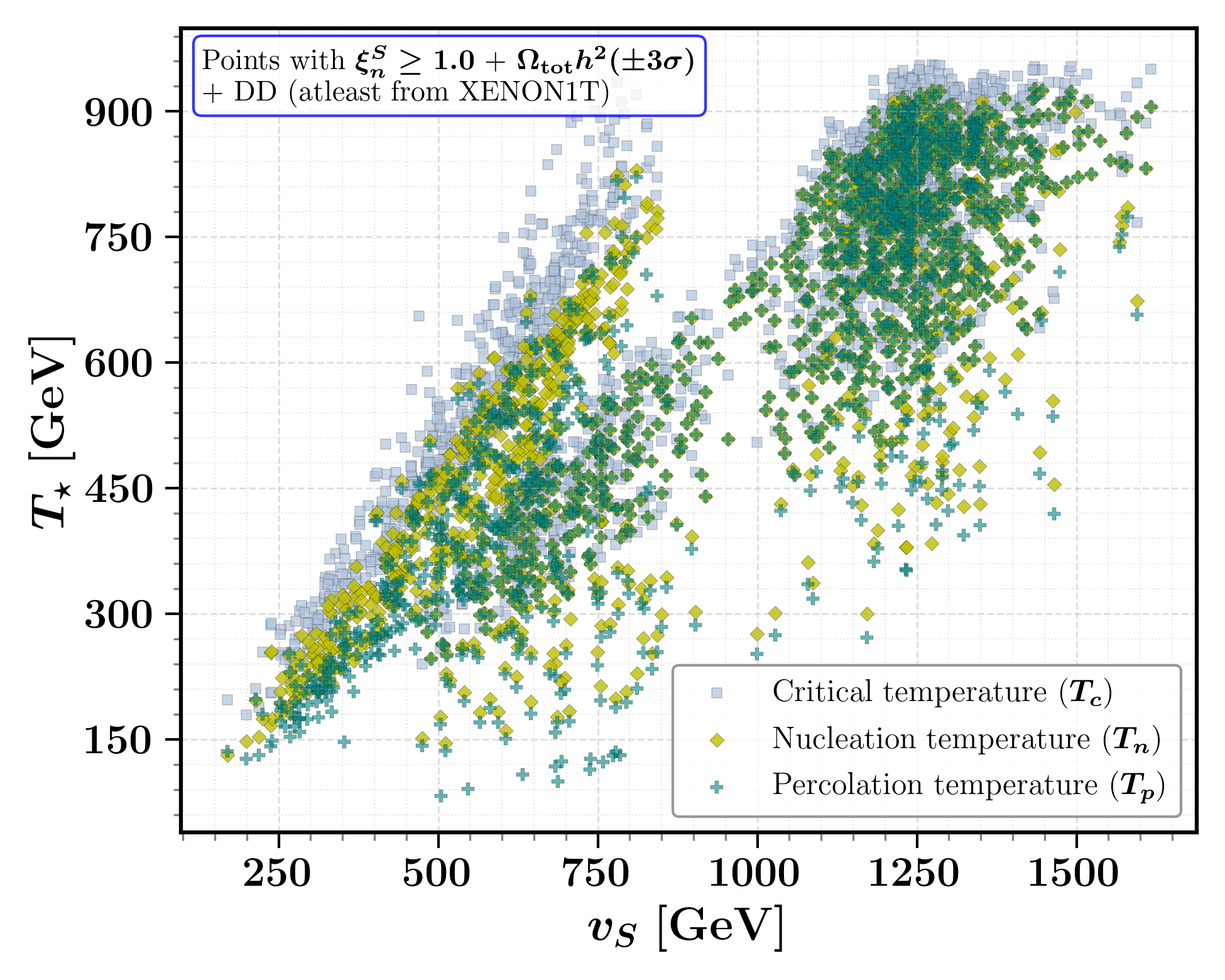}}
	\subfigure[\label{fig:result:PT-obs-model-param-b}]{\includegraphics[height=5.8cm,width=6.8cm]{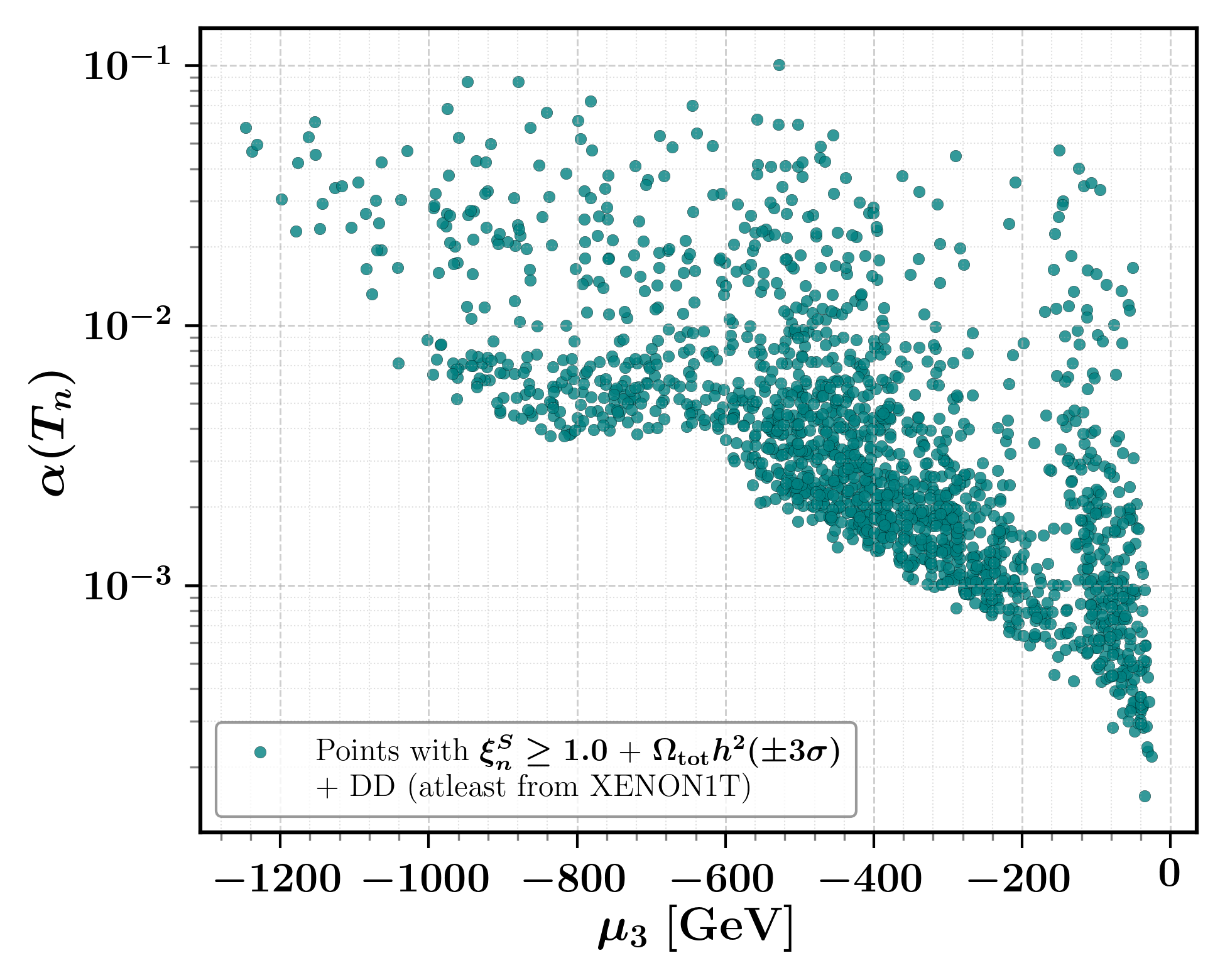}}
\caption{Dependence of the PT parameters $T_{\ast}$ and $ \alpha(T)$ on potential parameters $v_S$ (left panel) and $\mu_3$ (right panel), respectively. All sample points bypass constraints discussed in Sec.~\ref{sec:obs-cons}, obey the $3\sigma$ relic density bound, and evade DD (at least from the XENON1T)  and ID bounds.}
\label{fig:result:PT-obs-model-param}
\end{figure*}

Finally, Fig.~\ref{fig:result:PT-obs-model-param} illustrates the variation of two key PT parameters, $T_{\ast}$ and $\alpha$, which not only characterise the PT dynamics but are also important for the GW signal. Their dependency on the singlet VEV $v_S$, and the explicit $U(1)$ soft breaking parameter $\mu_3$, are examined. In Fig.~\ref{fig:result:PT-obs-model-param-a}, we show the variation of different transition temperatures -- critical ($T_c$), nucleation ($T_n$), and percolation ($T_p$) -- distinguished by different symbols and colour codes (see Fig. \ref{fig:result:PT-obs-model-param-a} legend). These temperatures generally increase with $v_S$, indicating that a larger singlet VEV induces a PT at a higher temperature. We observe that $T_{\ast}$ can be as low as 100 GeV and extend up to $\sim 1.0$ TeV. Fig.~\ref{fig:result:PT-obs-model-param}(a) further shows that the temperature hierarchy, $T_c > T_n > T_p$, is maintained. However, in most of the parameter space, the bubble percolation occurs almost immediately after nucleation, i.e., $T_p \approx T_n$. Exceptions arise in the singlet VEV regime, $v_S \lesssim 800$ GeV, or when the overall transition takes place at a relatively lower $T_{\ast}$ with increasing $v_S$. In these cases, the percolation can experience a significant delay. In Fig.~\ref{fig:result:PT-obs-model-param-a}, we observe that a non-zero $\mu_3$ allows for larger values of $\alpha$ (evaluated at $T_n$), indicating a stronger PT. This is expected, as $\mu_3$, being the cubic coupling in the tree-level potential (see Eq.~(\ref{eq:pot:tree-level})), enhances barrier formation at the tree-level itself. Our analysis reveals that, in our model, the viable parameter points — consistent with the correct $3\sigma$ relic density, satisfying DD (at least from XENON1T) and ID bounds, and other constraints outlined in Sec. \ref{sec:obs-cons} — yield PT strengths $\alpha(T_n) < 1.0$. This suggests the absence of significant supercooling in our model, with only intermediate transitions characterised by $\alpha(T_n) \sim \mathcal{O}(0.1)$ \cite{Athron:2024xrh}.

\subsubsection{Gravitational waves and its detection prospects} \label{subsubsec:PT-GW:GW-detection}
\noindent In subsection~\ref{subsec:PT-GW:GWs}, we discussed the possibility of a stochastic GW background generated by a FOPT in the early Universe. Additionally, in \ref{subsubsec:PT-GW:PT-s}, we established that a model space consistent with the viable DM phenomenology can successfully accommodate an SFOPT along the $s$-direction. This naturally leads to the prospect of observable GW signals arising from such transitions. These stochastic cosmic relics could be within the reach of various space-based GW interferometers, including SKA \cite{Janssen:2014dka}, $\mu$-Ares \cite{Sesana:2019vho}, LISA \cite{eLISA:2013xep,LISA:2017pwj}, BBO \cite{Crowder:2005nr,Corbin:2005ny,Harry:2006fi}, DECIGO and its variants \cite{Kudoh:2005as, Yagi:2011wg,Kawamura:2020pcg}, AEDGE \cite{AEDGE:2019nxb}, and AION \cite{Badurina:2019hst}, along with terrestrial detectors such as CE \cite{LIGOScientific:2016wof}, ET \cite{Hild:2008ng}, and future upgrades of LVK \cite{KAGRA:2021kbb, Jiang:2022uxp, LIGOScientific:2022sts}. A successful detection would not only confirm the occurrence of a FOPT but also provide crucial insights into the underlying new physics responsible for it. 

Before delving into the detectability prospects, it is essential to examine first the correlation between the transition strength parameter $\alpha$ and the inverse duration parameter $\beta$.
\begin{figure*}[!htpb]
	\centering
	{\includegraphics[height=7.0cm,width=8.5cm]{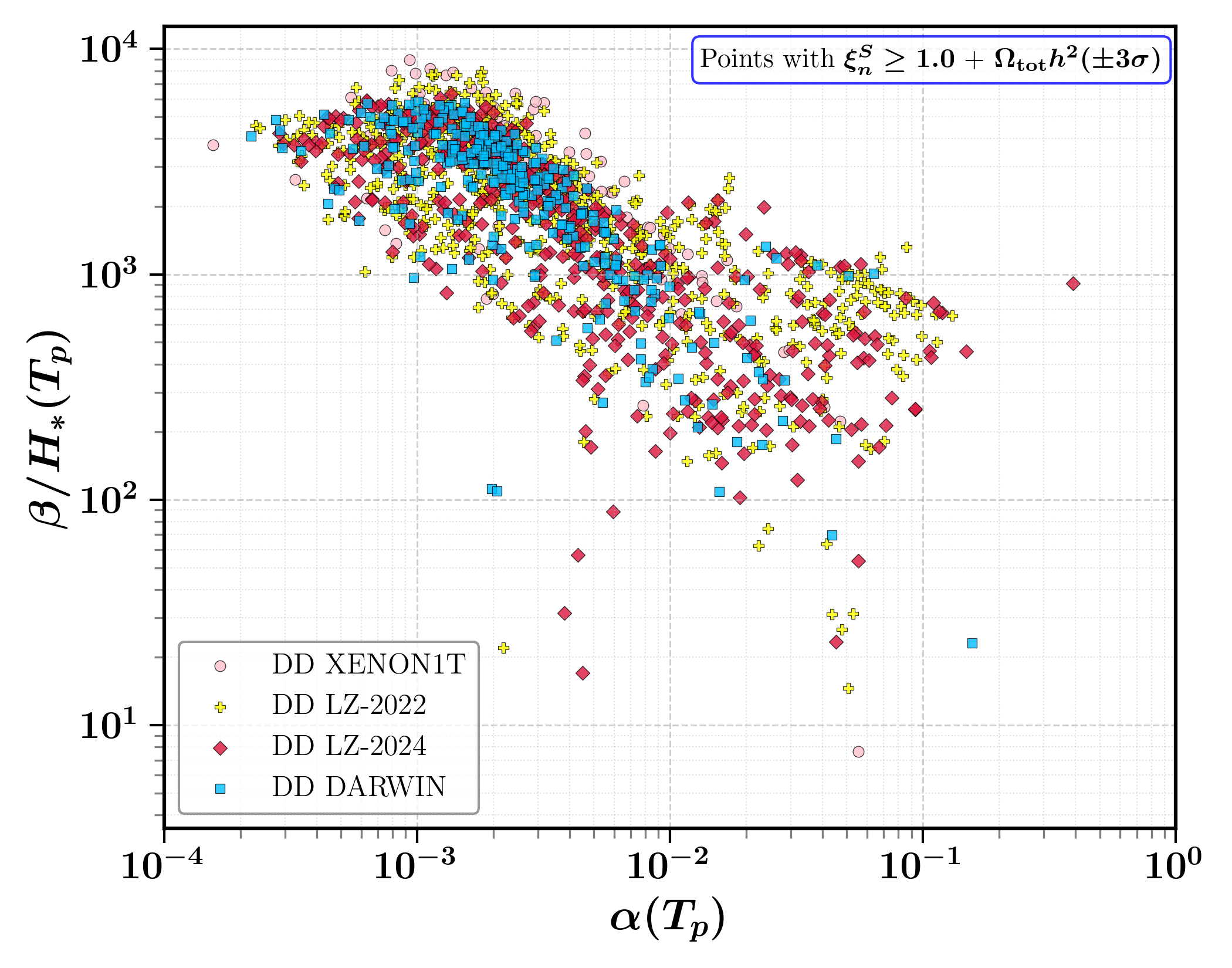}}
	\caption{The scatter points illustrate the correlation between $\alpha(T_p)$ and $\beta/H_{*}(T_p)$ for parameter samples that can generate an SFOPT while satisfying all constraints outlined in Sec.~\ref{sec:obs-cons}. These points also yield the correct $3 \sigma$ relic abundance and lie within the sensitivity reach of different DD experiments, as outlined in the legend.}
	\label{fig:result:GW-alpha-beta-Tp}
\end{figure*}
In Fig.~\ref{fig:result:GW-alpha-beta-Tp}, we show a scatter plot in the $\alpha$-$\beta/H_{*}$ plane, where both quantities are calculated at the percolation temperature $T_p$. As expected, we observe an inverse correlation between $\alpha(T_p)$ and $\beta/H_{*}(T_p)$. This behaviour arises because a larger $\alpha$, which quantifies the energy released during the PT, implies a greater separation between the false and true vacua, leading to a longer duration for the transition. In our model space, $\beta/H_{*}(T_p)$ ranges from a few tens to nearly $10^4$, while $\alpha(T_p)$ limited within $1.0$, similar to our findings from the nucleation analysis (see Fig.~\ref{fig:result:PT-obs-model-param} findings). The light-pink, yellow, red, and light-blue coloured points represent parameter points that satisfy the DD bounds from XENON1T, LZ-2022, LZ-2024, and DARWIN, respectively. Notably, data points within the reach of LZ-2022 (2024) or DARWIN are distributed across the $\alpha$–$\beta/H_{*}$ plane, indicating that most parameter points sensitive to the current or future DD experiments could also be probed at the upcoming GW detectors.

We now proceed to examine the detectability of the GW signals which can be obtained by the fit formulae (Eqs.~(\ref{eq:GW:soundwave-amp}) and (\ref{eq:GW:turb-amp})) presented in subsection \ref{subsec:PT-GW:GWs} for various sources, e.g., sound waves and MHD turbulence. The total GW signal amplitude can be obtained using Eq.~(\ref{eq:GW:total-contribution}), where the individual contributions are defined in Eqs.~(\ref{eq:GW:soundwave-amp}) and (\ref{eq:GW:turb-amp}). The resulting GW signal needs to be compared to the noise spectrum of the relevant experiment to determine the signal-to-noise ratio (SNR) \cite{Allen:1997ad,Maggiore:1999vm},
\bea
\label{eq:GW-result:SNR}
{\rm SNR} \equiv \rho = \left[\delta\,\times t_{\rm obs} \int\limits_{f_{\rm min}}^{f_{\rm max}} \frac{df}{\rm Hz} \left(\frac{\Omega_{\rm GW} h^2 (f)}{\Omega_{\rm noise} h^2 (f)}\right)^2 \right]^{1/2}.
\eea
Here, $\delta$ represents the number of independent channels used to distinguish detectors via auto-correlation ($\delta = 1$) or cross-correlation ($\delta = 2$) to confirm the stochastic nature of the GW. For LISA, $\delta = 1$, while for BBO, DECIGO, and its variants, $\delta = 2$. The observation time $t_{\rm obs}$ is set to 4 years in our calculations. The denominator in Eq.~(\ref{eq:GW-result:SNR}), $\Omega_{\rm noise} h^2(f)$, represents the strain noise power spectral density of a particular GW experiment, while the numerator, $\Omega_{\rm GW} h^2(f)$, corresponds to the total GW signal in a specific model, as given by Eq.~(\ref{eq:GW:total-contribution}). A successful detection requires the SNR to exceed the experiment-specific threshold $\rho_{\rm thr}$. For a four-link LISA, $\rho_{\rm thr} \approx 50$, whereas a six-link configuration reduces it to $\sim 10$ \cite{LISA:2017pwj, Robson:2018ifk}. In the present analysis, we consider a GW signal detectable if SNR $> 10$ for LISA, DECIGO, and BBO \cite{Sato:2017dkf, Crowder:2005nr, Yagi:2011yu, Yagi:2013du}.
\begin{figure*}[!htbp]
	\centering
	\subfigure[\label{fig:result:GW-total-SNR-a}]{\includegraphics[height=5.8cm,width=7.3cm]{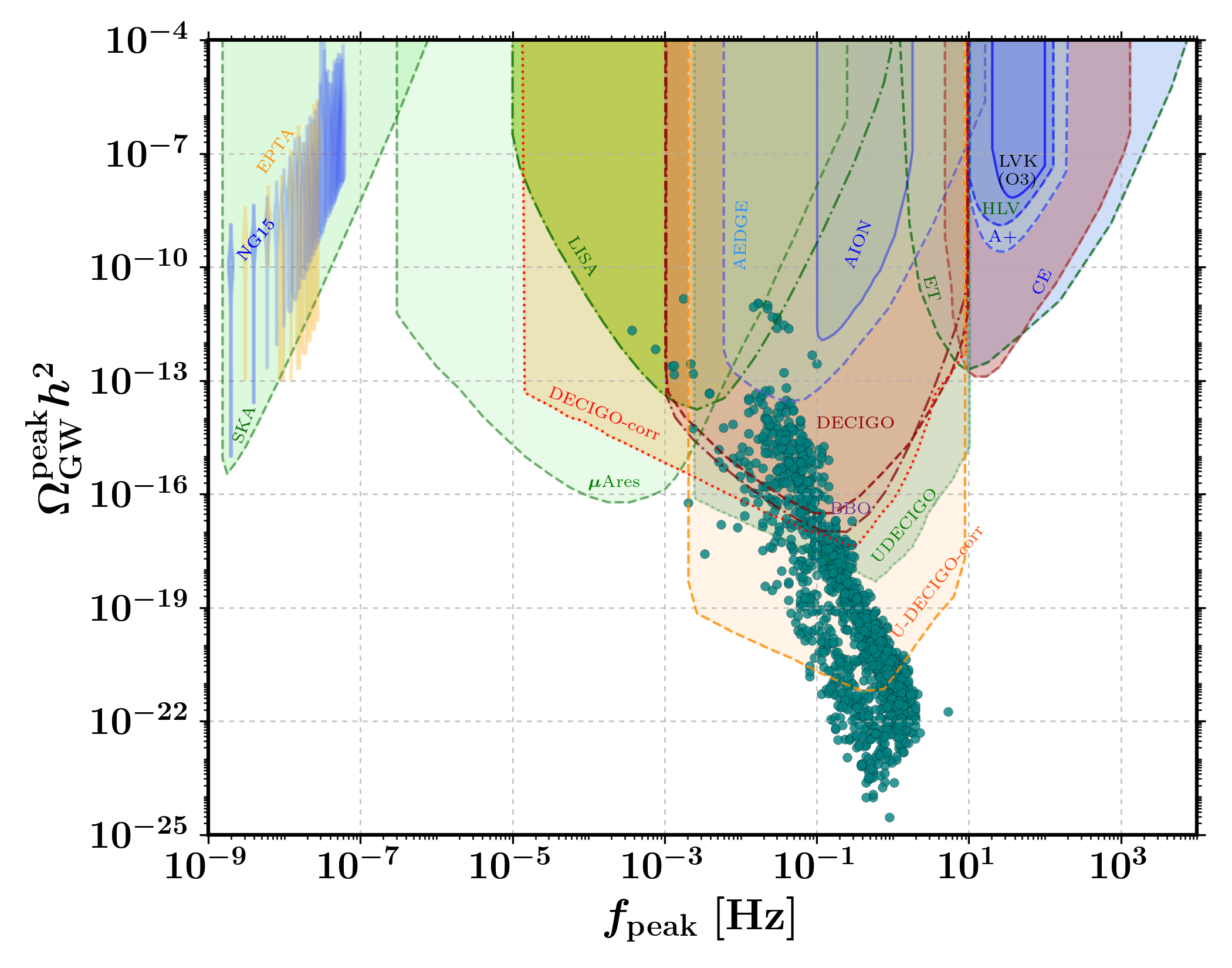}}
	\hspace{0.3cm}
	\subfigure[\label{fig:result:GW-total-SNR-b}]{\includegraphics[height=5.8cm,width=7.3cm]{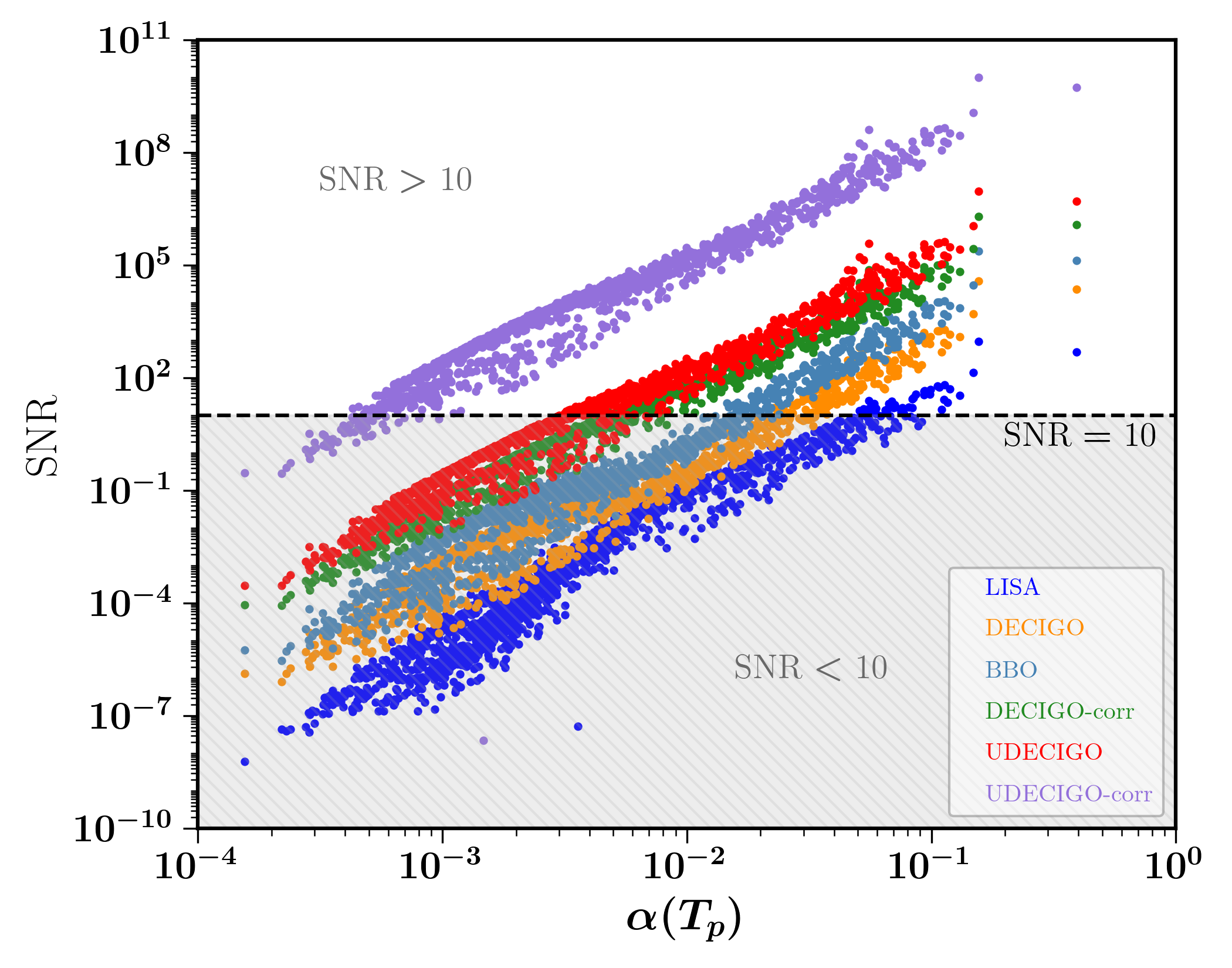}}
	\caption{The left plot illustrates the distribution of model points consistent with the DM phenomenology in the $f_{\rm peak}-\Omega_{\rm GW}^{\rm peak} h^2$ plane, overlaid with the sensitivity reach of various GW detectors, represented by differently styled and differently coloured lines. The right plot presents the estimated SNR values (see Eq.~(\ref{eq:GW-result:SNR})) for different GW detectors: LISA (blue coloured), DECIGO (orange coloured), BBO (light blue coloured), DECIGO-corr (green coloured), UDECIGO (red coloured), and UDECIGO-corr (purple coloured). The black horizontal dashed line marks the SNR threshold of 10, separating regions with $\rm{SNR} < 10$ and $\rm{SNR} > 10$.}
	\label{fig:result:GW-total-SNR}
\end{figure*}

In Fig.~\ref{fig:result:GW-total-SNR-a}\footnote{For this plot, the power-law sensitivity curve data for various GW experiments are taken from Refs.~\cite{Schmitz:2020syl,Fu:2024rsm}, except for DECIGO with correlation, U-DECIGO, and U-DECIGO with correlation, which are sourced from Ref.~\cite{Nakayama:2009ce}.}, we present the GW signals predicted in the chosen model in the $f_{\rm peak}$ - $\Omega_{\rm GW}^{\rm peak} h^2$ plane for the same set of points shown in Fig~\ref{fig:result:GW-alpha-beta-Tp}, which are consistent with the $3\sigma$ relic density constraint, different DD (XENON1T, LZ-2022/24, DARWIN) limits, ID bounds (as discussed in \ref{subsubsec:DM-results:indirect-detection}). We also show the power-law integrated sensitivity curves (PLI) \cite{Thrane:2013oya} of the upcoming and proposed GW detectors, such as SKA, $\mu$-Ares, LISA, BBO, DECIGO, U-DECIGO, and U-DECIGO-corr, AEDGE, AION, CE, ET, future upgrades of LVK, along with the recent results from pulsar timing arrays NANOGrav \cite{NANOGrav:2023gor,NANOGrav:2023hvm} and EPTA \cite{EPTA:2023sfo,EPTA:2023fyk} for comparison. The majority of the predicted GW signals are within the reach of LISA, BBO, DECIGO, and their variants. For these detectors, we further compute the SNR using Eq.~(\ref{eq:GW-result:SNR}) for the viable sample points and project them onto the $\alpha(T_p)$-SNR plane, as shown in Fig.~\ref{fig:result:GW-total-SNR}(b). The coloured points in this plot, as indicated in the legend, represent the SNR values for the respective detectors. The shaded grey coloured region is excluded based on the SNR threshold of 10.\footnote{The SNR thresholds for U-DECIGO and U-DECIGO-corr are not explicitly known (to the best of our knowledge). Hence, we adopt a conservative estimate of 10.} The SNR value rises for a stronger PT as $\alpha(T_p)$ increases. We notice that a small fraction of model points remain available to be tested at LISA. Whereas, DECIGO, BBO, and other variants of DECIGO have better prospects to probe the chosen model parameter space.

Although PLIs are used in Fig. \ref{fig:result:GW-total-SNR-a} to compare model predictions with experimental sensitivity reaches of the different GW interferometers and to estimate SNR via Eq.(\ref{eq:GW-result:SNR}), they have limitations \cite{Alanne:2019bsm, Schmitz:2020syl}. In particular, PLIs do not inherently encode SNR information and assume a power-law dependence, which is significantly violated near the GW spectral peak from an SFOPT.

To overcome these issues, we employ the recently introduced ``peak-integrated sensitivity curves'' (PISCs) \cite{Alanne:2019bsm, Schmitz:2020syl}. Unlike PLIs, PISCs incorporate the full spectral shape of the signal, allowing for direct integration over the spectral shape when the signal shape is known, as in the case of a FOPT. This approach ensures that the SNR is uniquely determined by the peak energy densities and frequencies, which depend solely on the model-specific PT parameters. Consequently, PISCs provide a more accurate representation of detectability, retaining the complete SNR information. In this formalism, the SNR in Eq.~(\ref{eq:GW-result:SNR}) can be re-written as,
\bea
\label{eq:GW-result:PICs-eqn}
\frac{\rho^2}{t_{\rm obs}} &=& \left(\frac{\Omega_{\rm b}^{\rm peak} h^2}{\Omega^{\rm b}_{\rm PISCs} h^2}\right)^2 + \left(\frac{\Omega_{\rm sw}^{\rm peak} h^2}{\Omega^{\rm sw}_{\rm PISCs} h^2}\right)^2 + \left(\frac{\Omega_{\rm t}^{\rm peak} h^2}{\Omega^{\rm t}_{\rm PISCs} h^2}\right)^2 \nn \\
&& + \left(\frac{\Omega_{\rm b/sw}^{\rm peak} h^2}{\Omega^{\rm b/sw}_{\rm PISCs} h^2}\right)^2 + \left(\frac{\Omega_{\rm sw/t}^{\rm peak} h^2}{\Omega^{\rm sw/t}_{\rm PISCs} h^2}\right)^2 + \left(\frac{\Omega_{\rm b/t}^{\rm peak} h^2}{\Omega^{\rm b/t}_{\rm PISCs} h^2}\right)^2.
\eea
Here, the frequency integration is implicitly assumed to have been performed,
\bea
\label{eq:GW-result:PICs-integration}
\Omega_{\rm PISCs}^{i/j} h^2 = \left[(2-\delta_{ij})\, \delta \times 4\,{\rm yr} \int_{f_{\rm min}}^{f_{\rm max}} df \frac{S_i(f)\,S_j(f)}{(\Omega_{\rm noise} h^2(f))^2} \right]^{-1/2},
\eea
where $i,j$ denotes $\{\rm b, sw, t\}$, corresponding to bubble collision, sound wave and MHD turbulence, respectively. However, as we have found that our model does not show any significant supercooling (see \ref{subsubsec:PT-GW:PT-s}), therefore, bubble collision has a negligible impact on the GW signal, as discussed in subsection \ref{subsec:PT-GW:GWs}. Hence, we omit bubble collision contribution in Eq.~(\ref{eq:GW-result:PICs-eqn}) without any loss of generality. Finally, the mixed peak amplitudes in the PISC formalism are defined as,
\bea
\label{eq:GW-result:PISC-geometric-mean}
\Omega_{i/j}^{\rm peak} = \left(\Omega_i^{\rm peak} h^2~ \Omega_j^{\rm peak} h^2\right)^{1/2}.
\eea
Note that, once Eq.(\ref{eq:GW-result:PICs-integration}) is integrated, the SNR is fully determined by the peak energy densities and frequencies. Unlike Refs.~\cite{Alanne:2019bsm,Schmitz:2020syl}, which used $t_{\rm obs} = 1~{\rm yr}$ and $\rho_{\rm thr} = 1$ for demonstration, we extend the analysis to $t_{\rm obs} = 4~{\rm yr}$ and $\rho_{\rm thr} = 10$, as stated earlier.

\begin{figure*}[!htbp]
	\centering
	\subfigure{\includegraphics[height=5.5cm,width=7.0cm]{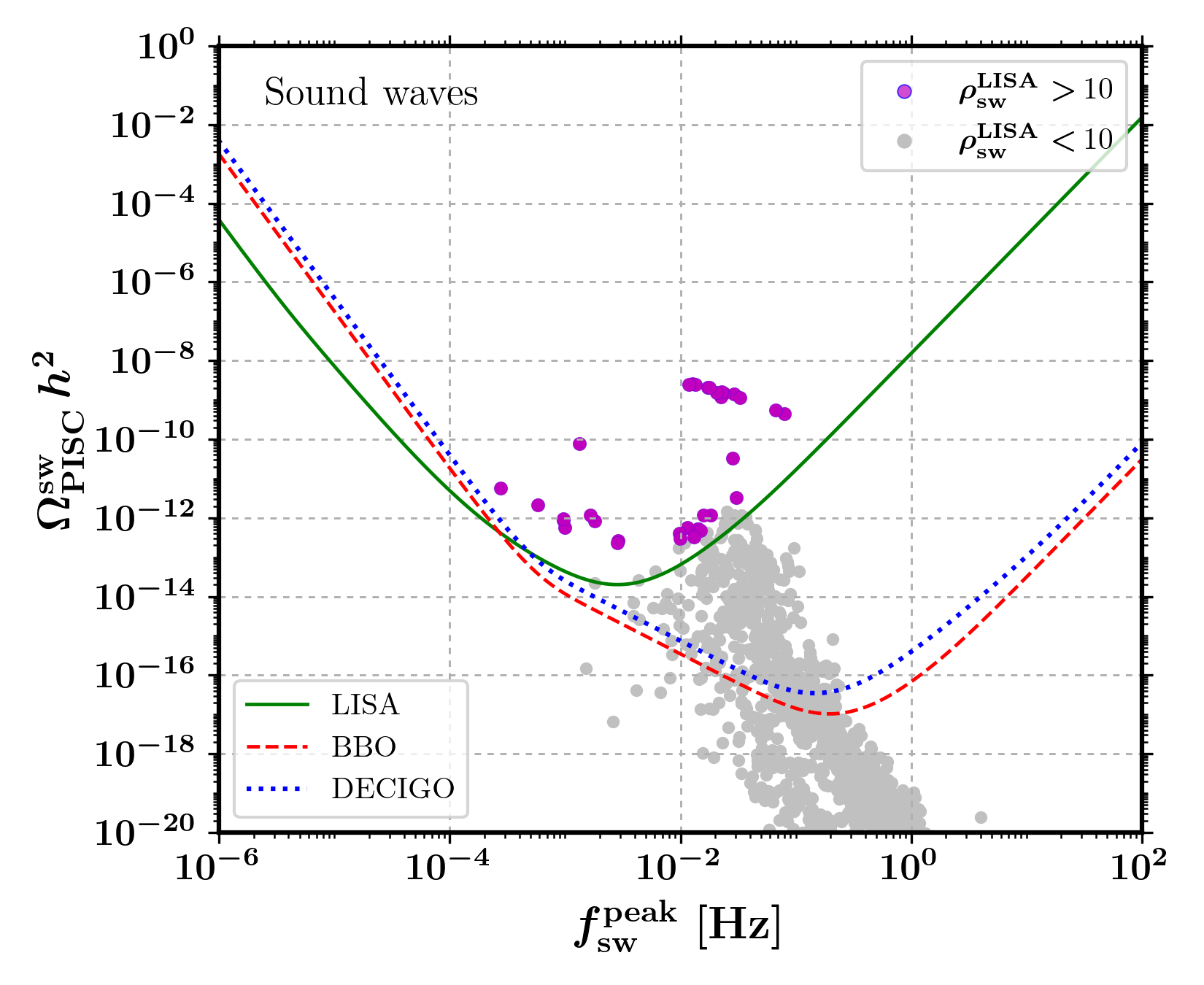}}
	\hspace{0.5cm}
	\subfigure{\includegraphics[height=5.5cm,width=7.0cm]{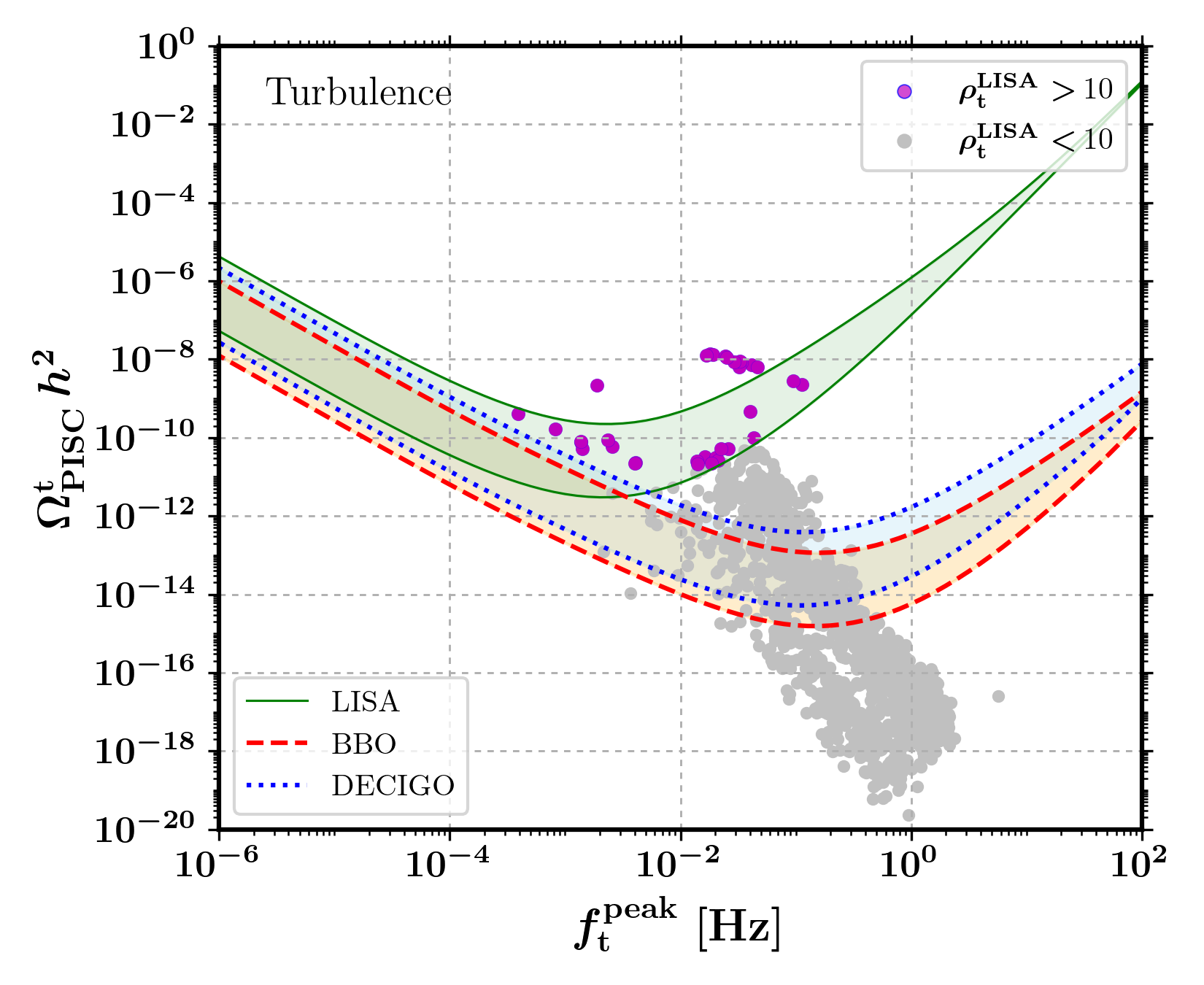}}
	\subfigure{\includegraphics[height=5.5cm,width=7.0cm]{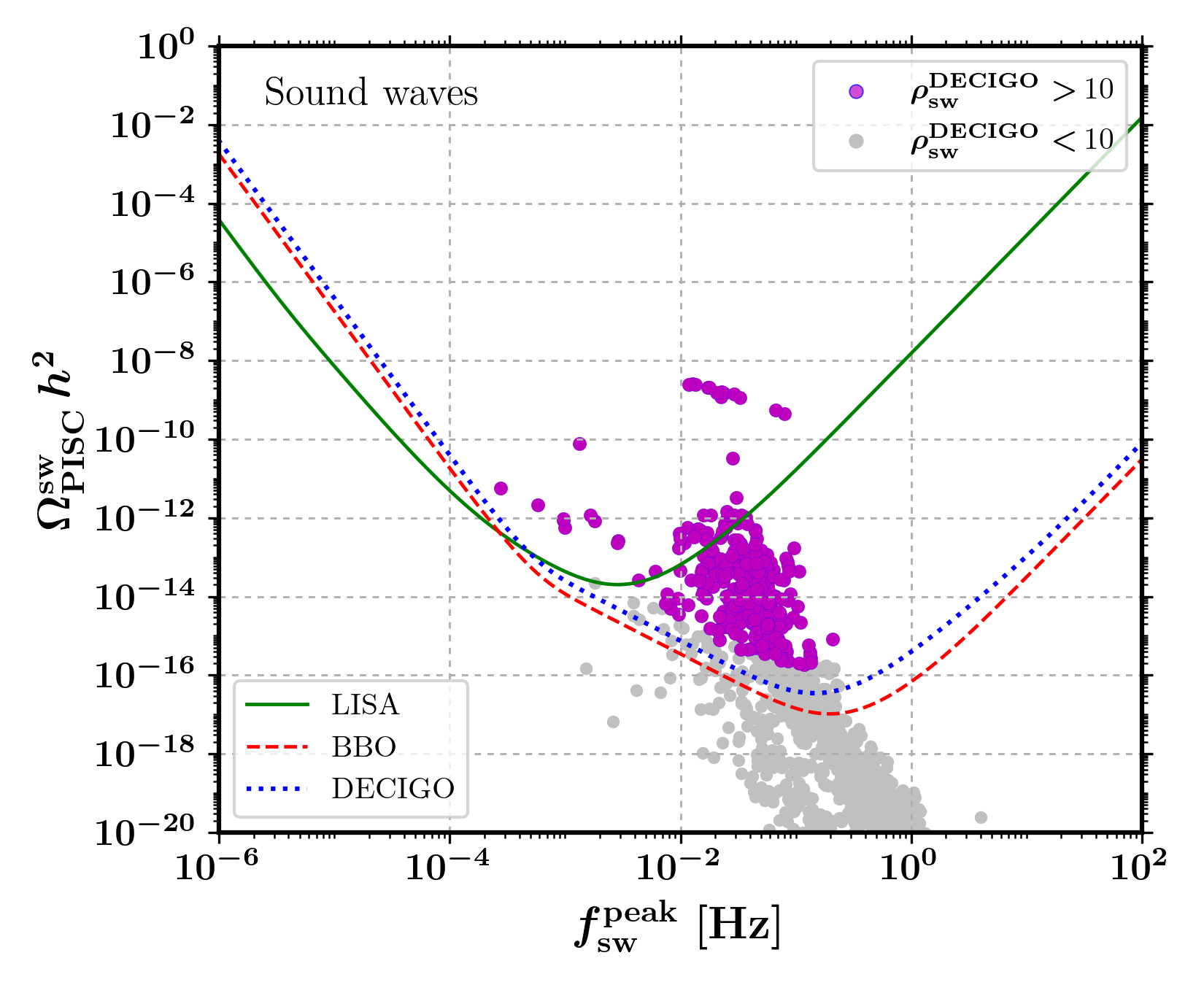}}
	\hspace{0.5cm}
	\subfigure{\includegraphics[height=5.5cm,width=7.0cm]{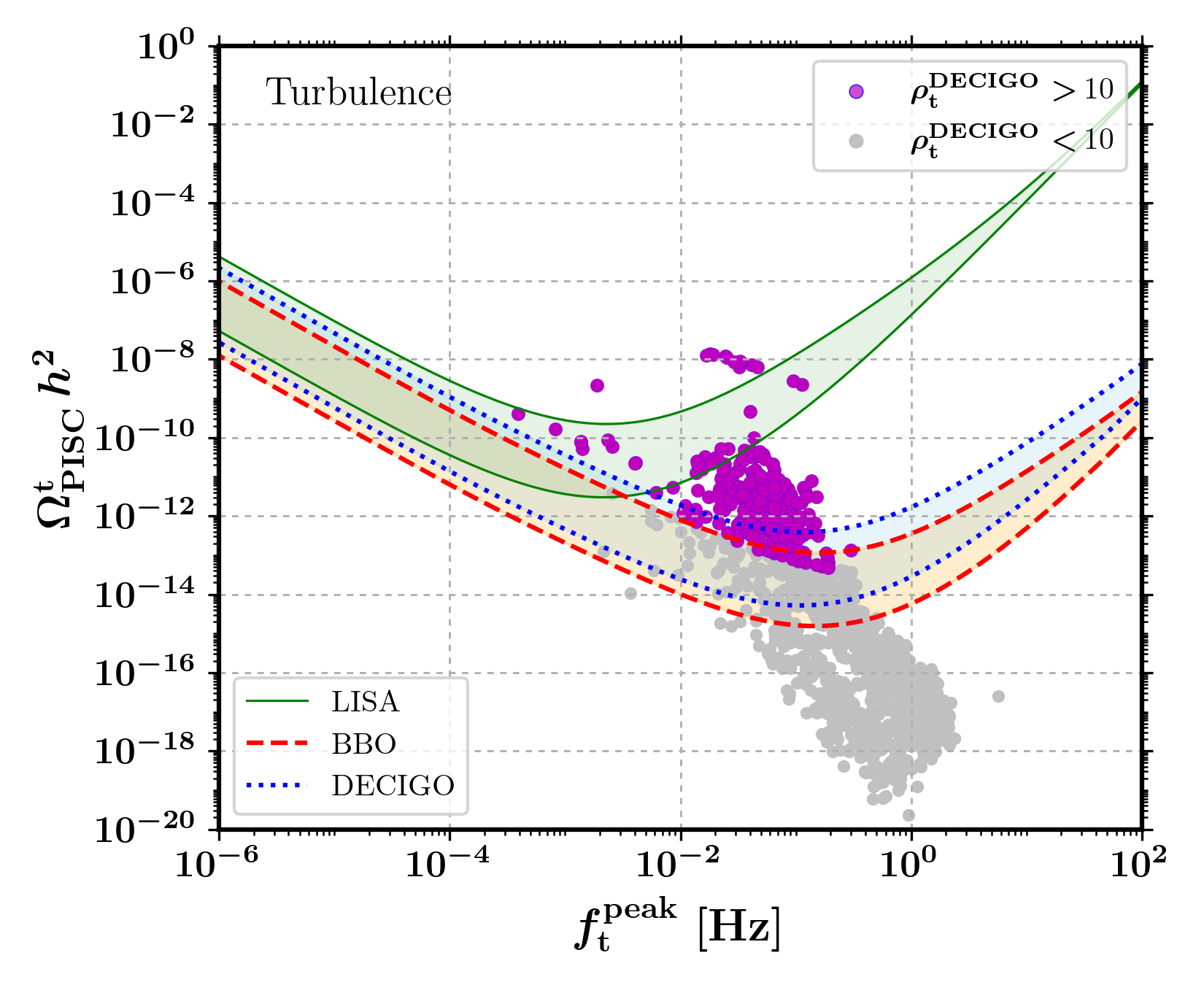}}
	\hspace{0.5cm}
	\subfigure{\includegraphics[height=5.5cm,width=7.0cm]{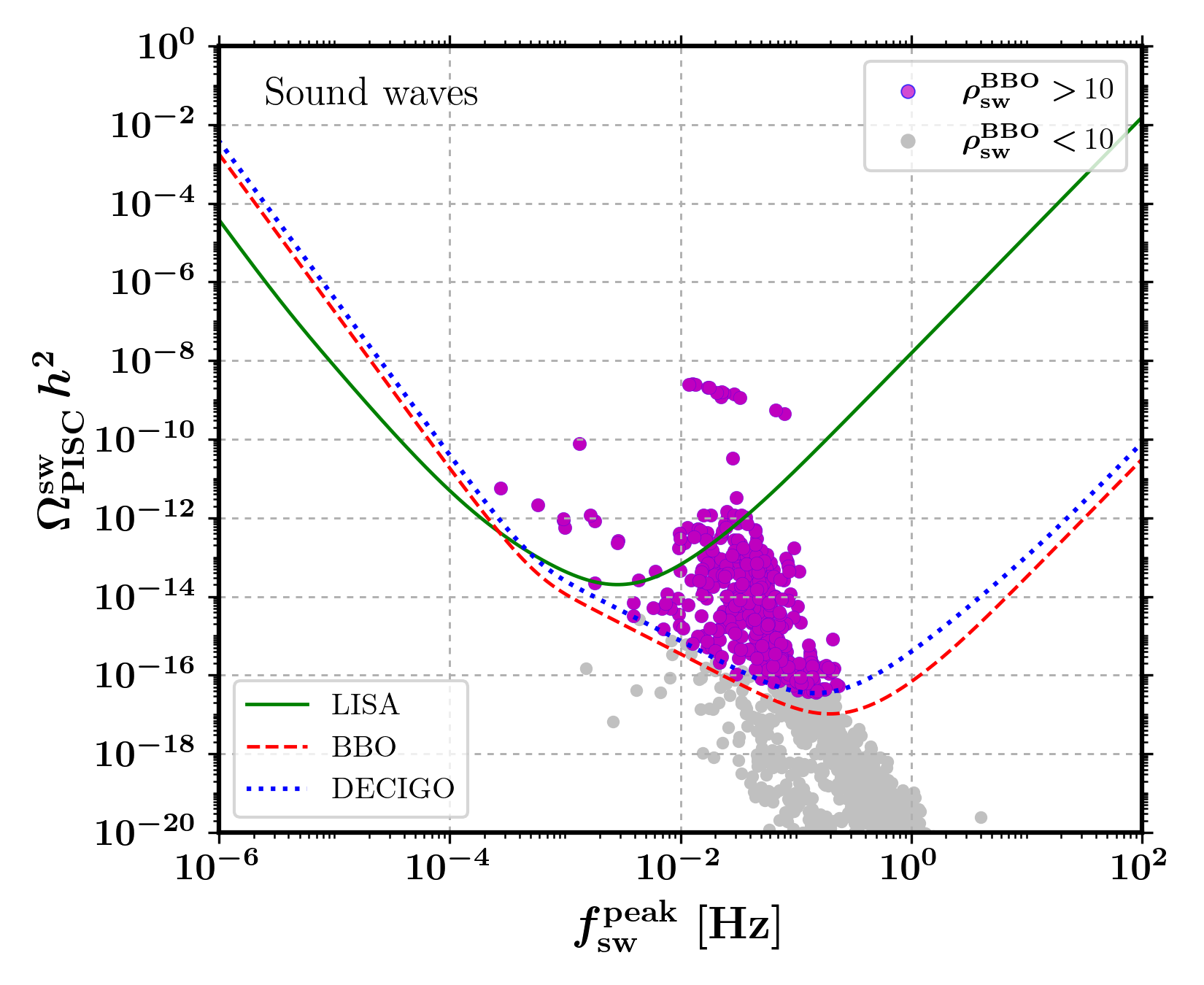}}
	\hspace{0.5cm}
	\subfigure{\includegraphics[height=5.5cm,width=7.0cm]{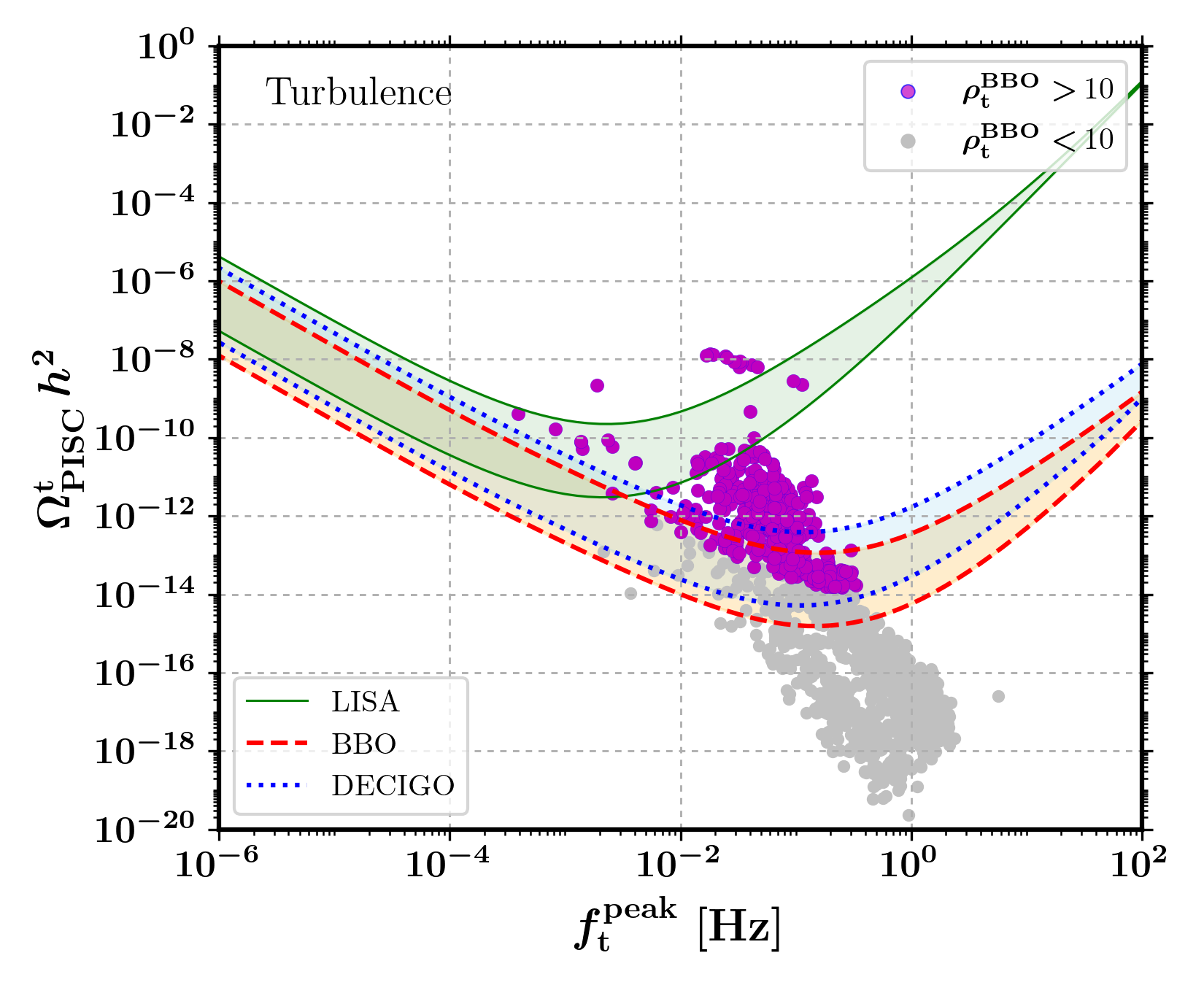}}
	\caption{Plots display the variation of PISC-driven peak amplitudes from Eq.(\ref{eq:GW-result:PICs-eqn}) with peak frequencies for LISA (top panel), DECIGO (middle panel), and BBO (bottom panel). The left (right) panel represents contributions from sound waves (turbulence). Magenta (light-grey) coloured points indicate SNR threshold values above (below) 10. Sensitivity curves for various GW detectors are shown with distinct styles and different colours. All model points satisfy the constraints of Sec.~\ref{sec:obs-cons}, accommodate the $3\sigma$ relic density bound, and evade the DD (at least from the XENON1T) and ID limits. }
	\label{fig:result:GW-PISC}
\end{figure*}

In Fig.~\ref{fig:result:GW-PISC}, we show PISC curves corresponding to contributions from sound waves (left panel) and turbulence (right panel) for various GW detectors, i.e., LISA (top panel), DECIGO (middle panel) and BBO (bottom panel). One can also obtain the remaining $\Omega^{i/j}_{\rm PISCs} h^2$ for a different combination of $i/j$ as a function of the corresponding peak frequency using Eq.~(\ref{eq:GW-result:PICs-integration}). However, our model predicts dominant contributions from sound waves and turbulence, which we highlight in Fig. \ref{fig:result:GW-PISC}. Unlike PLIs, where we need to estimate SNR separately, in the case of PISCs, we only need to verify if a given point lies above any of the PISCs of a particular experiment. In that case, the SNR will naturally exceed the predefined threshold of the respective experiments. Hence, the magenta coloured points represent model predictions surpassing the $\rho_{\rm thr} > 10$ threshold, making them strong candidates for a plausible detection in the future GW experiments. However, light-grey coloured points below the PISC curves require further scrutiny. Such points may still exceed the SNR threshold if the sum of all plausible contributions is greater than the concerned threshold.

Thus, employing both PLIs and PISCs approach, we demonstrate that, in our model, a significant region of the model parameter space, consistent with the viable DM phenomenology and other constraints detailed in Sec. \ref{sec:obs-cons}, predicts detectable GW signals arising from an SFOPT in the singlet field direction. These signals can be probed in the forthcoming GW detectors such as LISA, BBO, and DECIGO. Therefore, our study provides a complementary avenue for exploring the BSM physics, bridging the GW and DM frontiers alongside the collider searches.

\section{Summary and Conclusion}\label{sec:conclu}
In this work, we have explored a two-component DM framework by extending the SM with a hyperchargeless ($Y=0$) inert $SU(2)_L$ scalar triplet $\bm{T}$ and a $SU(2)_L$ complex scalar singlet $S$. The two DM candidates in this setup are the neutral component of the triplet, $T^0$, and a pNGB $\chi$, arising from $S$. The pNGB DM $\chi$, corresponding to the imaginary part of $S$, emerges from the explicit breaking of a global $U(1)$ symmetry by a $\mathbb{Z}_3$-symmetric term in $S$. This symmetry is subsequently broken spontaneously, leaving behind a remnant $\mathbb{Z}_2$-like symmetry, $S \to S^\dagger$, which ensures the stability of $\chi$. The presence of multiple DM components modifies both relic density calculations and constraints from the direct and indirect detections, often relaxing the bounds and improving compatibility with the experimental limits. While each DM species may be in tension with the current searches when considered alone, the combined two-component framework remains viable across a broad region of the parameter space.

To identify the viable parameter space of the model, we performed random scans over the independent input parameters and analysed their correlations and phenomenological consequences. Our study of the DM phenomenology is divided into two regimes, $\textbf{R-I}$ and $\textbf{R-II}$, categorized by the mass hierarchy between the two DM components: $m_{\chi} > m_{T^0}$ and $m_{\chi} < m_{T^0}$, respectively. In $\textbf{R-I}$, we find that DM-DM conversion processes can significantly enhance the relic abundance of $T^0$, allowing it to contribute up to $50 – 60\%$ of the total observed relic density within a sub-TeV mass range. This marks a substantial improvement over the pure $Y=0$ ITM scenario, where such contributions are severely suppressed, at most $10 - 20\%$. In contrast, in $\textbf{R-II}$, the scalar triplet DM behaves similarly to that in the pure ITM scenario, while the pNGB $\chi$ emerges as the dominant contributor to the relic density throughout the viable model parameter space. Notably, in both the regimes, the presence of the additional DM component helps to revive (although partially) the otherwise excluded ``desert'' region, i.e., $300~\text{GeV} \lesssim m_{T^0} \lesssim 1500~\text{GeV}$, in this chosen framework, with $T^0$ as the sole DM candidate. Moreover, a large range of parameter regions remain viable in the chosen model that are consistent with various current (or projected) DD and ID experiments. Needless to mention, the aforementioned parameter regions are also consistent with a range of theoretical requirements and experimental constraints, including those from collider searches, precision Higgs measurements, etc.

Having extended the scalar sector of the SM, we also computed the finite-temperature effective potential and examined its implications for an SFOEWPT in the early Universe. In the present study, we focused on a scenario where there is an SFOPT in the singlet field direction $s$, i.e., the CP-even component of $S$, and we further investigated its associated GW signals. The PT along the SM Higgs field direction, however, remain weaker in our phenomenologically viable model parameter space to generate a detectable GW spectrum. To analyse the dynamics of the PT, we employed two approaches (i) the HT expansion, which is the simplest gauge invariant manifestation of the effective potential and (ii) thermal evolution of the full effective potential $V^T_{\rm eff}$, which contains gauge dependency. However, the presence of a tree-level barrier in the chosen framework renders the gauge dependence of the PT observables negligibly small, when the full $V^T_{\rm eff}$ is computed in Landau gauge. Our analysis revealed that, considering the full $V^T_{\rm eff}$ yields more feasible parameter space over the HT expansion method that can accommodate an SFOPT along the $s$-direction. Moreover, the HT suggests $500~\text{GeV} \lesssim v_S \lesssim 1800~\text{GeV}$ for a successful PT, whereas, in the full $V^T_{\rm eff}$ analysis, the upper limit is reduced to $v_S \lesssim 900$ GeV and are allowed up to $\sim 200$ GeV. Additionally, we observed that the favourable model points that can support an SFOPT along the $s$-direction are typically concentrated within the mass window $m_{h_2}/2 \lesssim m_{\chi} \lesssim m_{h_2}$ and are clustered near to the $m_\chi = 2 m_{h_2}$ threshold. We further demonstrated that a significant number of model points supporting an SFOPT along the $s$-direction and yielding the correct DM relic abundance can be probed in the current or upcoming DD experiments, such as LZ-2024 and DARWIN. The associated GW signals from these points also lie within the projected sensitivity reaches of the future space-based interferometers, including LISA, BBO, and DECIGO.

The prospects for detecting GW signals can be significantly enhanced using the recently proposed PISCs, which evaluate the signal-to-noise ratio more effectively by incorporating both the peak amplitude and peak frequency based on an assumed signal shape. For completeness and comparison, we have, nevertheless, also presented results using the traditional PLIs-based analyses. Unlike PLIs, PISCs offer a more precise and informative assessment of the GW detectability at a specific interferometer, such as LISA, BBO, or DECIGO, while also enabling a clearer comparison between the different underlying GW generation mechanisms.

In summary, our analysis of the chosen two-component DM framework demonstrates that GWs originating from an SFOPT offer a complementary avenue for probing BSM physics permissible by various possible theoretical and experimental limits, beyond the conventional approaches of relic density measurements, direct/indirect detection, and collider searches.

\section*{Acknowledgements}
P. B. acknowledges the financial support from the Indian Institute of Technology, Delhi (IITD) and as a Senior Research Fellow and {\it{Hydra}} high-performance computing facility at the MS-516 HEP-PH Laboratory, Department of Physics, IITD.
P. G. acknowledges the IITD SEED grant support IITD/Plg/budget/2018-2019/21924, continued as IITD/Plg/budget/2019-2020/173965, IITD Equipment Matching Grant IITD/ IRD/MI02120/ 208794, Start-up Research Grant (SRG) support SRG/2019/000064, and Core Research Grant (CRG) support CRG/2022/002507 from the Science and Engineering Research Board (SERB), Department of Science and Technology, Government of India.


\appendix
\section{Field dependent and thermal masses}\label{appx:appendixA}
\subsection{Field dependent masses}\label{appx:A1}
The field-dependent scalar mass matrix at zero temperature is derived from the tree-level potential $V_0(\varphi)$ (see Eq.~(\ref{eq:pot:tree-level})),
\bea
\widetilde{\mathcal{M}}^2_{ij}(\varphi) = \frac{\partial^2 V_0(\varphi)}{\partial \varphi_i \partial \varphi_j}.
\eea
In the basis $\varphi = (\varphi_1,\,\varphi_2,\,\varphi_3) \equiv \{h, s, T^0\}$, the independent elements of the symmetric $3 \times 3$ scalar squared mass matrix are,
\begingroup
\allowdisplaybreaks
\bea
\label{eq:appx:field-dependent-scalar-neutral}
\widetilde{\mathcal{M}}^2_{11} &=& \mu_{H}^2 + 3 \l_{H} h^2 + \frac{1}{2} \l_{SH} s^2 + \frac{1}{2} \l_{HT} {T^0}^2, \nn \\
\widetilde{\mathcal{M}}^2_{12} &=& \l_{SH} s h, \nn \\
\widetilde{\mathcal{M}}^2_{13} &=& \l_{HT} h T^0, \nn \\
\widetilde{\mathcal{M}}^2_{22} &=& \mu_S^2 + \frac{3\sqrt{2}}{2} \mu_3 s + 3 \l_S s^2 + \frac{1}{2} \l_{SH} h^2 + \frac{1}{2} \l_{ST} {T^0}^2, \nn \\
\widetilde{\mathcal{M}}^2_{23} &=& \l_{ST} s T^0, \nn \\
\widetilde{\mathcal{M}}^2_{33} &=& \mu_{\bm{T}}^2 + 3 \l_{\bm{T}} {T^0}^2 + \frac{1}{2} \l_{ST} s^2 + \frac{1}{2} \l_{HT} h^2.
\eea
\endgroup
The above $3 \times 3$ squared mass matrix, at the zero temperature EW vacuum, i.e., $\{v, v_s, 0\}$, after using the corresponding tadpole equations reduces to a $2 \times 2$ matrix in the $\{h,s\}$ basis and matches the mass matrix $\mathcal{M}^2$ given in Eq.(\ref{eq:masses:scalar-mass-HS-new}). With the triplet VEV set to zero, $\widetilde{\mathcal{M}}^2_{33}$ decouples from the matrix in Eq.~(\ref{eq:appx:field-dependent-scalar-neutral}) and corresponds to $m^2_{T^0}$, as given in Eq.~(\ref{eq:mass-triplet}). The field-dependent scalar masses $m_{h_1}(\varphi)$, $m_{h_2}(\varphi)$, and $m_{T^0}(\varphi)$ follow from the eigenvalues of the $3 \times 3$ matrix. As the analytic expressions are lengthy and cumbersome, we refrain from presenting them here. Instead, the eigenvalues of the scalar mass matrix are computed numerically using {\tt Python} routines. The field-dependent masses of the pNGB DM ($\chi$), charged triplet scalar (${T^{\pm}}$) and the SM Goldstone bosons ($G^{0,\pm}$) are given as 
\bea
\label{eq:appx:field-dependent-pNGB-and-charged}
m^2_{\chi} (\varphi) &=& \mu_S^2 - \frac{3\sqrt{2}}{2} \mu_3 s + \l_S s^2 + \frac{1}{2} \l_{SH} h^2 + \frac{1}{2} \l_{ST} {T^0}^2, \nn \\
m^2_{T^{\pm}} (\varphi) &=& \mu_{\bm{T}}^2 + \l_{\bm{T}} {T^0}^2 + \frac{1}{2} \l_{ST} s^2 + \frac{1}{2} \l_{HT} h^2, \nn \\
m^2_{G^{0,\pm}} (\varphi) &=& \mu_{H}^2 + \l_{H} h^2 + \frac{1}{2} \l_{SH} s^2 + \frac{1}{2} \l_{HT} {T^0}^2.
\eea
Note that, with a vanishing triplet VEV, at the tree-level we have $\widetilde{\mathcal{M}}^2_{33} \equiv m^2_{T^0} = m^2_{T^\pm}$, as of Eq.~(\ref{eq:mass-triplet}). Among the SM fermions, the top quark provides the dominant contribution due to its large Yukawa coupling ($y_t$). Accordingly, we include only the top quark contribution in this analysis, neglecting all others. The corresponding zero-temperature field-dependent mass is given by:
\bea
m^2_t(\varphi) = \frac{y_t^2}{2} h^2.
\eea
Finally, the field-dependent masses of the EW gauge bosons are \cite{Niemi:2020hto}
\bea
m^2_W(\varphi) = \frac{g_2^2}{4} (h^2+4 {T^0}^2), ~~m^2_Z(\varphi) = \frac{g_1^2 + g_2^2}{4} h^2,
\eea
with $g_1,g_2$ being the SM $U(1)_Y,~SU(2)_L$ gauge couplings, respectively.
\subsection{Thermal masses}\label{appen:A2}
As mentioned earlier in subsection~\ref{subsec:PT-GW:thermal-pot}, resumming the leading self-energy (daisy) diagrams modifies the field-dependent masses as follows:
\bea\label{eq:field_mass:daisy-coeff}
m_i^2 (\varphi, T) = m^2_i(\varphi) + \Pi_i (T), ~~\text{where}~~ \Pi_i (T)=c_i T^2,
\eea
where $\Pi_i(T)$ denotes the thermal mass correction for the $i^{\text{th}}$ bosonic {\it d.o.f}, and $c_i$ are the corresponding Daisy coefficients \cite{Dolan:1973qd, Kirzhnits:1976ts, Parwani:1991gq, Espinosa:1992gq, Arnold:1992rz, Croon:2020cgk, Schicho:2021gca}. The non-zero thermal mass contributions in this chosen model are given by
\bea
\label{eq:appx:thermal-masses}
\Pi_{h_1}(T) &=& \left(\frac{g_1^2}{16} + \frac{3g_2^2}{16} + \frac{y_t^2}{4} + \frac{\l_{H}}{2} + \frac{\l_{SH}}{12} + \frac{\l_{HT}}{8}\right)~T^2, \nn \\
\Pi_{h_2}(T) &=& \left(\frac{\l_{S}}{3} + \frac{\l_{SH}}{6} + \frac{\l_{ST}}{8}\right)~T^2 = \Pi_{\chi}(T),  \nn \\
\Pi_{T^0}(T) &=& \left(\frac{\l_{HT}}{6} + \frac{\l_{ST}}{12} + \frac{5 \l_{\bm{T}}}{12}\right)~T^2 = \Pi_{T^{\pm}}(T), ~~{\rm and}, \nn \\
\Pi_{G^0}(T) &=& \Pi_{G^{\pm}}(T) = \Pi_{h_1}(T). 
\eea
The expressions within the parentheses in Eq.~(\ref{eq:appx:thermal-masses}) correspond to the Daisy coefficients defined in Eq.~(\ref{eq:field_mass:daisy-coeff}). At finite temperature ($T \neq 0$), the longitudinal components of the $W$ and $Z$ bosons, as well as the photon ($\gamma$), also acquire additional thermal mass corrections. For the $W$ boson, the thermal mass shift is given as \cite{Niemi:2018asa}:
\bea
m_W^2 (\varphi, T) = m_W^2 (\varphi) + \Pi_W (T), ~~~\text{with}~~\Pi_W (T) = \frac{13}{6} g_2^2 T^2,
\eea
whereas, for the $Z$ boson and the photon, the thermal corrections are determined by the eigenvalues of the squared mass matrix given below \cite{Niemi:2018asa, Beniwal:2018hyi, Niemi:2020hto}, 
\bea
m^2_{Z,\gamma}(\varphi, T) =
\begin{pmatrix}
	~\frac{1}{4}g_2^2 h^2 + \frac{11}{6} g_2^2 T^2~ & ~ -\frac{1}{4} g_1 g_2 h^2 ~ \\
	~-\frac{1}{4} g_1 g_2 h^2 ~ & ~ \frac{1}{4}g_1^2 h^2 + \frac{11}{6} g_1^2 T^2~
\end{pmatrix}.
\eea

	\bibliographystyle{JHEP}
	\bibliography{ref-v1}  
	
\end{document}